\documentclass[prd,preprintnumbers,twocolumn,amsmath,nofootinbib,amssymb]{revtex4}
\usepackage{graphicx,color,dcolumn,booktabs,bm}
\usepackage{longtable,lscape}
\usepackage{tabularx}
\usepackage{txfonts}
\usepackage{overpic}
\usepackage{amssymb}
\usepackage{epstopdf}
\usepackage{indentfirst}
\usepackage{feynmf}   %{feynmp}
\usepackage{slashed}  %for Feynman symbols
\usepackage{cases}
\usepackage{color}
\usepackage{float}
\usepackage{multirow}
\usepackage{ulem}
\usepackage{appendix}
\usepackage{graphicx,color,dcolumn,booktabs,bm}
\usepackage{epsfig,dsfont,amssymb,amsmath,amsfonts,amsbsy,mathrsfs}
\graphicspath{{Figures/}} %

\usepackage{hyperref}
\hypersetup{colorlinks,citecolor=blue,anchorcolor=red,menucolor=red, linkcolor=red,filecolor=red,runcolor=red,urlcolor=blue,frenchlinks=red}

\makeatletter
\@addtoreset{equation}{section}
\makeatother
\renewcommand{\theequation}{\arabic{section}.\arabic{equation}}
\allowdisplaybreaks

\begin{document}

\title{Hunting for the prospective $T_{cc}$ family based on the diquark-antidiquark configuration}
\author{Wen-Chao Dong$^{1,2}$}
\email{wen-chao$_$dong@qq.com}
\author{Zhi-Gang Wang$^{1}$\footnote{Corresponding author}}
\email{zgwang@aliyun.com}
\affiliation{$^1$Department of Physics, North China Electric Power University, Baoding 071003, China\\
$^2$School of Nuclear Science and Engineering, North China Electric Power University, Beijing 102206, China}

\begin{abstract}
Inspired by the first $T_{cc}$ observation at the LHCb Collaboration, the spectroscopic properties of the entire isoscalar and isovector $T_{cc}$ family are systematically investigated by means of multiple sorts of relativized and nonrelativistic diquark formalisms, which include the Godfrey-Isgur relativized diquark model, the modified Godfrey-Isgur relativized diquark model incorporating the color screening effects, the nonrelativistic diquark model with the Gaussian type hyperfine potential, and the nonrelativistic diquark model with the Yukawa type hyperfine potential. In terms of the $1S$-wave double-charm tetraquark state with $I(J^P)=0(1^+)$, the predicted masses of most diquark-antidiquark scenarios are somewhat higher than the observed value of the $T_{cc}(3875)^+$ structure. In light of the diquark-antidiquark configuration, this work unveils the mixing angles of the orbitally excited isovector $T_{cc}$ states and the magic mixing angles of the ideal heavy-light tetraquarks for the first time. As the advancement of the experimental detection capability, these phenomenological predictions will effectively boost the hunting for the prospective low-lying $T_{cc}$ states in the future.
\end{abstract}

\maketitle

\section{Introduction}\label{sec1}

Over the past two decades, a cornucopia of heavy flavored exotic hadrons gradually emerged from diverse particle detection facilities conducted by worldwide experimental collaborations \cite{ParticleDataGroup:2024cfk,Olsen:2017bmm}. Nonetheless, it is rather arduous to cognize the authentic nature of these novel hadron states. A well-known specimen is the hidden-charm $\chi_{c1}(3872)$ state, whose composition is identified as the conventional charmonium $c\bar c$, the hybrid charmonium $c\bar cg$, the charmoniumlike tetraquark/molecule $c\bar cq\bar q$, the mixture of them, or other configurations by discrepant arguments \cite{Olsen:2017bmm,Hosaka:2016pey,Ali:2017jda,Lebed:2016hpi,Esposito:2016noz,Guo:2017jvc,Karliner:2017qhf}. Conspicuously, the advent of the $\chi_{c1}(3872)$ state was a milestone in hadron physics, which incited multifarious phenomenological explorations beyond the conventional quark model. Hitherto, there are all sorts of theoretical propositions to construe the internal structure of exotic hadron states \cite{Olsen:2017bmm,Hosaka:2016pey,Ali:2017jda,Lebed:2016hpi,Esposito:2016noz,Guo:2017jvc,Karliner:2017qhf,Brambilla:2019esw,Chen:2022asf}, comprising hybrid hadrons, hadro-quarkonia, compact multiquarks, hadronic molecules, kinematical effects, and so forth. Regrettably, none of them are predominant to overwhelm the rest of the plausible perspectives on the so-called exotics, which substantiates that both experimental and theoretical further efforts are indispensable so as to disentangle the mysterious nature of nonstandard heavy hadrons.

Nowadays, quite a large portion of heavy flavored exotic hadrons are the states with hidden heavy flavors, e.g., the heavy quarkoniumlike $T_\psi/T_\Upsilon$ states, the fully charmed tetraquark $T_{\psi\psi}$ states, and the hidden charmed pentaquark $P_\psi$ states \cite{Brambilla:2019esw}. By contrast, the open heavy flavored exotic hadrons established by various experiments are fairly rare, merely touching upon the singly and doubly charmed tetraquark $T_{cs}/T_{c\bar s}/T_{cc}$ states \cite{Chen:2022asf}. Hence, the quest for the ground and excited states of the double-charm tetraquark $T_{cc}$ family is momentous to ascertain the properties of the exotic hadrons with open heavy flavors. In 2021, a very narrow peaking structure, marked as $T_{cc}(3875)^+$, was clearly detected in the $D^0D^0\pi^+$ invariant mass distribution by the LHCb Collaboration \cite{LHCb:2021vvq,LHCb:2021auc}, which signified the first observation of the exotic states with $cc\bar{u}\bar{d}$ quark component. In consideration of the absence of the double-charm signal in the $D^+D^0\pi^+$ invariant mass distribution, the $T_{cc}(3875)^+$ structure is presumably an isoscalar state. Manifestly, the observed mass of the narrow $T_{cc}(3875)^+$ state is in close proximity to the $D^{*+}D^0$ mass threshold. Thereupon, the spin-parity properties of the $T_{cc}(3875)^+$ state are speculated as $J^P=1^+$ by postulating that the relative angular momentum in the $D^*D$ pair is $S$-wave. The experimental mass value of the $T_{cc}(3875)^+$ state is expressed as
\begin{eqnarray}
m_{T_{cc}(3875)^+}&=&m_{D^{*+}}+m_{D^0}+\delta m,\label{eq11}
\end{eqnarray}
where $m_{D^{*+}}$, $m_{D^0}$, and $\delta m$ denote the observed mass of the $D^{*+}$ meson, the observed mass of the $D^0$ meson, and the binding energy of the $T_{cc}(3875)^+$ state with respect to the $D^{*+}D^0$ mass threshold, respectively. For the sake of the determination of the binding energy and the decay width of the $T_{cc}(3875)^+$ structure, the LHCb Collaboration employed two sorts of Breit-Wigner (BW) parametrization schemes, i.e., the generic BW parametrization and the unitarised BW parametrization. In generic case \cite{LHCb:2021vvq}, a relativistic $P$-wave two-body BW function with a Blatt-Weisskopf form factor was adopted to attain the binding energy $\delta m_\text{BW}$ and the decay width $\Gamma_\text{BW}$, i.e.,
\begin{eqnarray}
\delta m_\text{BW}&=&-273\pm61\pm5^{+11}_{-14}~\text{keV},\label{eq12}\\
\Gamma_\text{BW}&=&410\pm165\pm43^{+18}_{-38}~\text{keV}.\label{eq13}
\end{eqnarray}
Moreover, the LHCb Collaboration extracted the position of the amplitude pole of the narrow $T_{cc}(3875)^+$ peaking structure on the second Riemann sheet, by making use of a unitarised three-body BW function and taking into account a novel model of the detector mass resolution \cite{LHCb:2021auc}. The parameters $\delta m_\text{pole}$ and $\Gamma_\text{pole}$ of the $T_{cc}(3875)^+$ pole were determined as
\begin{eqnarray}
\delta m_\text{pole}&=&-360\pm40^{+4}_{-0}~\text{keV},\label{eq14}\\
\Gamma_\text{pole}&=&48\pm2^{+0}_{-14}~\text{keV}.\label{eq15}
\end{eqnarray}
It is evident to be conscious of that the decay width turned up in the imaginary part of the $T_{cc}(3875)^+$ pole is much tinier than its counterpart acquired by the generic BW parametrization, although the binding energies obtained by two sorts of BW parametrization schemes are somewhat near. Thus, the visible information of the $T_{cc}(3875)^+$ structure is kind of scant compared to the famed $XYZ$ states, which entails further experimental explorations and theoretical investigations of the exotic $cc\bar{u}\bar{d}$ states.

Taking the perspectives of phenomenological theories, the delving of the multiquark configuration is commenced by the seminal works of the quark model \cite{Gell-Mann:1964ewy,Zweig:1964ruk}. In terms of the open-flavor tetraquark states that carry two heavy quarks, a few pioneering surveys unraveled their structural stabilities and existential feasibilities \cite{Ader:1981db,Ballot:1983iv,Zouzou:1986qh,Manohar:1992nd,Lipkin:1986dw}. In 1986, the earliest theoretical inquiry with regard to $T_{cc}$ spectroscopy rendered an upper bound on the mass ambit of the lowest isoscalar $cc\bar{u}\bar{d}$ state with $J^P=1^+$, by virtue of a model-independent approach \cite{Lipkin:1986dw}. Subsequently, the various properties of the $cc\bar{u}\bar{d}$ states, e.g., mass spectra, decay properties, and production mechanisms, were extensively probed by Refs. \cite{Carlson:1987hh,Semay:1994ht,Pepin:1996id,Gelman:2002wf,Vijande:2003ki,Janc:2004qn,Ebert:2007rn,Navarra:2007yw,Vijande:2009kj,Lee:2009rt,Yang:2009zzp,
Ohkoda:2012hv,Li:2012ss,Du:2012wp,Feng:2013kea,Ikeda:2013vwa,Luo:2017eub,Mehen:2017nrh,Karliner:2017qjm,Eichten:2017ffp,Hyodo:2017hue,Wang:2017uld,Wang:2017dtg,
Yan:2018gik,Xing:2018bqt,Zhu:2019iwm,Liu:2019stu,Maiani:2019lpu,Park:2018wjk,Agaev:2019qqn,Junnarkar:2018twb,Yang:2019itm,Tang:2019nwv,Lu:2020rog,Ding:2020dio,
Deng:2018kly,Tan:2020ldi,Wallbott:2020jzh,Jin:2021cxj,Feijoo:2021ppq,Qin:2020zlg,Cheng:2020wxa,Noh:2021lqs,Wang:2021yld,Azizi:2021aib,Zhang:2021yul,Braaten:2020nwp,
Chen:2021vhg,Huang:2021urd,Meng:2020knc,Fleming:2021wmk,Meng:2021jnw,Hu:2021gdg,Dong:2021bvy,Wang:2021ajy,Xin:2021wcr,Ling:2021bir,Du:2021zzh,Kim:2022mpa,
Weng:2021hje,Chen:2021tnn,Agaev:2022ast,Liu:2019yye,Albuquerque:2022weq,Zhao:2021cvg,Guo:2021yws,Gao:2020bvl,Agaev:2021vur,Ke:2021rxd,Dai:2021vgf,Yan:2021wdl,
Chen:2021cfl,Deng:2021gnb,Albaladejo:2021vln,Wang:2022clw,Braaten:2022elw,Deng:2022cld,Dai:2021wxi,Padmanath:2022cvl,Liu:2023vrk,Maiani:2022qze,Meng:2023for,
Song:2023izj,Ortega:2022efc,Noh:2023zoq,Albuquerque:2023rrf,He:2023ucd,Wang:2022jop,Dai:2023mxm,Meng:2023jqk,Lyu:2023xro,Wu:2022gie,Asanuma:2023atv,Sakai:2023syt,
Qiu:2023uno,Ma:2023int,Kucab:2024nkv,Mutuk:2023oyz,Chen:2023web,Wang:2024vjc,Park:2024cic,Li:2023wug,Mutuk:2024vzv,Meng:2024yhu,Sonnenschein:2024rzw,Lebed:2024zrp} with a lot of theoretical prescriptions, including the MIT bag model \cite{Carlson:1987hh,Zhang:2021yul}, the constituent quark model (CQM) \cite{Semay:1994ht,Pepin:1996id,Gelman:2002wf,Vijande:2003ki,Janc:2004qn,Ebert:2007rn,Vijande:2009kj,Lee:2009rt,Yang:2009zzp,Luo:2017eub,Karliner:2017qjm,
Hyodo:2017hue,Yan:2018gik,Xing:2018bqt,Zhu:2019iwm,Park:2018wjk,Yang:2019itm,Lu:2020rog,Deng:2018kly,Tan:2020ldi,Cheng:2020wxa,Noh:2021lqs,Meng:2020knc,Kim:2022mpa,
Weng:2021hje,Chen:2021tnn,Guo:2021yws,Deng:2021gnb,Wang:2022clw,Deng:2022cld,Liu:2023vrk,Meng:2023for,Ortega:2022efc,Noh:2023zoq,He:2023ucd,Meng:2023jqk,Wu:2022gie,
Ma:2023int,Mutuk:2023oyz,Wang:2024vjc,Park:2024cic,Li:2023wug,Meng:2024yhu}, the QCD sum rules (QSR) \cite{Navarra:2007yw,Du:2012wp,Wang:2017uld,Wang:2017dtg,Agaev:2019qqn,Tang:2019nwv,Azizi:2021aib,Xin:2021wcr,Agaev:2022ast,Albuquerque:2022weq,Gao:2020bvl,
Agaev:2021vur,Albuquerque:2023rrf}, the adiabatic Born-Oppenheimer (BO) approximation \cite{Maiani:2019lpu,Maiani:2022qze,Mutuk:2024vzv,Lebed:2024zrp}, the one-boson exchange (OBE) model \cite{Ohkoda:2012hv,Li:2012ss,Liu:2019stu,Wang:2021yld,Chen:2021vhg,Dong:2021bvy,Wang:2021ajy,Asanuma:2023atv,Sakai:2023syt,Qiu:2023uno}, the Bethe-Salpeter (BS) equation \cite{Feng:2013kea,Ding:2020dio,Wallbott:2020jzh,Zhao:2021cvg,Ke:2021rxd}, the heavy quark symmetry (HQS) \cite{Mehen:2017nrh,Eichten:2017ffp,Braaten:2020nwp}, the lattice QCD (LQCD) \cite{Ikeda:2013vwa,Junnarkar:2018twb,Padmanath:2022cvl,Lyu:2023xro}, the coupled-channel approach \cite{Feijoo:2021ppq,Huang:2021urd,Ling:2021bir,Du:2021zzh,Dai:2021vgf,Albaladejo:2021vln,Dai:2021wxi}, the effective field theory (EFT) framework \cite{Fleming:2021wmk,Meng:2021jnw,Yan:2021wdl,Chen:2021cfl,Braaten:2022elw,Wang:2022jop,Dai:2023mxm}, the chiral quark-soliton model ($\chi$QSM) \cite{Kucab:2024nkv}, the Regge trajectory relation \cite{Song:2023izj,Chen:2023web}, the holographic QCD (HQCD) \cite{Liu:2019yye,Sonnenschein:2024rzw}, and so on \cite{Jin:2021cxj,Qin:2020zlg,Hu:2021gdg}. Lately, following the experimental establishment of the $T_{cc}(3875)^+$ state \cite{LHCb:2021vvq,LHCb:2021auc}, sundry phenomenological expositions for its intrinsic configuration were put forward, containing the compact tetraquark (with the diquark-antidiquark configuration) \cite{Jin:2021cxj,Qin:2020zlg,Zhang:2021yul,Weng:2021hje,Albuquerque:2022weq,Guo:2021yws,Gao:2020bvl,Agaev:2021vur,Liu:2023vrk,Maiani:2022qze,Meng:2023for,Noh:2023zoq,
Wu:2022gie,Mutuk:2023oyz,Wang:2024vjc,Park:2024cic,Li:2023wug,Sonnenschein:2024rzw}, the hadronic molecule (with the meson-meson configuration) \cite{Feijoo:2021ppq,Chen:2021vhg,Huang:2021urd,Fleming:2021wmk,Meng:2021jnw,Hu:2021gdg,Dong:2021bvy,Xin:2021wcr,Ling:2021bir,Du:2021zzh,Chen:2021tnn,Zhao:2021cvg,
Ke:2021rxd,Chen:2021cfl,Deng:2021gnb,Albaladejo:2021vln,Braaten:2022elw,Deng:2022cld,Ortega:2022efc,Wang:2022jop,Dai:2023mxm,Meng:2023jqk,Asanuma:2023atv,
Sakai:2023syt,Qiu:2023uno,Ma:2023int}, the tetraquark-molecule mixing (with an admixture of diquark-antidiquark and meson-meson configurations) \cite{Yan:2021wdl,Song:2023izj,He:2023ucd,Lebed:2024zrp}, the virtual state (with the pole in the real axis below threshold) \cite{Dai:2021wxi,Padmanath:2022cvl,Lyu:2023xro}, the Efimov state (with the meson-meson-sphaleron configuration) \cite{Liu:2019yye}, etc. Consequently, the authentic nature of the $T_{cc}(3875)^+$ structure has been equivocal till now due to the dearth of the phenomenological smoking gun.

As the generalised case of the $cc\bar{u}\bar{d}$ states, the double-heavy tetraquarks $Q_1Q_2\bar{q}_1\bar{q}_2$ constituted by two heavy quarks $Q$ and two light antiquarks $\bar{q}$ are the principal candidates of the long-lived exotic states, since they are relatively stable against the strong decays \cite{Karliner:2017qjm}. As is well known, the doubly heavy diquark $Q_1Q_2$ with antitriplet color can be approximated as a point-like heavy antiquark $\bar{Q}$ when the mass of the heavy quark is large enough, which implies the potential existence of such doubly heavy tetraquark states \cite{Eichten:2017ffp}. Recently, the QCD version of the hydrogen bond was employed to regard the $cc$ quark pair as the heavy color sources, successfully acquiring the mass that conformed to the exotic $T_{cc}(3875)^+$ structure \cite{Maiani:2022qze}. In addition, the isoscalar $T_{cc}$ tetraquark composed of the color-antitriplet diquark $cc$ and the color-triplet antidiquark $\bar{u}\bar{d}$ was considered a state whose mass well reproduced the experimental result of the $T_{cc}(3875)^+$, in light of the heavy antiquark-diquark symmetry (HADS) \cite{Wu:2022gie}. Therefore, spurred on by the first experimental discovery of the doubly charmed tetraquark \cite{LHCb:2021vvq,LHCb:2021auc}, the theoretical study on the ground and excited $cc\bar{u}\bar{d}$ states with the diquark-antidiquark configuration is undoubtedly of great phenomenological significance \cite{Lipkin:1986dw,Karliner:2017qjm,Eichten:2017ffp,Maiani:2022qze,Wu:2022gie}. In order to shed light on the spectroscopic properties of the $T_{cc}$ family, this work takes advantage of several diquark-antidiquark scenarios, involving the Godfrey-Isgur (GI) relativized diquark model, the modified Godfrey-Isgur (MGI) relativized diquark model (incorporating the color screening effects), and the nonrelativistic (NR) diquark models. The synopsis of this paper is organized as follows. At the beginning, the experimental and theoretical status quo of the double-charm tetraquark states is revisited in Section \ref{sec1}. Next, the corresponding diquark-antidiquark scenarios are explicated in Section \ref{sec2}. Whereafter, Section \ref{sec3} exhibits the predicted outcomes with regard to the low-lying $T_{cc}$ spectroscopy. Furthermore, concerning the mass spectra, Regge trajectories, and mixing angles of the $T_{cc}$ states, Section \ref{sec4} discusses the connection between this work and other approaches. In the end, Section \ref{sec5} lays out a concise summary of this work.

\section{Formalism}\label{sec2}

In this section, the Godfrey-Isgur (GI) relativized diquark model, the modified Godfrey-Isgur (MGI) relativized diquark model with the color screening effects, and the nonrelativistic (NR) diquark models whose hyperfine interaction potentials are of the Gaussian or Yukawa forms are limned for the sake of pinning down the spectroscopic properties of the low-lying $cc\bar{u}\bar{d}$ states.

\subsection{Godfrey-Isgur (GI) relativized diquark model}\label{subsec21}

As one of the most eminent version among all sorts of quark models, the Godfrey-Isgur relativized quark model (GI model) proposed by S. Godfrey and N. Isgur is capable of successfully depicting the mass spectra of nearly all types of mesons that consist of light or heavy (anti)quarks \cite{Godfrey:1985xj}. Fabulously, the GI model not only attained the universality of the parameters in the one-gluon-exchange-plus-linear-confinement potential stimulated by QCD, but also embraced pivotal relativistic effects. There was no doubt that the GI model manifested that ``\textit{all mesons---from the pion to the upsilon---can be described in a unified framework}'', as mentioned by Ref. \cite{Godfrey:1985xj}. Up to now, the spectroscopic properties of light mesons \cite{Godfrey:1985xj}, singly heavy mesons \cite{Godfrey:1985xj,Godfrey:2016nwn,Godfrey:2015dva}, doubly heavy mesons \cite{Godfrey:1985xj,Godfrey:2004ya,Barnes:2005pb,Godfrey:2015dia}, light baryons \cite{Capstick:1986ter}, singly heavy baryons \cite{Capstick:1986ter,Lu:2016ctt}, doubly heavy baryons \cite{Lu:2017meb}, triply heavy baryons \cite{Bedolla:2019zwg}, light tetraquarks \cite{Lu:2019ira}, open heavy tetraquarks \cite{Lu:2016zhe}, hidden heavy tetraquarks \cite{Lu:2016cwr}, fully heavy tetraquarks \cite{Dong:2022sef}, and diquarks \cite{Lu:2016ctt,Bedolla:2019zwg,Lu:2019ira,Lu:2016zhe,Lu:2016cwr,Dong:2022sef,Ferretti:2019zyh} have been amply surveyed by the GI model. The particular introduction and pertinent details with respect to the GI model are elucidated in Ref. \cite{Godfrey:1985xj}. In terms of mesons, the Hamiltonian of the GI model is decomposed into
\begin{eqnarray}
H_{\rm GI}&=&H_{\rm GI}^0+V_{\rm GI}^{\rm si}+V_{\rm GI}^{\rm sd}\nonumber\\
&=&H_{\rm GI}^0+V_{\rm GI}^{\rm conf}+V_{\rm GI}^{\rm cont}+V_{\rm GI}^{\rm ten}+V_{\rm GI}^{\rm so},\label{eq21}
\end{eqnarray}
where $H_{\rm GI}^0$, $V_{\rm GI}^{\rm si}$, and $V_{\rm GI}^{\rm sd}$ denote the relativistic energy of all (anti)quarks, the spin-independent interaction potential (made up of the confinement potential $V_{\rm GI}^{\rm conf}$), and the spin-dependent interaction potential (made up of the contact potential $V_{\rm GI}^{\rm cont}$, the tensor potential $V_{\rm GI}^{\rm ten}$, and the spin-orbit potential $V_{\rm GI}^{\rm so}$), respectively. Concretely, these terms are
\begin{eqnarray}
H_{\rm GI}^0&=&\sum_{i=1}^2E_{i}(p),\label{eq22}\\
V_{\rm GI}^{\rm conf}&=&\tilde G_{12}^{\rm Coul}(p,r)+\tilde S_{12}(r),\label{eq23}\\
V_{\rm GI}^{\rm cont}&=&\frac{2}{3m_1m_2r^2}\frac{\partial}{\partial r}\left[r^2\frac{\partial\tilde G_{12}^{\rm cont}(p,r)}{\partial r}\right]\bm S_1\cdot\bm S_2,\label{eq24}\\
V_{\rm GI}^{\rm ten}&=&\frac{1}{m_1m_2}\left(\frac{1}{r}-\frac{\partial}{\partial r}\right)\frac{\partial\tilde G_{12}^{\rm ten}(p,r)}{\partial r}\mathbb T,\label{eq25}\\
V_{\rm GI}^{\rm so}&=&\frac{1}{2m_1^2r}\left[\frac{\partial\tilde G_{11}^{\rm so(v)}(p,r)}{\partial r}-\frac{\partial\tilde S_{11}^{\rm so(s)}(p,r)}{\partial r}\right]\bm L\cdot\bm S_1\nonumber\\
&&+\frac{1}{2m_2^2r}\left[\frac{\partial\tilde G_{22}^{\rm so(v)}(p,r)}{\partial r}-\frac{\partial\tilde S_{22}^{\rm so(s)}(p,r)}{\partial r}\right]\bm L\cdot\bm S_2\nonumber\\
&&+\frac{1}{m_1m_2r}\frac{\partial\tilde G_{12}^{\rm so(v)}(p,r)}{\partial r}\bm L\cdot\bm S,\label{eq26}
\end{eqnarray}
with
\begin{eqnarray*}
&&E_{i}(p)=\left(p^2+m_i^2\right)^\frac{1}{2},\\
&&\tilde G_{ij}^{\rm Coul}(p,r)\nonumber\\
&=&\left[1+\frac{p^2}{E_i(p)E_j(p)}\right]^\frac{1}{2}\tilde G_{ij}(r)\left[1+\frac{p^2}{E_i(p)E_j(p)}\right]^\frac{1}{2},\\
&&\tilde G_{ij}^{\rm cont/ten/so(v)}(p,r)\nonumber\\
&=&\left[\frac{m_im_j}{E_i(p)E_j(p)}\right]^{\frac{1}{2}+\epsilon_{\rm cont/ten/so(v)}}\tilde G_{ij}(r)\left[\frac{m_im_j}{E_i(p)E_j(p)}\right]^{\frac{1}{2}+\epsilon_{\rm cont/ten/so(v)}},\\
&&\tilde S_{ij}^{\rm so(s)}(p,r)\nonumber\\
&=&\left[\frac{m_im_j}{E_i(p)E_j(p)}\right]^{\frac{1}{2}+\epsilon_{\rm so(s)}}\tilde S_{ij}(r)\left[\frac{m_im_j}{E_i(p)E_j(p)}\right]^{\frac{1}{2}+\epsilon_{\rm so(s)}},\\
&&\mathbb T=\frac{\mathbb S_{12}}{12}=\frac{\left(\bm S_1\cdot\bm r\right)\left(\bm S_2\cdot\bm r\right)}{r^2}-\frac{1}{3}\bm S_1\cdot\bm S_2.
\end{eqnarray*}
Here, $E_{i}$, $m_{i}$, and $\mathbb T$ denote the relativistic energy of each (anti)quark $i$, the mass of each (anti)quark $i$, and the operator of the tensor coupling interaction (calculated by utilizing the identity from Landau and Lifshitz \cite{Landau:1991wop,Ali:2017wsf} or the Wigner-Eckart theorem \cite{Varshalovich:1988krb}), respectively. Taking into account the momentum dependence of the potentials, several sorts of factors are affixed to the smeared Coulomb and linear confinement potentials ($\tilde G_{ij}$ and $\tilde S_{ij}$), conducing to the momentum-dependent Coulomb, contact, tensor, vector spin-orbit, and scalar spin-orbit potentials ($\tilde G_{ij}^{\rm Coul}$, $\tilde G_{ij}^{\rm cont}$, $\tilde G_{ij}^{\rm ten}$, $\tilde G_{ij}^{\rm so(v)}$, and $\tilde S_{ij}^{\rm so(s)}$) which contain universal parameters ($\epsilon_{\rm cont}$, $\epsilon_{\rm ten}$, $\epsilon_{\rm so(v)}$, and $\epsilon_{\rm so(s)}$). As far as the smeared potential $\tilde f_{ij}(r)$ is concerned \cite{Godfrey:1985xj}, the definition of the smearing procedure is
\begin{eqnarray}
\tilde f_{ij}(r)&\equiv&\int d^3r'\rho_{ij}({\bf r}-{\bf r}')f(r'),\label{eq27}
\end{eqnarray}
with
\begin{gather}
\rho_{ij}({\bf r}-{\bf r}')=\frac{\sigma_{ij}^3}{\pi^\frac{3}{2}}e^{-\sigma_{ij}^2({\bf r}-{\bf r}')^2},\label{eq28}\\
\sigma_{ij}=\sqrt{\sigma_0^2\left[\frac{1}{2}+\frac{1}{2}\left(\frac{4m_im_j}{(m_i+m_j)^2}\right)^4\right]+s^2\left(\frac{2m_im_j}{m_i+m_j}\right)^2}.\label{eq29}
\end{gather}
Here, both $\sigma_0$ and $s$ are the universal parameters of the GI model \cite{Godfrey:1985xj}. Subtly, the Coulomb potential $G(r)$ with the $\gamma^\mu\otimes\gamma_\mu$ short-range interaction and the linear potential $S(r)$ with the $1\otimes1$ long-range interaction are recast into the smeared Coulomb potential $\tilde G_{ij}(r)$ and the smeared linear potential $\tilde S_{ij}(r)$, respectively, by means of the smearing function $\rho_{ij}({\bf r}-{\bf r}')$. The Coulomb and linear confinement potentials are expressed as
\begin{eqnarray}
G(r)&=&\frac{\alpha_s(r)}{4r}\bm\lambda_1\cdot\bm\lambda_2,\label{eq210}\\
S(r)&=&-\frac{3}{16}(br+c)\bm\lambda_1\cdot\bm\lambda_2,\label{eq211}
\end{eqnarray}
with
\begin{eqnarray}
\alpha_s(r)&=&\sum_{k=1}^3\alpha_k{\rm erf}(\gamma_kr),\label{eq212}\\
{\rm erf}(x)&=&\frac{2}{\pi^\frac{1}{2}}\int_0^xe^{-t^2}dt.\label{eq213}
\end{eqnarray}
Here, all of the arisen parameters ($\alpha_{k=1,2,3}$, $\gamma_{k=1,2,3}$, $b$, and $c$) are universal in the GI model \cite{Godfrey:1985xj}. In addition, $\bm\lambda_i$, $\alpha_s(r)$, and ${\rm erf}(x)$ denote the Gell-Mann matrices acting on each (anti)quark $i$, the running coupling constant of QCD, and the error function, respectively. Evidently, it is convenient to evaluate the color operator $\bm\lambda_i\cdot\bm\lambda_j$ by dint of the eigenvalue of the quadratic Casimir operator \cite{Vijande:2009ac}. For the case of heavy-light tetraquarks, what is striking is that the off-diagonal part of the tensor potential $V_{\rm GI}^{\rm ten}$ can not only cause $^3L_J$$\leftrightarrow$$^3L^\prime_J$ and $^5L_J$$\leftrightarrow$$^5L^\prime_J$ mixings but also cause $^1L_J$$\leftrightarrow$$^5L_J$ mixing, quite dissimilar from the counterpart of heavy-light mesons that can only cause $^3L_J$$\leftrightarrow$$^3L^\prime_J$ mixing. Moreover, $^1L_J$$\leftrightarrow$$^3L_J$ and $^3L_J$$\leftrightarrow$$^5L_J$ mixings are caused by the off-diagonal part of the spin-orbit potential $V_{\rm GI}^{\rm so}$. Following the steps in Refs. \cite{Godfrey:1985xj,Godfrey:2016nwn,Godfrey:2015dva,Godfrey:2004ya}, this work tackles $^1L_J$$\leftrightarrow$$^3L_J$, $^3L_J$$\leftrightarrow$$^5L_J$, and $^1L_J$$\leftrightarrow$$^5L_J$ mixings perturbatively, and omits $^3L_J$$\leftrightarrow$$^3L^\prime_J$ and $^5L_J$$\leftrightarrow$$^5L^\prime_J$ mixings for convenience. With regard to mixing angles of the doubly charmed tetraquarks and magic mixing angles of the ideal heavy-light tetraquarks, the detailed analyses are displayed in Section \ref{sec4}.

Currently, the hypothetical diquark is extensively employed to the theoretical interpretations on the miscellaneous properties of baryons and multiquarks \cite{Anselmino:1992vg,Jaffe:2004ph,Barabanov:2020jvn}, for instance, spectroscopy, production, magnetic moments, form factors, and decay properties. On the basis of the color-triplet representation of the quark, the color SU(3) representation of the diquark is antitriplet or sextet, i.e.,
\begin{eqnarray}
{\bf 3}_q\otimes{\bf 3}_q&=&\bar{\bf 3}_{qq}\oplus{\bf 6}_{qq},\label{eq214}\\
\bar{\bf 3}_{\bar q}\otimes\bar{\bf 3}_{\bar q}&=&{\bf 3}_{\bar q\bar q}\oplus\bar{\bf 6}_{\bar q\bar q}.\label{eq215}
\end{eqnarray}
Remarkably, in regard to the color-(anti)sextet (anti)diquark, the matrix element of the color operator $\bm\lambda_i\cdot\bm\lambda_j$ is positive $4/3$, signifying that the repulsive interquark force deters the forming of the color-(anti)sextet (anti)diquark in the diquark model \cite{Lipkin:1986dw,Karliner:2017qjm,Eichten:2017ffp,Bedolla:2019zwg,Lu:2019ira,Lu:2016zhe,Lu:2016cwr,Dong:2022sef,Ferretti:2019zyh}. In consequence, the diquark-antidiquark scenarios utilized by this work merely treat the color-antitriplet diquark and the color-triplet antidiquark as the genuinely effective (anti)diquark to study the spectroscopy of the $cc\bar{u}\bar{d}$ states. Admittedly, the color-antitriplet diquark (color-triplet antidiquark) can be approximated as the color-antitriplet antiquark (color-triplet quark) owing to the equivalent color between them. Accordingly, the GI relativized diquark model is effectuated in two steps. Initially, the masses of the doubly charmed diquark, the isoscalar light diquark, and the isovector light diquark are procured. Subsequently, the spectroscopic properties of the low-lying $T_{cc}$ tetraquark states are investigated by regarding the diquark (antidiquark) as the antiquark (quark). The outcomes of the GI relativized diquark model are revealed in Section \ref{sec3}.

\subsection{Modified Godfrey-Isgur (MGI) relativized diquark model with the color screening effects}\label{subsec22}

Albeit the long-range interaction between (anti)quarks can be delineated by the Lorentz-scalar linear potential $S(r)$ \cite{Godfrey:1985xj}, the vacuum polarization effects of dynamical fermions may induce the fracture of the color flux tube at large distances \cite{Born:1989iv}, i.e., the color screening effects (also known as the string breaking effects). A successful employment of the nonrelativistic potential model with the color screening effects is the spectroscopic investigation of heavy quarkonia \cite{Li:2009zu}, via the replacement of the linear potential $S(r)$ by the screened linear potential $S^{\rm scr}(r)$ whose form is
\begin{eqnarray}
S^{\rm scr}(r)&=&-\frac{3}{16}\left(b\frac{1-e^{-\mu r}}{\mu}+c\right)\bm\lambda_1\cdot\bm\lambda_2.\label{eq216}
\end{eqnarray}
Here, $\mu$ denotes the screening factor, flattening the linear confinement potential with the $1\otimes1$ long-range interaction when the interquark distance $r$ is large enough. In the tiny and large $r$ limits, the screened linear potential $S^{\rm scr}(r)$ is approximated as the following forms, i.e.,
\begin{eqnarray}
S^{\rm scr}(r)\rightarrow\left\{
\begin{aligned}
S(r)=&-\frac{3}{16}(br+c)\bm\lambda_1\cdot\bm\lambda_2,&&r\rightarrow0,\\
c_\mu=&-\frac{3}{16}\left(\frac{b}{\mu}+c\right)\bm\lambda_1\cdot\bm\lambda_2,&&r\rightarrow\infty,\label{eq217}
\end{aligned}
\right.
\end{eqnarray}
which demonstrates that the screened linear potential $S^{\rm scr}(r)$ is reverted to the linear potential $S(r)$ in the tiny $r$ limit, and reduced to a particular constant $c_\mu$ that contains a certain saturation distance $\mu^{-1}$ in the large $r$ limit \cite{Mezoir:2008vx}.

The modified Godfrey-Isgur relativized quark model (MGI model) is obtained by incorporating the color screening effects into the orthodox GI model \cite{Song:2015nia}. Specifically, the smeared linear potential $\tilde S_{ij}(r)$ in the GI model is substituted by the smeared screened linear potential $\tilde S_{ij}^{\rm scr}(r)$ in the MGI model, acquired by embedding the screened linear potential $S^{\rm scr}(r)$ in Eq. (\ref{eq27}). The concrete introduction and relevant details with respect to the MGI model are expounded in Ref. \cite{Song:2015nia}. Thus far, the spectroscopic properties of light mesons \cite{Pang:2017dlw}, singly heavy mesons \cite{Song:2015nia,Song:2015fha}, doubly heavy mesons \cite{Wang:2018rjg}, light tetraquarks \cite{Lu:2019ira}, heavy tetraquarks \cite{Lu:2016cwr,Dong:2022sef}, and diquarks \cite{Lu:2019ira,Lu:2016cwr,Dong:2022sef} have been well explored by the MGI model. Mimicking the effectuation of the GI relativized diquark model, the MGI relativized diquark model with the color screening effects is carried out in two aforementioned steps. The outcomes of the MGI relativized diquark model are laid out in Section \ref{sec3}.

\subsection{Nonrelativistic (NR) diquark models}\label{subsec23}

As is well known, a prestigious paragon of the nonrelativistic quark model (NR model) is the Cornell potential model which superbly predicted the mass spectra of heavy quarkonia \cite{Eichten:1979ms}. Conventionally, the Hamiltonian of the NR model is decomposed into
\begin{eqnarray}
H_{\rm NR}&=&H_{\rm NR}^0+V_{\rm NR}^{\rm si}+V_{\rm NR}^{\rm sd}\nonumber\\
&=&H_{\rm NR}^0+V_{\rm NR}^{\rm conf}+V_{\rm NR}^{\rm cont}+V_{\rm NR}^{\rm ten}+V_{\rm NR}^{\rm so},\label{eq218}
\end{eqnarray}
where $H_{\rm NR}^0$, $V_{\rm NR}^{\rm si}$, and $V_{\rm NR}^{\rm sd}$ denote the nonrelativistic energy of all (anti)quarks, the spin-independent term (composed of the confinement potential $V_{\rm NR}^{\rm conf}$), and the spin-dependent term (composed of the contact potential $V_{\rm NR}^{\rm cont}$, the tensor potential $V_{\rm NR}^{\rm ten}$, and the spin-orbit potential $V_{\rm NR}^{\rm so}$), respectively. The form of $H_{\rm NR}^0$ is expressed as
\begin{eqnarray}
H_{\rm NR}^0&=&\sum_{i=1}^2\mathcal E_{i}(p),\label{eq219}
\end{eqnarray}
with
\begin{eqnarray}
\mathcal E_{i}(p)&=&m_i+\frac{p^2}{2m_i}.\label{eq220}
\end{eqnarray}
Here, $\mathcal E_{i}$ is the nonrelativistic energy of each (anti)quark $i$. As far as the forms of the spin-independent term $V_{\rm NR}^{\rm si}$ and the spin-dependent term $V_{\rm NR}^{\rm sd}$ are concerned, there are two scenarios capable of characterizing the spectroscopy of heavy-light hadrons primely \cite{Kim:2020imk,Kim:2021ywp}. Based on the distinctive functional forms of the hyperfine interaction potentials, these two scenarios are elucidated as follows.

\subsubsection{Scenario I: NR-G diquark model}\label{subsec231}

Following Refs. \cite{Godfrey:1985xj,Barnes:2005pb,Kim:2020imk}, the nonrelativistic quark model with the Gaussian contact hyperfine interaction (NR-G model) is employed in the first scenario. To be specific, the terms $V_\text{NR-G}^{\rm conf}$, $V_\text{NR-G}^{\rm cont}$, $V_\text{NR-G}^{\rm ten}$, and $V_\text{NR-G}^{\rm so}$ possess the forms of
\begin{eqnarray}
V_\text{NR-G}^{\rm conf}&=&G^{\rm Coul}(r)+S(r),\label{eq221}\\
V_\text{NR-G}^{\rm cont}&=&\frac{2}{3m_1m_2r^2}\frac{\rm d}{{\rm d}r}\left[r^2\frac{{\rm d}G^{\rm Gauss}(r)}{{\rm d}r}\right]\bm S_1\cdot\bm S_2,\label{eq222}\\
V_\text{NR-G}^{\rm ten}&=&\frac{1}{m_1m_2}\left(\frac{1}{r}-\frac{\rm d}{{\rm d}r}\right)\frac{{\rm d}G^{\rm Coul}(r)}{{\rm d}r}\mathbb T,\label{eq223}\\
V_\text{NR-G}^{\rm so}&=&\frac{1}{2m_1^2r}\left[\frac{{\rm d}G^{\rm Coul}(r)}{{\rm d}r}-\frac{{\rm d}S(r)}{{\rm d}r}\right]\bm L\cdot\bm S_1\nonumber\\
&&+\frac{1}{2m_2^2r}\left[\frac{{\rm d}G^{\rm Coul}(r)}{{\rm d}r}-\frac{{\rm d}S(r)}{{\rm d}r}\right]\bm L\cdot\bm S_2\nonumber\\
&&+\frac{1}{m_1m_2r}\frac{{\rm d}G^{\rm Coul}(r)}{{\rm d}r}\bm L\cdot\bm S,\label{eq224}
\end{eqnarray}
with
\begin{eqnarray}
G^{\rm Coul}(r)&=&\frac{\alpha_c}{4r}\bm\lambda_1\cdot\bm\lambda_2,\label{eq225}\\
G^{\rm Gauss}(r)&=&\frac{\mathfrak{\alpha}_g(r)}{4r}\bm\lambda_1\cdot\bm\lambda_2,\label{eq226}\\
\alpha_g(r)&=&\alpha_c{\rm erf}(\gamma_cr).\label{eq227}
\end{eqnarray}
Here, all of the corresponding parameters of the NR-G model ($\alpha_c$, $\gamma_c$, $b$, and $c$) stem from the spectroscopic inquiries of conventional heavy hadrons \cite{Barnes:2005pb,Kim:2020imk}. A salient feature of the NR-G model is the emergence of a Gaussian function in the Laplace operator of $G^{\rm Gauss}(r)$ which contains the error function ${\rm erf}(\gamma_cr)$. Manifestly, the form of $\alpha_g(r)$ in $G^{\rm Gauss}(r)$ is embodied as the approximation of $\alpha_s(r)$ in $G(r)$. Additionally, the form of $G^{\rm Coul}(r)$ prevalently utilized in the Cornell potential model \cite{Eichten:1979ms} can be looked upon as the approximate form of $G(r)$, if the distance-dependent $\alpha_s(r)$ is simplified as the constant $\alpha_c$. By comparing Eqs. (\ref{eq221})-(\ref{eq224}) with Eqs. (\ref{eq23})-(\ref{eq26}), there is a visible similitude of the forms of the interaction potentials between the NR-G model and the GI model \cite{Godfrey:1985xj}. Imitating Ref. \cite{Barnes:2005pb}, the leading-order perturbation theory is employed to regard the tensor potential $V_\text{NR-G}^{\rm ten}$ and the spin-orbit potential $V_\text{NR-G}^{\rm so}$ as mass shifts, which touch upon the diagonal terms and the off-diagonal parts. The particular introduction and pertinent details with respect to the NR-G model are elucidated in Refs. \cite{Godfrey:1985xj,Barnes:2005pb,Kim:2020imk}. In consideration of the identical color SU(3) representation between the diquark (antidiquark) and the antiquark (quark), the NR-G diquark model is performed in the framework of the diquark-antidiquark configuration. Concerning the spectroscopic properties of the doubly charmed tetraquark system, the predicted outcomes of the NR-G diquark model are revealed in Section \ref{sec3}.

\subsubsection{Scenario II: NR-Y diquark model}\label{subsec232}

The second scenario adopts the nonrelativistic quark model whose contact hyperfine interaction is of the Yukawa form (NR-Y model) in light of Refs. \cite{Kim:2020imk,Kim:2021ywp,Yoshida:2015tia}. Further, the potentials $V_\text{NR-Y}^{\rm conf}$, $V_\text{NR-Y}^{\rm cont}$, $V_\text{NR-Y}^{\rm ten}$, and $V_\text{NR-Y}^{\rm so}$ are represented as
\begin{eqnarray}
V_\text{NR-Y}^{\rm conf}&=&\frac{m_1+m_2}{m_1m_2}G^{\rm Coul}(r)+S(r),\label{eq228}\\
V_\text{NR-Y}^{\rm cont}&=&\frac{2}{3m_1m_2r^2}\frac{\rm d}{{\rm d}r}\left[r^2\frac{{\rm d}G^{\rm Yukawa}(r)}{{\rm d}r}\right]\bm S_1\cdot\bm S_2,\label{eq229}\\
V_\text{NR-Y}^{\rm ten}&=&\frac{(1-e^{-\gamma_cr})^2}{m_1m_2}\left(\frac{1}{r}-\frac{\rm d}{{\rm d}r}\right)\frac{{\rm d}G^{\rm ten}(r)}{{\rm d}r}\mathbb T,\label{eq230}\\
V_\text{NR-Y}^{\rm so}&=&\frac{(1-e^{-\gamma_cr})^2}{2m_1^2r}\frac{{\rm d}G^{\rm so}(r)}{{\rm d}r}\bm L\cdot\bm S_1\nonumber\\
&&+\frac{(1-e^{-\gamma_cr})^2}{2m_2^2r}\frac{{\rm d}G^{\rm so}(r)}{{\rm d}r}\bm L\cdot\bm S_2\nonumber\\
&&+\frac{(1-e^{-\gamma_cr})^2}{m_1m_2r}\frac{{\rm d}G^{\rm so}(r)}{{\rm d}r}\bm L\cdot\bm S,\label{eq231}
\end{eqnarray}
with
\begin{eqnarray}
G^{\rm ten/so}(r)&=&\frac{\alpha_{\rm ten/so}}{4r}\bm\lambda_1\cdot\bm\lambda_2,\label{eq232}\\
G^{\rm Yukawa}(r)&=&\frac{\mathfrak{\alpha}_y(r)}{4r}\bm\lambda_1\cdot\bm\lambda_2,\label{eq233}\\
\alpha_y(r)&=&-\alpha_{\rm cont}e^{-\gamma_cr}.\label{eq234}
\end{eqnarray}
It is notable that the form of $G^{\rm Yukawa}(r)$ which includes the exponential function ${\rm exp}(-\gamma_cr)$ supremely resembles the canonical form of the Yukawa potential. In the NR-Y model, the total relevant parameters ($\alpha_{\rm cont}$, $\alpha_{\rm ten}$, $\alpha_{\rm so}$, $\alpha_c$, $\gamma_c$, $b$, and $c$) are determined by the mass spectra of heavy-light hadrons \cite{Kim:2020imk,Kim:2021ywp}. In view of the latent quark-mass dependence of the Coulombic parameter alluded by a lattice QCD survey \cite{Kawanai:2011xb}, the reciprocal of the reduced mass was introduced into the Coulomb term of the NR-Y model \cite{Kim:2020imk,Kim:2021ywp,Yoshida:2015tia}, as exhibited in Eq. (\ref{eq228}). The concrete introduction and relevant details with respect to the NR-Y model are expounded in Refs. \cite{Kim:2020imk,Kim:2021ywp,Yoshida:2015tia}. When it comes to the $T_{cc}$ tetraquark comprised of the double-charm diquark and the light antidiquark, the advent of five sorts of mixings ($^1L_J$$\leftrightarrow$$^3L_J$, $^3L_J$$\leftrightarrow$$^5L_J$, $^1L_J$$\leftrightarrow$$^5L_J$, $^3L_J$$\leftrightarrow$$^3L^\prime_J$, and $^5L_J$$\leftrightarrow$$^5L^\prime_J$) caused by the tensor and spin-orbit potentials ($V_\text{NR-Y}^{\rm ten}$ and $V_\text{NR-Y}^{\rm so}$) is inevitable. Following Ref. \cite{Kim:2021ywp}, $^1L_J$$\leftrightarrow$$^3L_J$, $^3L_J$$\leftrightarrow$$^5L_J$, and $^1L_J$$\leftrightarrow$$^5L_J$ mixings are treated perturbatively by the NR-Y diquark model, then remnant $^3L_J$$\leftrightarrow$$^3L^\prime_J$ and $^5L_J$$\leftrightarrow$$^5L^\prime_J$ mixings are left out for convenience. As expounded in the GI relativized diquark model, the color-antitriplet diquark $cc$ and the color-triplet antidiquark $\bar{u}\bar{d}$ are deemed as the genuinely effective (anti)diquark to unravel the spectroscopy of the exotic $cc\bar{u}\bar{d}$ states in the NR-Y diquark model. Subsequently, the particular results of the mass spectrum of the low-lying $T_{cc}$ tetraquark family obtained by the NR-Y diquark model are displayed in Section \ref{sec3}.

\section{Results}\label{sec3}

\renewcommand\tabcolsep{0.55cm}
\renewcommand{\arraystretch}{1.5}
\begin{table*}[!htbp]
\caption{Parameters of the GI (MGI) relativized diquark model \cite{Godfrey:1985xj}, NR-G diquark model \cite{Barnes:2005pb,Kim:2020imk}, and NR-Y diquark model \cite{Kim:2020imk,Kim:2021ywp}.}\label{para}
\begin{tabular}{cccc|cc|cc}
\toprule[1.0pt]\toprule[1.0pt]
\multicolumn{4}{c|}{GI (MGI) \cite{Godfrey:1985xj}} & \multicolumn{2}{c|}{NR-G \cite{Barnes:2005pb,Kim:2020imk}} & \multicolumn{2}{c}{NR-Y \cite{Kim:2020imk,Kim:2021ywp}} \\
Parameter & Value & Parameter & Value & Parameter & Value & Parameter & Value \\
\midrule[1.0pt]
$m_c$ (GeV) & $1.628$ & $m_{u,d}$ (GeV) & $0.220$ & $m_c$ (GeV) & $1.4794$ & $m_c$ (GeV) & $1.750$ \\
$b$ (GeV$^2$) & $0.18$ & $c$ (GeV) & $-0.253$ & $b$ (GeV$^2$) & $0.1425$ & $b$ (GeV$^2$) & $0.165$ \\
$\gamma_1$ (GeV) & $\sqrt{1/4}$ & $\alpha_1$ & $0.25$ & $c$ (GeV) & $-0.191$ & $c$ (GeV) & $-0.831$ \\
$\gamma_2$ (GeV) & $\sqrt{10/4}$ & $\alpha_2$ & $0.15$ & $\gamma_c$ (GeV) & $1.0946$ & $\gamma_c$ (GeV) & $0.691$ \\
$\gamma_3$ (GeV) & $\sqrt{1000/4}$ & $\alpha_3$ & $0.20$ & $\alpha_c$ & $0.5461$ & $\alpha_c$ (GeV) & $0.045$ \\
$\sigma_0$ (GeV) & $1.80$ & $s$ & $1.55$ & $\cdots$ & $\cdots$ & $\alpha_{\rm cont}$ & $0.9659$ \\
$\epsilon_{\rm cont}$ & $-0.168$ & $\epsilon_{\rm so(v)}$ & $-0.035$ & $\cdots$ & $\cdots$ & $\alpha_{\rm ten}$ & $0.3741$ \\
$\epsilon_{\rm ten}$ & $0.025$ & $\epsilon_{\rm so(s)}$ & $0.055$ & $\cdots$ & $\cdots$ & $\alpha_{\rm so}$ & $0.7482$ \\
\bottomrule[1.0pt]\bottomrule[1.0pt]
\end{tabular}
\end{table*}

This section presents the predicted outcomes on the spectroscopic properties of the low-lying doubly charmed tetraquark family by utilizing four sorts of aforementioned diquark-antidiquark scenarios, i.e., GI relativized diquark model, MGI relativized diquark model, NR-G diquark model, and NR-Y diquark model.

\subsection{Parameters}\label{subsec31}

As enumerated in Table \ref{para}, the parameters of the GI (MGI) relativized diquark model, NR-G diquark model, and NR-Y diquark model with respect to the mass spectra of doubly charmed tetraquark states are designated to keep consistency with the ones employed by Ref. \cite{Godfrey:1985xj}, Refs. \cite{Barnes:2005pb,Kim:2020imk}, and Refs. \cite{Kim:2020imk,Kim:2021ywp}, respectively, in order to retain the model universality between conventional and exotic hadrons.

\subsection{Diquarks}\label{subsec32}

As the essential constituents of the doubly charmed tetraquark, doubly charmed diquark $cc$ and light diquark $ud$ play a crucial role in comprehending $T_{cc}$ spectroscopy. Constrained by the Pauli exclusion principle \cite{Amsler:2018zkm}, the spin quantum number of the ground state diquark $cc$ with antitriplet color is endowed with $1$. Hence, the $S$-wave doubly charmed diquark employed by this work is an axial-vector diquark \cite{Amsler:2018zkm}. As delineated in Section \ref{sec2}, the GI (MGI) model procures the mass of the color-antitriplet diquark $cc$ via the universal parameters \cite{Godfrey:1985xj}. Analogously, the NR-G model reaps the mass of the diquark $cc$ by dint of the parameters of the charmonium family \cite{Barnes:2005pb}. Taking into account the absence of the pertinent employment of charmonium spectroscopy, the NR-Y model makes use of the mass relations of heavy-light hadrons proposed by the heavy quark symmetry (HQS) to garner the diquark $cc$ mass \cite{Eichten:2017ffp}, i.e.,
\begin{eqnarray}
m_{\{cc\}[\bar{u}\bar{d}]}-m_{\{cc\}u}&=&m_{c[ud]}-m_{c\bar{u}},\label{eq31}\\
m_{\{cc\}[\bar{u}\bar{d}]}-m_{c[ud]}&=&m_{\{cc\}u}-m_{c\bar{u}},\label{eq32}
\end{eqnarray}
where the braces $\{qq\}$ and the brackets $[qq]$ denote the axial-vector diquark and the scalar diquark \cite{Eichten:2017ffp}, respectively. By averaging the LHS and RHS of Eq. (\ref{eq32}), the mass gap between doubly charmed diquark and charm quark in the NR-Y model is acquired appropriately, i.e.,
\begin{eqnarray}
m_{\{cc\}}-m_{c}&=&\frac{1}{2}(m_{\{cc\}[\bar{u}\bar{d}]}-m_{c[ud]}+m_{\{cc\}u}-m_{c\bar{u}}),\label{eq33}
\end{eqnarray}
where the value of $m_{c}$ is rendered in the last column of Table \ref{para}. Inserting the NR-Y model mass of charm quark and the experimental masses of corresponding heavy-light hadrons into Eq. (\ref{eq33}), the diquark $cc$ mass adopted by the NR-Y model is determined legitimately. Alternatively, on the basis of the heavy antiquark-diquark symmetry (HADS), the mass formula of the doubly charmed diquark $cc$ in the NR-Y model is offered by Ref. \cite{Wu:2022gie}, i.e.,
\begin{eqnarray}
m_{\{cc\}}&=&2m_{c}+(A_{cc}\bm S_1\cdot\bm S_2+\frac{B_{cc}}{4})\bm\lambda_1\cdot\bm\lambda_2,\label{eq34}
\end{eqnarray}
where the parameters $A_{cc}=-21.2$ MeV and $B_{cc}=217.7$ MeV are determined from experimental data of the charmonium spectrum and the doubly charmed baryon spectrum \cite{Wu:2022gie}, respectively. After the substitution of the NR-Y model mass of charm quark, the NR-Y model mass of the color-antitriplet diquark $cc$ obtained by HADS is magically identical with the one previously attained by HQS, which effectively illustrates the resemblance between HQS and HADS. The particular introduction and pertinent details with respect to HQS and HADS are elucidated in Refs. \cite{Eichten:2017ffp,Wu:2022gie}. Subsequently, the masses of the ground state axial-vector doubly charmed diquark from these four sorts of diquark-antidiquark scenarios are explicitly enumerated in Table \ref{dq}. Concretely, the values of the diquark $cc$ masses from the GI, MGI ($\mu=30$), MGI ($\mu=50$), MGI ($\mu=70$), NR-G, and NR-Y scenarios are 3329, 3320, 3314, 3309, 3152, and 3369 MeV, respectively.

\renewcommand\tabcolsep{0.63cm}
\renewcommand{\arraystretch}{1.5}
\begin{table}[!htbp]
\caption{The masses of the doubly charmed and light diquarks from the GI, MGI, NR-G, and NR-Y scenarios (in unit of MeV).}\label{dq}
\begin{tabular}{cccc}
\toprule[1.0pt]\toprule[1.0pt]
Scenario & $m_{\{cc\}}$ & $m_{[ud]}$ & $m_{\{ud\}}$ \\
\midrule[1.0pt]
GI & $3329$ & $691$ & $840$ \\
MGI ($\mu=30$) & $3320$ & $673$ & $814$ \\
MGI ($\mu=50$) & $3314$ & $662$ & $796$ \\
MGI ($\mu=70$) & $3309$ & $650$ & $778$ \\
NR-G & $3152$ & $725$ & $1019$ \\
NR-Y & $3369$ & $725$ & $973$ \\
\bottomrule[1.0pt]\bottomrule[1.0pt]
\end{tabular}
\end{table}

As far as the light diquark $ud$ is concerned, both scalar and axial-vector cases are eligible owing to the symmetry of wave functions \cite{Amsler:2018zkm}. In accordance with the universal parameters stemmed from a variety of mesons with discrepant flavors \cite{Godfrey:1985xj}, the masses of scalar and axial-vector light diquarks are obtained by the GI (MGI) model. On account of the deficiency of adequate spectroscopic applications of light mesons, the NR models take advantage of the light diquark masses deduced by unquenched lattice QCD and chiral effective theory, whose concrete introduction and relevant details are expounded in Refs. \cite{Kim:2020imk,Kim:2021ywp,Bi:2015ifa,Harada:2019udr}. Whereafter, the scalar and axial-vector light diquark masses from aforementioned diquark-antidiquark scenarios are particularized in Table \ref{dq}. Specifically, the values of scalar light diquark masses carried out by the GI, MGI ($\mu=30$), MGI ($\mu=50$), MGI ($\mu=70$), NR-G, and NR-Y scenarios are 691, 673, 662, 650, 725, and 725 MeV, respectively. In the case of axial-vector light diquark, the masses performed by the GI, MGI ($\mu=30$), MGI ($\mu=50$), MGI ($\mu=70$), NR-G, and NR-Y scenarios are 840, 814, 796, 778, 1019, and 973 MeV, respectively.

\subsection{Doubly charmed tetraquarks}\label{subsec33}

\renewcommand\tabcolsep{0.24cm}
\renewcommand{\arraystretch}{1.5}
\begin{table*}[!htbp]
\caption{The mass spectrum of the $1S$-, $1P$-, $2S$-, and $1D$-wave doubly charmed tetraquark states procured by this work (in unit of MeV).}\label{ccud1}
\begin{tabular}{p{1.77cm}<{\centering}p{1.77cm}<{\centering}|p{1.77cm}<{\centering}p{1.77cm}<{\centering}p{1.77cm}<{\centering}p{1.77cm}<{\centering}
p{1.77cm}<{\centering}p{1.77cm}<{\centering}}
\toprule[1.0pt]\toprule[1.0pt]
\multicolumn{2}{c|}{State} & \multicolumn{6}{c}{Mass} \\
$T_{cc}(n^{2S+1}L_J)$ & $I(J^P)$ & GI & MGI ($\mu=30$) & MGI ($\mu=50$) & MGI ($\mu=70$) & NR-G & NR-Y \\
\midrule[1.0pt]
$T_{cc1}^f(1^3S_1)$ & $0(1^+)$ & $3948$ & $3917$ & $3897$ & $3877$ & $3884$ & $3876$ \\
$T_{cc0}^a(1^1S_0)$ & $1(0^+)$ & $3842$ & $3809$ & $3787$ & $3766$ & $3894$ & $4029$ \\
$T_{cc1}^a(1^3S_1)$ & $1(1^+)$ & $3960$ & $3925$ & $3902$ & $3879$ & $3991$ & $4057$ \\
$T_{cc2}^a(1^5S_2)$ & $1(2^+)$ & $4121$ & $4083$ & $4058$ & $4032$ & $4135$ & $4105$ \\
$T_{cc0}^\eta(1^3P_0)$ & $0(0^-)$ & $4349$ & $4307$ & $4279$ & $4252$ & $4289$ & $4161$ \\
$T_{cc1}^\eta(1^3P_1)$ & $0(1^-)$ & $4370$ & $4327$ & $4299$ & $4271$ & $4311$ & $4174$ \\
$T_{cc2}^\eta(1^3P_2)$ & $0(2^-)$ & $4408$ & $4364$ & $4335$ & $4305$ & $4356$ & $4199$ \\
$T_{cc0}^\pi(1^3P_0)$ & $1(0^-)$ & $4404$ & $4357$ & $4326$ & $4295$ & $4408$ & $4310$ \\
$T_{cc1}^\pi(1P_1)$ & $1(1^-)$ & $4477$ & $4432$ & $4400$ & $4369$ & $4534$ & $4370$ \\
$T_{cc1}^\pi(1P_1^\prime)$ & $1(1^-)$ & $4480$ & $4427$ & $4394$ & $4361$ & $4501$ & $4331$ \\
$T_{cc1}^\pi(1P_1^{\prime\prime})$ & $1(1^-)$ & $4410$ & $4362$ & $4331$ & $4300$ & $4425$ & $4308$ \\
$T_{cc2}^\pi(1P_2)$ & $1(2^-)$ & $4501$ & $4452$ & $4419$ & $4388$ & $4554$ & $4386$ \\
$T_{cc2}^\pi(1P_2^\prime)$ & $1(2^-)$ & $4503$ & $4453$ & $4420$ & $4385$ & $4535$ & $4354$ \\
$T_{cc3}^\pi(1^5P_3)$ & $1(3^-)$ & $4529$ & $4479$ & $4446$ & $4412$ & $4577$ & $4409$ \\
$T_{cc1}^f(2^3S_1)$ & $0(1^+)$ & $4534$ & $4478$ & $4441$ & $4405$ & $4489$ & $4388$ \\
$T_{cc0}^a(2^1S_0)$ & $1(0^+)$ & $4554$ & $4497$ & $4460$ & $4423$ & $4609$ & $4523$ \\
$T_{cc1}^a(2^3S_1)$ & $1(1^+)$ & $4595$ & $4537$ & $4498$ & $4459$ & $4637$ & $4539$ \\
$T_{cc2}^a(2^5S_2)$ & $1(2^+)$ & $4674$ & $4612$ & $4570$ & $4529$ & $4694$ & $4568$ \\
$T_{cc1}^f(1^3D_1)$ & $0(1^+)$ & $4689$ & $4631$ & $4592$ & $4554$ & $4607$ & $4428$ \\
$T_{cc2}^f(1^3D_2)$ & $0(2^+)$ & $4701$ & $4643$ & $4603$ & $4564$ & $4616$ & $4440$ \\
$T_{cc3}^f(1^3D_3)$ & $0(3^+)$ & $4718$ & $4659$ & $4619$ & $4579$ & $4631$ & $4457$ \\
$T_{cc0}^a(1^5D_0)$ & $1(0^+)$ & $4776$ & $4712$ & $4670$ & $4628$ & $4790$ & $4549$ \\
$T_{cc1}^a(1D_1)$ & $1(1^+)$ & $4793$ & $4729$ & $4686$ & $4643$ & $4789$ & $4575$ \\
$T_{cc1}^a(1D_1^\prime)$ & $1(1^+)$ & $4783$ & $4720$ & $4678$ & $4636$ & $4803$ & $4558$ \\
$T_{cc2}^a(1D_2)$ & $1(2^+)$ & $4792$ & $4730$ & $4688$ & $4646$ & $4790$ & $4616$ \\
$T_{cc2}^a(1D_2^\prime)$ & $1(2^+)$ & $4822$ & $4756$ & $4712$ & $4668$ & $4828$ & $4588$ \\
$T_{cc2}^a(1D_2^{\prime\prime})$ & $1(2^+)$ & $4799$ & $4735$ & $4693$ & $4650$ & $4801$ & $4575$ \\
$T_{cc3}^a(1D_3)$ & $1(3^+)$ & $4802$ & $4739$ & $4696$ & $4654$ & $4797$ & $4628$ \\
$T_{cc3}^a(1D_3^\prime)$ & $1(3^+)$ & $4821$ & $4756$ & $4713$ & $4669$ & $4820$ & $4604$ \\
$T_{cc4}^a(1^5D_4)$ & $1(4^+)$ & $4813$ & $4750$ & $4707$ & $4664$ & $4805$ & $4644$ \\
\bottomrule[1.0pt]\bottomrule[1.0pt]
\end{tabular}
\end{table*}

\renewcommand\tabcolsep{0.24cm}
\renewcommand{\arraystretch}{1.5}
\begin{table*}[!htbp]
\caption{The mass spectrum of the $2P$-, $3S$-, and $1F$-wave doubly charmed tetraquark states procured by this work (in unit of MeV).}\label{ccud2}
\begin{tabular}{p{1.77cm}<{\centering}p{1.77cm}<{\centering}|p{1.77cm}<{\centering}p{1.77cm}<{\centering}p{1.77cm}<{\centering}p{1.77cm}<{\centering}
p{1.77cm}<{\centering}p{1.77cm}<{\centering}}
\toprule[1.0pt]\toprule[1.0pt]
\multicolumn{2}{c|}{State} & \multicolumn{6}{c}{Mass} \\
$T_{cc}(n^{2S+1}L_J)$ & $I(J^P)$ & GI & MGI ($\mu=30$) & MGI ($\mu=50$) & MGI ($\mu=70$) & NR-G & NR-Y \\
\midrule[1.0pt]
$T_{cc0}^\eta(2^3P_0)$ & $0(0^-)$ & $4796$ & $4725$ & $4677$ & $4629$ & $4734$ & $4605$ \\
$T_{cc1}^\eta(2^3P_1)$ & $0(1^-)$ & $4810$ & $4738$ & $4689$ & $4641$ & $4755$ & $4615$ \\
$T_{cc2}^\eta(2^3P_2)$ & $0(2^-)$ & $4837$ & $4762$ & $4713$ & $4663$ & $4795$ & $4634$ \\
$T_{cc0}^\pi(2^3P_0)$ & $1(0^-)$ & $4872$ & $4798$ & $4748$ & $4698$ & $4833$ & $4727$ \\
$T_{cc1}^\pi(2P_1)$ & $1(1^-)$ & $4908$ & $4832$ & $4781$ & $4730$ & $4949$ & $4774$ \\
$T_{cc1}^\pi(2P_1^\prime)$ & $1(1^-)$ & $4916$ & $4838$ & $4786$ & $4734$ & $4913$ & $4743$ \\
$T_{cc1}^\pi(2P_1^{\prime\prime})$ & $1(1^-)$ & $4881$ & $4805$ & $4755$ & $4704$ & $4851$ & $4725$ \\
$T_{cc2}^\pi(2P_2)$ & $1(2^-)$ & $4926$ & $4849$ & $4797$ & $4745$ & $4968$ & $4786$ \\
$T_{cc2}^\pi(2P_2^\prime)$ & $1(2^-)$ & $4934$ & $4855$ & $4802$ & $4749$ & $4947$ & $4761$ \\
$T_{cc3}^\pi(2^5P_3)$ & $1(3^-)$ & $4950$ & $4871$ & $4818$ & $4765$ & $4989$ & $4805$ \\
$T_{cc1}^f(3^3S_1)$ & $0(1^+)$ & $4956$ & $4868$ & $4809$ & $4750$ & $4913$ & $4801$ \\
$T_{cc0}^a(3^1S_0)$ & $1(0^+)$ & $4999$ & $4912$ & $4853$ & $4795$ & $5027$ & $4911$ \\
$T_{cc1}^a(3^3S_1)$ & $1(1^+)$ & $5026$ & $4937$ & $4877$ & $4817$ & $5046$ & $4923$ \\
$T_{cc2}^a(3^5S_2)$ & $1(2^+)$ & $5082$ & $4988$ & $4925$ & $4862$ & $5085$ & $4946$ \\
$T_{cc2}^\eta(1^3F_2)$ & $0(2^-)$ & $4958$ & $4882$ & $4831$ & $4780$ & $4852$ & $4664$ \\
$T_{cc3}^\eta(1^3F_3)$ & $0(3^-)$ & $4965$ & $4888$ & $4837$ & $4785$ & $4857$ & $4674$ \\
$T_{cc4}^\eta(1^3F_4)$ & $0(4^-)$ & $4973$ & $4896$ & $4845$ & $4792$ & $4864$ & $4687$ \\
$T_{cc1}^\pi(1^5F_1)$ & $1(1^-)$ & $5065$ & $4984$ & $4929$ & $4874$ & $5049$ & $4777$ \\
$T_{cc2}^\pi(1F_2)$ & $1(2^-)$ & $5056$ & $4976$ & $4922$ & $4868$ & $5029$ & $4800$ \\
$T_{cc2}^\pi(1F_2^\prime)$ & $1(2^-)$ & $5074$ & $4992$ & $4937$ & $4881$ & $5056$ & $4785$ \\
$T_{cc3}^\pi(1F_3)$ & $1(3^-)$ & $5043$ & $4964$ & $4912$ & $4858$ & $5003$ & $4830$ \\
$T_{cc3}^\pi(1F_3^\prime)$ & $1(3^-)$ & $5088$ & $5005$ & $4949$ & $4893$ & $5066$ & $4809$ \\
$T_{cc3}^\pi(1F_3^{\prime\prime})$ & $1(3^-)$ & $5064$ & $4983$ & $4929$ & $4874$ & $5034$ & $4798$ \\
$T_{cc4}^\pi(1F_4)$ & $1(4^-)$ & $5047$ & $4969$ & $4916$ & $4862$ & $5005$ & $4840$ \\
$T_{cc4}^\pi(1F_4^\prime)$ & $1(4^-)$ & $5074$ & $4993$ & $4938$ & $4882$ & $5042$ & $4822$ \\
$T_{cc5}^\pi(1^5F_5)$ & $1(5^-)$ & $5053$ & $4974$ & $4921$ & $4867$ & $5009$ & $4852$ \\
\bottomrule[1.0pt]\bottomrule[1.0pt]
\end{tabular}
\end{table*}

\renewcommand\tabcolsep{0.24cm}
\renewcommand{\arraystretch}{1.5}
\begin{table*}[!htbp]
\caption{The mass spectrum of the $2D$-, $3P$-, and $4S$-wave doubly charmed tetraquark states procured by this work (in unit of MeV).}\label{ccud3}
\begin{tabular}{p{1.77cm}<{\centering}p{1.77cm}<{\centering}|p{1.77cm}<{\centering}p{1.77cm}<{\centering}p{1.77cm}<{\centering}p{1.77cm}<{\centering}
p{1.77cm}<{\centering}p{1.77cm}<{\centering}}
\toprule[1.0pt]\toprule[1.0pt]
\multicolumn{2}{c|}{State} & \multicolumn{6}{c}{Mass} \\
$T_{cc}(n^{2S+1}L_J)$ & $I(J^P)$ & GI & MGI ($\mu=30$) & MGI ($\mu=50$) & MGI ($\mu=70$) & NR-G & NR-Y \\
\midrule[1.0pt]
$T_{cc1}^f(2^3D_1)$ & $0(1^+)$ & $5058$ & $4967$ & $4906$ & $4844$ & $4989$ & $4825$ \\
$T_{cc2}^f(2^3D_2)$ & $0(2^+)$ & $5068$ & $4976$ & $4914$ & $4852$ & $4998$ & $4836$ \\
$T_{cc3}^f(2^3D_3)$ & $0(3^+)$ & $5081$ & $4988$ & $4926$ & $4863$ & $5013$ & $4851$ \\
$T_{cc0}^a(2^5D_0)$ & $1(0^+)$ & $5148$ & $5053$ & $4989$ & $4925$ & $5139$ & $4922$ \\
$T_{cc1}^a(2D_1)$ & $1(1^+)$ & $5160$ & $5064$ & $5000$ & $4935$ & $5151$ & $4945$ \\
$T_{cc1}^a(2D_1^\prime)$ & $1(1^+)$ & $5152$ & $5057$ & $4993$ & $4929$ & $5142$ & $4929$ \\
$T_{cc2}^a(2D_2)$ & $1(2^+)$ & $5156$ & $5062$ & $4999$ & $4935$ & $5149$ & $4979$ \\
$T_{cc2}^a(2D_2^\prime)$ & $1(2^+)$ & $5182$ & $5084$ & $5018$ & $4952$ & $5175$ & $4955$ \\
$T_{cc2}^a(2D_2^{\prime\prime})$ & $1(2^+)$ & $5164$ & $5068$ & $5004$ & $4939$ & $5154$ & $4944$ \\
$T_{cc3}^a(2D_3)$ & $1(3^+)$ & $5165$ & $5070$ & $5006$ & $4942$ & $5156$ & $4990$ \\
$T_{cc3}^a(2D_3^\prime)$ & $1(3^+)$ & $5181$ & $5084$ & $5018$ & $4953$ & $5172$ & $4970$ \\
$T_{cc4}^a(2^5D_4)$ & $1(4^+)$ & $5175$ & $5080$ & $5015$ & $4950$ & $5165$ & $5004$ \\
$T_{cc0}^\eta(3^3P_0)$ & $0(0^-)$ & $5161$ & $5055$ & $4984$ & $4912$ & $5105$ & $4986$ \\
$T_{cc1}^\eta(3^3P_1)$ & $0(1^-)$ & $5172$ & $5065$ & $4993$ & $4921$ & $5124$ & $4995$ \\
$T_{cc2}^\eta(3^3P_2)$ & $0(2^-)$ & $5193$ & $5084$ & $5011$ & $4937$ & $5162$ & $5012$ \\
$T_{cc0}^\pi(3^3P_0)$ & $1(0^-)$ & $5241$ & $5132$ & $5060$ & $4987$ & $5182$ & $5084$ \\
$T_{cc1}^\pi(3P_1)$ & $1(1^-)$ & $5263$ & $5154$ & $5081$ & $5008$ & $5291$ & $5123$ \\
$T_{cc1}^\pi(3P_1^\prime)$ & $1(1^-)$ & $5272$ & $5161$ & $5087$ & $5012$ & $5255$ & $5097$ \\
$T_{cc1}^\pi(3P_1^{\prime\prime})$ & $1(1^-)$ & $5249$ & $5140$ & $5066$ & $4993$ & $5196$ & $5082$ \\
$T_{cc2}^\pi(3P_2)$ & $1(2^-)$ & $5279$ & $5169$ & $5094$ & $5020$ & $5309$ & $5134$ \\
$T_{cc2}^\pi(3P_2^\prime)$ & $1(2^-)$ & $5287$ & $5175$ & $5099$ & $5023$ & $5287$ & $5113$ \\
$T_{cc3}^\pi(3^5P_3)$ & $1(3^-)$ & $5300$ & $5188$ & $5112$ & $5036$ & $5331$ & $5150$ \\
$T_{cc1}^f(4^3S_1)$ & $0(1^+)$ & $5305$ & $5180$ & $5096$ & $5012$ & $5270$ & $5164$ \\
$T_{cc0}^a(4^1S_0)$ & $1(0^+)$ & $5357$ & $5234$ & $5151$ & $5069$ & $5368$ & $5250$ \\
$T_{cc1}^a(4^3S_1)$ & $1(1^+)$ & $5377$ & $5252$ & $5169$ & $5084$ & $5382$ & $5261$ \\
$T_{cc2}^a(4^5S_2)$ & $1(2^+)$ & $5421$ & $5292$ & $5205$ & $5118$ & $5414$ & $5280$ \\
\bottomrule[1.0pt]\bottomrule[1.0pt]
\end{tabular}
\end{table*}

In the framework of the diquark-antidiquark configuration, the total angular momentum $\bm J$ of the doubly charmed tetraquark $cc\bar u\bar d$ is expressed as
\begin{eqnarray}
\bm J&=&\bm J_{cc}\otimes\bm J_{\bar u\bar d}\otimes\bm L_\lambda,\label{eq35}
\end{eqnarray}
with
\begin{eqnarray}
\bm J_{cc}&=&\bm L_{cc}\otimes\bm S_{cc},\label{eq36}\\
\bm J_{\bar u\bar d}&=&\bm L_{\bar u\bar d}\otimes\bm S_{\bar u\bar d},\label{eq37}\\
\bm S_{cc}&=&\bm S_{c}\otimes\bm S_{c},\label{eq38}\\
\bm S_{\bar u\bar d}&=&\bm S_{\bar u}\otimes\bm S_{\bar d}.\label{eq39}
\end{eqnarray}
Here, the total angular momentum $\bm J_{cc}$ of the doubly charmed diquark $cc$ is acquired by coupling the relative orbital angular momentum $\bm L_{cc}$ between two charm quarks and the spin quantum number $\bm S_{cc}$ of the doubly charmed diquark $cc$. Likewise, the total angular momentum $\bm J_{\bar u\bar d}$ of the light antidiquark $\bar u\bar d$ is procured by coupling the relative orbital angular momentum $\bm L_{\bar u\bar d}$ between two light antiquarks and the spin quantum number $\bm S_{\bar u\bar d}$ of the light antidiquark $\bar u\bar d$. In addition, $\bm L_\lambda$ denotes the relative orbital angular momentum between the doubly charmed diquark $cc$ and the light antidiquark $\bar u\bar d$. For convenience, this work leaves out the relative orbital excitations between two (anti)quarks within the (anti)diquark, i.e.,
\begin{eqnarray}
L_{cc}=L_{\bar u\bar d}&=&0,\label{eq310}\\
J_{cc}=S_{cc}&=&1,\label{eq311}\\
J_{\bar u\bar d}=S_{\bar u\bar d}&=&0\ {\rm or}\ 1.\label{eq312}
\end{eqnarray}
In order to discriminate multifarious low-lying states of the doubly charmed tetraquark $cc\bar u\bar d$, the conventional mesonic notation $n^{2S+1}L_J$ is utilized in this work, as Tables \ref{ccud1}-\ref{ccud3} lay out. Accordingly, the doubly charmed tetraquark states with the total angular momentum $\bm J$ are categorized by the principal quantum number $n$, the orbital angular momentum $L$, and the spin quantum number $\bm S$, i.e.,
\begin{eqnarray}
n&=&n_{cc}+n_{\bar u\bar d}+n_\lambda+1,\label{eq313}\\
L&=&L_{cc}+L_{\bar u\bar d}+L_\lambda,\label{eq314}\\
\bm S&=&\bm S_{cc}\otimes\bm S_{\bar u\bar d}.\label{eq315}
\end{eqnarray}
Here, the principal quantum number $n$ of the doubly charmed tetraquark $cc\bar u\bar d$ is made up of the radial quantum number $n_{cc}$ between two charm quarks, the radial quantum number $n_{\bar u\bar d}$ between two light antiquarks, and the radial quantum number $n_\lambda$ between the diquark $cc$ and the antidiquark $\bar u\bar d$. For the sake of brevity, this work omits the relative radial excitations between two (anti)quarks within the (anti)diquark, leading to
\begin{eqnarray}
n&=&n_\lambda+1.\label{eq316}
\end{eqnarray}

Currently, there are several orthodox angular momentum coupling schemes, including the $L-S$ and $j-j$ coupling schemes. On the basis of the $L-S$ coupling scheme, the total angular momentum $\bm J$ of the doubly charmed tetraquark $cc\bar u\bar d$ is written as \cite{Varshalovich:1988krb}
\begin{eqnarray}
&&\left|\left[\left(\bm L_{cc}\otimes\bm L_{\bar u\bar d}\right)_{\bm L_\rho}\otimes\bm L_\lambda\right]_{\bm L_t}\otimes\left(\bm S_{cc}\otimes\bm S_{\bar u\bar d}\right)_{\bm S}\right\rangle_{\bm J}\nonumber\\
&=&\sum_{J_\rho}\sum_{J_{cc}}\sum_{J_{\bar u\bar d}}(-1)^{L_\lambda+S+L_t+J_\rho}\sqrt{(2L_t+1)(2J_\rho+1)}\nonumber\\
&&\times\sqrt{(2L_\rho+1)(2S+1)(2J_{cc}+1)(2J_{\bar u\bar d}+1)}\nonumber\\
&&\times\left\{
\begin{array}{ccc}
L_\lambda&L_\rho&L_t\\
S&J&J_\rho
\end{array}
\right\}\left\{
\begin{array}{ccc}
L_{cc}&L_{\bar u\bar d}&L_\rho\\
S_{cc}&S_{\bar u\bar d}&S\\
J_{cc}&J_{\bar u\bar d}&J_\rho
\end{array}
\right\}\nonumber\\
&&\times\left|\left[\left(\bm L_{cc}\otimes\bm S_{cc}\right)_{\bm J_{cc}}\otimes\left(\bm L_{\bar u\bar d}\otimes\bm S_{\bar u\bar d}\right)_{\bm J_{\bar u\bar d}}\right]_{\bm J_\rho}\otimes\bm L_\lambda\right\rangle_{\bm J},\qquad\qquad\label{eq317}
\end{eqnarray}
with
\begin{eqnarray}
\bm L_t&=&\bm L_\rho\otimes\bm L_\lambda,\label{eq318}\\
\bm L_\rho&=&\bm L_{cc}\otimes\bm L_{\bar u\bar d},\label{eq319}\\
\bm J_\rho&=&\bm J_{cc}\otimes\bm J_{\bar u\bar d}.\label{eq320}
\end{eqnarray}
Evidently, the $L-S$ coupling scheme is equivalent to the coupling scheme employed by this work, since the internal orbital excitations inside the (anti)diquark are left out, i.e.,
\begin{eqnarray}
\bm J&=&\bm L_t\otimes\bm S=\bm L_\lambda\otimes\bm S=\bm L_\lambda\otimes\bm J_\rho.\label{eq321}
\end{eqnarray}
Therefore, following the exemplary spectroscopic inquiries of heavy-light hadrons \cite{Godfrey:1985xj,Godfrey:2016nwn,Godfrey:2015dva,Godfrey:2004ya,Lu:2017meb,Ali:2017wsf,Song:2015nia,Pang:2017dlw,Song:2015fha,Kim:2021ywp}, this work actually adopts the $L-S$ coupling scheme to unriddle the exotic $cc\bar u\bar d$ states. Moreover, in light of the $j-j$ coupling scheme, the total angular momentum $\bm J$ of the doubly charmed tetraquark $cc\bar u\bar d$ is expressed as \cite{Varshalovich:1988krb}
\begin{eqnarray}
&&\left|\left(\bm L_{cc}\otimes\bm S_{cc}\right)_{\bm J_{cc}}\otimes\left[\left(\bm L_{\bar u\bar d}\otimes\bm S_{\bar u\bar d}\right)_{\bm J_{\bar u\bar d}}\otimes\bm L_\lambda\right]_{\bm J_l}\right\rangle_{\bm J}\nonumber\\
&=&\sum_{J_\rho}(-1)^{J_{cc}+J_{\bar u\bar d}+L_\lambda+J}\sqrt{(2J_\rho+1)(2J_l+1)}\left\{
\begin{array}{ccc}
J_{cc}&J_{\bar u\bar d}&J_\rho\\
L_\lambda&J&J_l
\end{array}
\right\}\nonumber\\
&&\times\left|\left[\left(\bm L_{cc}\otimes\bm S_{cc}\right)_{\bm J_{cc}}\otimes\left(\bm L_{\bar u\bar d}\otimes\bm S_{\bar u\bar d}\right)_{\bm J_{\bar u\bar d}}\right]_{\bm J_\rho}\otimes\bm L_\lambda\right\rangle_{\bm J},\qquad\qquad\label{eq322}
\end{eqnarray}
with
\begin{eqnarray}
\bm J_l&=&\bm J_{\bar u\bar d}\otimes\bm L_\lambda.\label{eq323}
\end{eqnarray}

In the doubly charmed tetraquark system, the isospin properties are determined by the flavor wave function of the light antidiquark $\bar u\bar d$. On the premise of the omission of the internal orbital excitations inside the (anti)diquark, the isospin quantum number $I$ of the doubly charmed tetraquark $cc\bar u\bar d$ is equal to the spin quantum number $S_{\bar u\bar d}$ of the color-triplet ground state light antidiquark $\bar u\bar d$ \cite{Amsler:2018zkm}, i.e.,
\begin{eqnarray}
I=S_{\bar u\bar d}&=&0\ {\rm or}\ 1.\label{eq324}
\end{eqnarray}
Additionally, in terms of the doubly charmed tetraquark $cc\bar u\bar d$ constituted by diquark $cc$ and antidiquark $\bar u\bar d$, the internal parity $P$ is expressed as \cite{Amsler:2018zkm}
\begin{eqnarray}
P&=&(-1)^{L_\lambda}P_{cc}P_{\bar u\bar d}=(-1)^L,\label{eq325}
\end{eqnarray}
with
\begin{eqnarray}
P_{cc}=(-1)^{L_{cc}},&&P_{\bar u\bar d}=(-1)^{L_{\bar u\bar d}}.\label{eq326}
\end{eqnarray}
Here, $P_{cc}$ and $P_{\bar u\bar d}$ denote the internal parity of the diquark $cc$ and the internal parity of the antidiquark $\bar u\bar d$, respectively. Whereafter, in compliance with the definite $I(J^P)$ characteristics of $cc\bar u\bar d$ states, the spectroscopic properties of the low-lying $T_{cc}$ family established by the GI, MGI ($\mu=30$), MGI ($\mu=50$), MGI ($\mu=70$), NR-G, and NR-Y diquark-antidiquark scenarios are fully exhibited in Tables \ref{ccud1}-\ref{ccud3}.

\section{Discussion}\label{sec4}

Concerning the low-lying doubly charmed tetraquark states, this section discusses the root-mean square distance, Regge trajectories, spectroscopic comparison, and mixing angles. What is more, the magic mixing angles of ideal heavy-light tetraquarks are analyzed as well.

\renewcommand\tabcolsep{0.42cm}
\renewcommand{\arraystretch}{1.5}
\begin{table}[!htbp]
\caption{The root-mean square distance of the $1S$-wave $T_{cc}$ states from the GI, MGI, NR-G, and NR-Y scenarios (in unit of fm).}\label{rms}
\begin{tabular}{ccccc}
\toprule[1.0pt]\toprule[1.0pt]
$I(J^P)$ & $0(1^+)$ & $1(0^+)$ & $1(1^+)$ & $1(2^+)$ \\
\midrule[1.0pt]
GI & $0.36$ & $0.28$ & $0.31$ & $0.38$ \\
MGI ($\mu=30$) & $0.37$ & $0.28$ & $0.32$ & $0.39$ \\
MGI ($\mu=50$) & $0.38$ & $0.28$ & $0.32$ & $0.39$ \\
MGI ($\mu=70$) & $0.38$ & $0.29$ & $0.33$ & $0.40$ \\
NR-G & $0.46$ & $0.33$ & $0.36$ & $0.43$ \\
NR-Y & $0.56$ & $0.48$ & $0.50$ & $0.54$ \\
\bottomrule[1.0pt]\bottomrule[1.0pt]
\end{tabular}
\end{table}

\renewcommand\tabcolsep{0.17cm}
\renewcommand{\arraystretch}{1.5}
\begin{table*}[!htbp]
\caption{Spin-averaged masses of low-lying doubly charmed tetraquarks obtained by corresponding Regge trajectories (in unit of MeV).}\label{Regge}
\begin{tabular}{ccccccccccccc}
\toprule[1.0pt]\toprule[1.0pt]
State & \multicolumn{2}{c}{GI} & \multicolumn{2}{c}{MGI ($\mu=30$)} & \multicolumn{2}{c}{MGI ($\mu=50$)} & \multicolumn{2}{c}{MGI ($\mu=70$)} & \multicolumn{2}{c}{NR-G} & \multicolumn{2}{c}{NR-Y} \\
$I(nL)$ & $m_\text{RT}$ & $\left|m_\text{RT}-m_\text{th}\right|$ & $m_\text{RT}$ & $\left|m_\text{RT}-m_\text{th}\right|$ & $m_\text{RT}$ & $\left|m_\text{RT}-m_\text{th}\right|$ & $m_\text{RT}$ & $\left|m_\text{RT}-m_\text{th}\right|$ & $m_\text{RT}$ & $\left|m_\text{RT}-m_\text{th}\right|$ & $m_\text{RT}$ & $\left|m_\text{RT}-m_\text{th}\right|$ \\
\midrule[1.0pt]
$0(1S)$ & $3909$ & $39$ & $3904$ & $13$ & $3900$ & $3$ & $3897$ & $20$ & $3882$ & $2$ & $3746$ & $130$ \\
$0(2S)$ & $4555$ & $21$ & $4490$ & $12$ & $4446$ & $5$ & $4402$ & $3$ & $4508$ & $19$ & $4435$ & $47$ \\
$0(3S)$ & $4964$ & $8$ & $4868$ & $0$ & $4803$ & $6$ & $4739$ & $11$ & $4925$ & $12$ & $4829$ & $28$ \\
$0(4S)$ & $5289$ & $16$ & $5170$ & $10$ & $5090$ & $6$ & $5010$ & $2$ & $5261$ & $9$ & $5137$ & $27$ \\
$0(1P)$ & $4383$ & $6$ & $4342$ & $4$ & $4314$ & $3$ & $4286$ & $1$ & $4308$ & $26$ & $4180$ & $7$ \\
$0(2P)$ & $4839$ & $15$ & $4759$ & $9$ & $4706$ & $5$ & $4652$ & $1$ & $4777$ & $2$ & $4654$ & $30$ \\
$0(3P)$ & $5186$ & $3$ & $5080$ & $6$ & $5009$ & $7$ & $4938$ & $9$ & $5138$ & $5$ & $4996$ & $8$ \\
$0(1D)$ & $4703$ & $4$ & $4642$ & $6$ & $4601$ & $7$ & $4560$ & $9$ & $4614$ & $7$ & $4452$ & $6$ \\
$0(2D)$ & $5077$ & $5$ & $4986$ & $6$ & $4925$ & $7$ & $4863$ & $7$ & $5007$ & $4$ & $4841$ & $0$ \\
$0(1F)$ & $4961$ & $6$ & $4885$ & $5$ & $4835$ & $4$ & $4784$ & $3$ & $4867$ & $8$ & $4668$ & $9$ \\
$1(1S)$ & $4028$ & $9$ & $4012$ & $12$ & $4001$ & $25$ & $3991$ & $39$ & $4108$ & $48$ & $4026$ & $55$ \\
$1(2S)$ & $4647$ & $13$ & $4578$ & $3$ & $4531$ & $3$ & $4485$ & $9$ & $4665$ & $0$ & $4581$ & $28$ \\
$1(3S)$ & $5057$ & $3$ & $4958$ & $4$ & $4892$ & $9$ & $4825$ & $14$ & $5067$ & $1$ & $4952$ & $18$ \\
$1(4S)$ & $5387$ & $12$ & $5266$ & $6$ & $5184$ & $3$ & $5103$ & $1$ & $5398$ & $1$ & $5251$ & $19$ \\
$1(1P)$ & $4479$ & $11$ & $4432$ & $8$ & $4401$ & $7$ & $4369$ & $6$ & $4494$ & $35$ & $4361$ & $6$ \\
$1(2P)$ & $4933$ & $9$ & $4850$ & $4$ & $4794$ & $0$ & $4738$ & $4$ & $4934$ & $9$ & $4791$ & $19$ \\
$1(3P)$ & $5284$ & $5$ & $5175$ & $8$ & $5102$ & $9$ & $5029$ & $11$ & $5286$ & $0$ & $5118$ & $4$ \\
$1(1D)$ & $4798$ & $7$ & $4733$ & $9$ & $4689$ & $10$ & $4644$ & $12$ & $4791$ & $14$ & $4608$ & $2$ \\
$1(2D)$ & $5175$ & $6$ & $5080$ & $7$ & $5016$ & $8$ & $4952$ & $9$ & $5168$ & $8$ & $4973$ & $2$ \\
$1(1F)$ & $5059$ & $3$ & $4979$ & $2$ & $4926$ & $2$ & $4872$ & $1$ & $5042$ & $14$ & $4814$ & $6$ \\
\midrule[1.0pt]
Coefficient & $I=0$ & $I=1$ & $I=0$ & $I=1$ & $I=0$ & $I=1$ & $I=0$ & $I=1$ & $I=0$ & $I=1$ & $I=0$ & $I=1$ \\
\midrule[1.0pt]
$\beta_n$ (GeV$^2$) & $1.17$ & $1.25$ & $1.03$ & $1.10$ & $0.94$ & $1.01$ & $0.85$ & $0.92$ & $1.30$ & $1.38$ & $0.99$ & $1.04$ \\
$\beta_l$ (GeV$^2$) & $0.78$ & $0.83$ & $0.70$ & $0.76$ & $0.66$ & $0.71$ & $0.61$ & $0.66$ & $0.80$ & $0.89$ & $0.52$ & $0.55$ \\
$\beta_0$ (GeV$^2$) & $0.34$ & $0.49$ & $0.34$ & $0.48$ & $0.34$ & $0.47$ & $0.35$ & $0.46$ & $0.53$ & $0.91$ & $0.14$ & $0.43$ \\
\bottomrule[1.0pt]\bottomrule[1.0pt]
\end{tabular}
\end{table*}

\renewcommand\tabcolsep{0.32cm}
\renewcommand{\arraystretch}{1.5}
\begin{table*}[!htbp]
\caption{A comparison of the ground state masses of the doubly charmed and light diquarks from this work (GI, MGI, and NR models) and other phenomenological approaches, categorized by the $I(J^P)$ feature (in unit of MeV).}\label{diquark}
\begin{tabular}{cccccccc}
\toprule[1.0pt]\toprule[1.0pt]
\{cc\} & $0(1^+)$ & \{cc\} & $0(1^+)$ & \{cc\} & $0(1^+)$ & \{cc\} & $0(1^+)$ \\
\midrule[1.0pt]
Regge \cite{Song:2023izj} & $2865.5$ & NR-G & $3152$ & CQM \cite{Wu:2022gie} & $3300.8$ & CQM \cite{Yan:2018gik} & $3340$ \\
$\chi$QSM \cite{Kucab:2024nkv} & $2912$ & BS \cite{Feng:2013kea} I & $3170$ & MGI ($\mu=70$) & $3309$ & NR-Y & $3369$ \\
BS \cite{Feng:2013kea} III & $3020$ & CQM \cite{Weng:2021hje} & $3171.51$ & MGI ($\mu=50$) & $3314$ & BS \cite{Wallbott:2020jzh} & $3423\pm8$ \\
BO \cite{Lebed:2024zrp} & $3026.0$ & CQM \cite{Guo:2021yws} & $3182.67\pm30$ & MGI ($\mu=30$) & $3320$ & BO \cite{Mutuk:2024vzv} & $3510\pm350$ \\
BS \cite{Feng:2013kea} II & $3100$ & CQM \cite{Ebert:2007rn} & $3226$ & GI \cite{Ferretti:2019zyh} & $3329$ & CQM \cite{Noh:2021lqs} I & $3609$ \\
\midrule[1.0pt]
[ud] & $0(0^+)$ & [ud] & $0(0^+)$ & \{ud\} & $1(1^+)$ & \{ud\} & $1(1^+)$ \\
\midrule[1.0pt]
CQM \cite{Yan:2018gik} & $395$ & GI \cite{Ferretti:2019zyh} & $691$ & CQM \cite{Yan:2018gik} & $395$ & MGI ($\mu=30$) & $814$ \\
Regge \cite{Song:2023izj} & $535$ & CQM \cite{Ebert:2007rn} & $710$ & CQM \cite{Guo:2021yws} & $723.86$ & HQS \cite{Braaten:2020nwp} & $835.7\pm11.1$ \\
HQS \cite{Braaten:2020nwp} & $627.4\pm11.2$ & NR-G \cite{Bi:2015ifa} & $725$ & CQM \cite{Weng:2021hje} & $724.85$ & GI \cite{Ferretti:2019zyh} & $840$ \\
BS \cite{Feng:2013kea} III & $650$ & NR-Y \cite{Bi:2015ifa} & $725$ & Regge \cite{Song:2023izj} & $745$ & CQM \cite{Ebert:2007rn} & $909$ \\
MGI ($\mu=70$) & $650$ & BS \cite{Feng:2013kea} II & $750$ & MGI ($\mu=70$) & $778$ & NR-Y \cite{Kim:2021ywp} & $973$ \\
MGI ($\mu=50$) & $662$ & BS \cite{Feng:2013kea} I & $800$ & MGI ($\mu=50$) & $796$ & BS \cite{Wallbott:2020jzh} & $999\pm60$ \\
MGI ($\mu=30$) & $673$ & BS \cite{Wallbott:2020jzh} & $802\pm77$ & BO \cite{Mutuk:2024vzv} & $810\pm60$ & NR-G \cite{Kim:2021ywp} & $1019$ \\
\bottomrule[1.0pt]\bottomrule[1.0pt]
\end{tabular}
\end{table*}

\renewcommand\tabcolsep{0.15cm}
\renewcommand{\arraystretch}{1.5}
\begin{table*}[!htbp]
\caption{The first part of the comparison of the $1S$-wave doubly charmed tetraquark masses from this work (GI, MGI, and NR models) and other phenomenological approaches, categorized by the $I(J^P)$ feature (in unit of MeV).}\label{Tcc1}
\begin{tabular}{cccccccccc}
\toprule[1.0pt]\toprule[1.0pt]
$1S$ & $0(1^+)$ & $1(0^+)$ & $1(1^+)$ & $1(2^+)$ & $1S$ & $0(1^+)$ & $1(0^+)$ & $1(1^+)$ & $1(2^+)$ \\
\midrule[1.0pt]
CQM \cite{Pepin:1996id} II & $3580$ & $\cdots$ & $\cdots$ & $\cdots$ & CQM \cite{Gelman:2002wf} I & $3845$ & $\cdots$ & $\cdots$ & $\cdots$ \\
BS \cite{Feng:2013kea} III & $3660$ & $\cdots$ & $\cdots$ & $\cdots$ & BO \cite{Maiani:2022qze} IV & $3846$ & $3832$ & $3848$ & $3879$ \\
CQM \cite{He:2023ucd} II & $3696$ & $\cdots$ & $\cdots$ & $\cdots$ & CQM \cite{Deng:2018kly} III & $3847$ & $\cdots$ & $\cdots$ & $\cdots$ \\
CQM \cite{Yan:2018gik} & $3700$ & $4140$ & $4170$ & $4240$ & QSR \cite{Wang:2017uld} II & $3850\pm90$ & $\cdots$ & $\cdots$ & $\cdots$ \\
CQM \cite{Yang:2009zzp} II & $3701.6$ & $4212.9$ & $4131.5$ & $4158.1$ & BO \cite{Maiani:2022qze} II & $3851$ & $\cdots$ & $\cdots$ & $\cdots$ \\
CQM \cite{Chen:2021tnn} & $3704.8$ & $4086.6$ & $4133.3$ & $4159.4$ & LQCD \cite{Junnarkar:2018twb} & $3853\pm11$ & $3761\pm11$ & $\cdots$ & $\cdots$ \\
CQM \cite{Tan:2020ldi} & $3709.3$ & $3725.9$ & $3844.3$ & $3962.5$ & CQM \cite{Meng:2020knc} & $3853$ & $\cdots$ & $\cdots$ & $\cdots$ \\
CQM \cite{Meng:2023jqk} III & $3724.2$ & $\cdots$ & $\cdots$ & $\cdots$ & CQM \cite{Vijande:2009kj} VI & $3856$ & $3862$ & $3914$ & $3991$ \\
BO \cite{Maiani:2019lpu} II & $3725$ & $\cdots$ & $\cdots$ & $\cdots$ & CQM \cite{Park:2024cic} II & $3858.5$ & $\cdots$ & $\cdots$ & $\cdots$ \\
CQM \cite{Pepin:1996id} I & $3727$ & $\cdots$ & $\cdots$ & $\cdots$ & CQM \cite{Vijande:2009kj} III & $3860$ & $3911$ & $3975$ & $4031$ \\
CQM \cite{Deng:2018kly} I & $3731\pm12$ & $3962\pm8$ & $4017\pm7$ & $4013\pm7$ & CQM \cite{Xing:2018bqt} I & $3860$ & $\cdots$ & $\cdots$ & $\cdots$ \\
BO \cite{Maiani:2019lpu} I & $3742$ & $\cdots$ & $\cdots$ & $\cdots$ & CQM \cite{Zhu:2019iwm} & $\cdots$ & $3923\pm59$ & $3957^{+63}_{-62}$ & $4026\pm69$ \\
QSR \cite{Gao:2020bvl} & $3742^{+50}_{-40}$ & $\cdots$ & $\cdots$ & $\cdots$ & CQM \cite{Vijande:2009kj} VII & $3861$ & $3877$ & $3952$ & $\cdots$ \\
CQM \cite{Weng:2021hje} I & $3749.8$ & $3833.2$ & $3946.4$ & $4017.1$ & CQM \cite{Vijande:2009kj} V & $3861$ & $3905$ & $3972$ & $4025$ \\
CQM \cite{Deng:2018kly} IV & $3758$ & $\cdots$ & $\cdots$ & $\cdots$ & CQM \cite{Park:2024cic} I & $3861.5$ & $\cdots$ & $\cdots$ & $\cdots$ \\
CQM \cite{Ma:2023int} III & $3759$ & $\cdots$ & $\cdots$ & $\cdots$ & CQM \cite{Meng:2024yhu} & $3863$ & $\cdots$ & $\cdots$ & $\cdots$ \\
CQM \cite{Guo:2021yws} & $3761.6$ & $3836.7$ & $3950.9$ & $4021.2$ & CQM \cite{Meng:2023for} I & $3863.1$ & $\cdots$ & $\cdots$ & $\cdots$ \\
CQM \cite{Vijande:2003ki} I & $3764$ & $4150$ & $4186$ & $4211$ & CQM \cite{Meng:2023jqk} I & $3864.5$ & $\cdots$ & $\cdots$ & $\cdots$ \\
CQM \cite{Li:2023wug} II & $3769.2$ & $3839.7$ & $3963.7$ & $4034.1$ & QSR \cite{Agaev:2021vur} & $3868\pm124$ & $\cdots$ & $\cdots$ & $\cdots$ \\
CQM \cite{Xing:2018bqt} II & $3770$ & $\cdots$ & $\cdots$ & $\cdots$ & QSR \cite{Agaev:2019qqn} & $\cdots$ & $3845\pm175$ & $\cdots$ & $\cdots$ \\
CQM \cite{Cheng:2020wxa} II & $3773.8$ & $3844.3$ & $3968.4$ & $4038.8$ & CQM \cite{Weng:2021hje} II & $3868.7$ & $3969.2$ & $4053.2$ & $4123.8$ \\
CQM \cite{Yang:2019itm} & $3778$ & $\cdots$ & $\cdots$ & $\cdots$ & CQM \cite{Meng:2023for} II & $3869.9$ & $\cdots$ & $\cdots$ & $\cdots$ \\
CQM \cite{Luo:2017eub} II & $3779$ & $3850$ & $3973$ & $4044$ & BS \cite{Feng:2013kea} I & $3870$ & $\cdots$ & $\cdots$ & $\cdots$ \\
CQM \cite{Lee:2009rt} & $3796.5$ & $\cdots$ & $\cdots$ & $\cdots$ & HQCD \cite{Sonnenschein:2024rzw} II & $3870$ & $\cdots$ & $3926\pm74$ & $\cdots$ \\
BS \cite{Feng:2013kea} II & $3800$ & $\cdots$ & $\cdots$ & $\cdots$ & CQM \cite{He:2023ucd} I & $3870.9$ & $\cdots$ & $\cdots$ & $\cdots$ \\
CQM \cite{Hyodo:2017hue} & $3805$ & $3934$ & $3966$ & $4030$ & BO \cite{Maiani:2022qze} III & $3871$ & $3868$ & $3872$ & $3881$ \\
CQM \cite{Luo:2017eub} III & $3813$ & $\cdots$ & $\cdots$ & $\cdots$ & BO \cite{Maiani:2022qze} I & $3872$ & $\cdots$ & $\cdots$ & $\cdots$ \\
CQM \cite{Deng:2018kly} II & $3817$ & $\cdots$ & $\cdots$ & $\cdots$ & CQM \cite{Noh:2023zoq} & $3872$ & $\cdots$ & $\cdots$ & $\cdots$ \\
CQM \cite{Deng:2022cld} & $3820$ & $\cdots$ & $\cdots$ & $\cdots$ & CQM \cite{Noh:2021lqs} II & $3872.8$ & $\cdots$ & $\cdots$ & $\cdots$ \\
Bag \cite{Carlson:1987hh} & $3835$ & $3805$ & $3845$ & $3955$ & CQM \cite{Wu:2022gie} & $3875.8\pm7.6$ & $4035.4\pm13.6$ & $4058.0\pm9.5$ & $4103.2\pm9.5$ \\
\bottomrule[1.0pt]\bottomrule[1.0pt]
\end{tabular}
\end{table*}

\renewcommand\tabcolsep{0.15cm}
\renewcommand{\arraystretch}{1.5}
\begin{table*}[!htbp]
\caption{The second part of the comparison of the $1S$-wave doubly charmed tetraquark masses from this work (GI, MGI, and NR models) and other phenomenological approaches, categorized by the $I(J^P)$ feature (in unit of MeV).}\label{Tcc2}
\begin{tabular}{cccccccccc}
\toprule[1.0pt]\toprule[1.0pt]
$1S$ & $0(1^+)$ & $1(0^+)$ & $1(1^+)$ & $1(2^+)$ & $1S$ & $0(1^+)$ & $1(0^+)$ & $1(1^+)$ & $1(2^+)$ \\
\midrule[1.0pt]
CQM \cite{Janc:2004qn} II & $3875.9$ & $\cdots$ & $\cdots$ & $\cdots$ & CQM \cite{Ma:2023int} II & $3916$ & $\cdots$ & $\cdots$ & $\cdots$ \\
NR-Y & $3876$ & $4029$ & $4057$ & $4105$ & MGI ($\mu=30$) & $3917$ & $3809$ & $3925$ & $4083$ \\
MGI ($\mu=70$) & $3877$ & $3766$ & $3879$ & $4032$ & CQM \cite{Noh:2021lqs} I & $3920$ & $\cdots$ & $\cdots$ & $\cdots$ \\
CQM \cite{Karliner:2017qjm} II & $3877.1\pm12$ & $\cdots$ & $\cdots$ & $\cdots$ & Bag \cite{Zhang:2021yul} & $3925$ & $4032$ & $4117$ & $4179$ \\
HQS \cite{Mehen:2017nrh} & $\cdots$ & $4043$ & $4064$ & $4107$ & CQM \cite{Vijande:2009kj} II & $3926$ & $4154$ & $4175$ & $4193$ \\
CQM \cite{Li:2023wug} I & $3878.2$ & $3948.8$ & $4072.8$ & $4143.2$ & CQM \cite{Vijande:2009kj} I & $3927$ & $4155$ & $4176$ & $4195$ \\
CQM \cite{Liu:2023vrk} & $3879.2$ & $3975.2$ & $4053.7$ & $4124.7$ & CQM \cite{Semay:1994ht} I & $3931$ & $\cdots$ & $\cdots$ & $\cdots$ \\
CQM \cite{Ma:2023int} I & $3880$ & $\cdots$ & $\cdots$ & $\cdots$ & CQM \cite{Ebert:2007rn} & $3935$ & $4056$ & $4079$ & $4118$ \\
CQM \cite{Karliner:2017qjm} I & $3882.2\pm12$ & $\cdots$ & $\cdots$ & $\cdots$ & HQS \cite{Braaten:2020nwp} & $3947\pm11$ & $4111\pm11$ & $4133\pm11$ & $4177\pm11$ \\
NR-G & $3884$ & $3894$ & $3991$ & $4135$ & GI & $3948$ & $3842$ & $3960$ & $4121$ \\
QSR \cite{Albuquerque:2022weq} III & $\cdots$ & $3878\pm5$ & $\cdots$ & $\cdots$ & $\chi$QSM \cite{Kucab:2024nkv} & $3948$ & $4121$ & $4140$ & $4177$ \\
QSR \cite{Albuquerque:2022weq} I & $3885\pm123$ & $3882\pm129$ & $\cdots$ & $\cdots$ & CQM \cite{Cheng:2020wxa} I & $3961$ & $4087$ & $4122$ & $4194$ \\
QSR \cite{Albuquerque:2022weq} IV & $\cdots$ & $3883\pm3$ & $\cdots$ & $\cdots$ & CQM \cite{Kim:2022mpa} & $3961$ & $4132$ & $4151$ & $4185$ \\
QSR \cite{Albuquerque:2022weq} II & $3886\pm4$ & $3885\pm4$ & $\cdots$ & $\cdots$ & CQM \cite{Park:2018wjk} I & $3965$ & $4104$ & $4158$ & $\cdots$ \\
CQM \cite{Wang:2024vjc} & $3889$ & $4107$ & $4131$ & $4176$ & CQM \cite{Park:2018wjk} II & $3971$ & $4110$ & $4164$ & $\cdots$ \\
CQM \cite{Semay:1994ht} II & $3892$ & $\cdots$ & $\cdots$ & $\cdots$ & CQM \cite{Park:2018wjk} III & $3973.3$ & $\cdots$ & $\cdots$ & $\cdots$ \\
BO \cite{Mutuk:2024vzv} & $\cdots$ & $3904$ & $3952$ & $4001$ & HQS \cite{Eichten:2017ffp} & $3978$ & $4146$ & $4167$ & $4210$ \\
CQM \cite{Mutuk:2023oyz} & $3892$ & $4062$ & $4104$ & $4207$ & CQM \cite{Yang:2009zzp} I & $3996.3$ & $4137.5$ & $4211.7$ & $4253.5$ \\
CQM \cite{Meng:2023jqk} II & $3892.2$ & $\cdots$ & $\cdots$ & $\cdots$ & Regge \cite{Song:2023izj} & $3997$ & $4163$ & $4185$ & $4229$ \\
MGI ($\mu=50$) & $3897$ & $3787$ & $3902$ & $4058$ & CQM \cite{Wang:2022clw} & $3998.90$ & $4069.69$ & $4092.34$ & $4134.59$ \\
CQM \cite{Semay:1994ht} IV & $3899$ & $\cdots$ & $\cdots$ & $\cdots$ & CQM \cite{Cheng:2020wxa} III & $3999.4$ & $4069.9$ & $4194.1$ & $4264.5$ \\
CQM \cite{Vijande:2009kj} IV & $3899$ & $\cdots$ & $\cdots$ & $\cdots$ & QSR \cite{Navarra:2007yw} & $4000\pm200$ & $\cdots$ & $\cdots$ & $\cdots$ \\
BS \cite{Wallbott:2020jzh} & $3900\pm80$ & $3800\pm100$ & $4220\pm440$ & $\cdots$ & CQM \cite{Li:2023wug} III & $4000.2$ & $4070.7$ & $4194.7$ & $4265.1$ \\
QSR \cite{Wang:2017uld} I & $3900\pm90$ & $\cdots$ & $\cdots$ & $\cdots$ & CQM \cite{Noh:2021lqs} III & $4002.2$ & $\cdots$ & $\cdots$ & $\cdots$ \\
QSR \cite{Wang:2017dtg} & $\cdots$ & $3870\pm90$ & $3900\pm90$ & $3950\pm90$ & CQM \cite{Luo:2017eub} I & $4007$ & $4078$ & $4201$ & $4271$ \\
CQM \cite{Park:2018wjk} IV & $3904.3$ & $\cdots$ & $\cdots$ & $\cdots$ & CQM \cite{Lipkin:1986dw} & $4012$ & $\cdots$ & $\cdots$ & $\cdots$ \\
CQM \cite{Janc:2004qn} I & $3904.7$ & $\cdots$ & $\cdots$ & $\cdots$ & CQM \cite{Yang:2009zzp} III & $4037.3$ & $4235.4$ & $4175.5$ & $4201.2$ \\
CQM \cite{Gelman:2002wf} II & $3905$ & $\cdots$ & $\cdots$ & $\cdots$ & HQCD \cite{Sonnenschein:2024rzw} I & $4050.5\pm67.5$ & $\cdots$ & $4048\pm67$ & $\cdots$ \\
CQM \cite{Semay:1994ht} V & $3915$ & $\cdots$ & $\cdots$ & $\cdots$ & CQM \cite{Lu:2020rog} & $4053$ & $4241$ & $4268$ & $4318$ \\
CQM \cite{Semay:1994ht} III & $3916$ & $\cdots$ & $\cdots$ & $\cdots$ & CQM \cite{Vijande:2003ki} II & $4101$ & $4175$ & $4231$ & $4254$ \\
\bottomrule[1.0pt]\bottomrule[1.0pt]
\end{tabular}
\end{table*}

\renewcommand\tabcolsep{0.09cm}
\renewcommand{\arraystretch}{1.5}
\begin{table*}[!htbp]
\caption{A comparison of the $2S$-, $3S$-, $4S$-, and $1D$-wave doubly charmed tetraquark masses from this work (GI, MGI, and NR models) and other phenomenological approaches, categorized by the $I(J^P)$ feature (in unit of MeV).}\label{Tcc3}
\begin{tabular}{ccccccccccccc}
\toprule[1.0pt]\toprule[1.0pt]
$2S$ & $\cdots$ & $0(1^+)$ & $1(0^+)$ & $1(1^+)$ & $1(2^+)$ & $\cdots$ & $2S$ & $\cdots$ & $0(1^+)$ & $1(0^+)$ & $1(1^+)$ & $1(2^+)$ \\
\midrule[1.0pt]
CQM \cite{Meng:2024yhu} & $\cdots$ & $4028$ & $4717$ & $4667$ & $4775$ & $\cdots$ & MGI ($\mu=70$) & $\cdots$ & $4405$ & $4423$ & $4459$ & $4529$ \\
HQCD \cite{Sonnenschein:2024rzw} II & $\cdots$ & $4271$ & $\cdots$ & $4318.5\pm62.5$ & $\cdots$ & $\cdots$ & MGI ($\mu=50$) & $\cdots$ & $4441$ & $4460$ & $4498$ & $4570$ \\
CQM \cite{Chen:2021tnn} & $\cdots$ & $4304$ & $\cdots$ & $4639$ & $4687$ & $\cdots$ & MGI ($\mu=30$) & $\cdots$ & $4478$ & $4497$ & $4537$ & $4612$ \\
CQM \cite{Yang:2019itm} & $\cdots$ & $4310$ & $\cdots$ & $\cdots$ & $\cdots$ & $\cdots$ & NR-G & $\cdots$ & $4489$ & $4609$ & $4637$ & $4694$ \\
CQM \cite{Kim:2022mpa} & $\cdots$ & $4363$ & $4546$ & $4560$ & $4585$ & $\cdots$ & GI & $\cdots$ & $4534$ & $4554$ & $4595$ & $4674$ \\
NR-Y & $\cdots$ & $4388$ & $4523$ & $4539$ & $4568$ & $\cdots$ & HQCD \cite{Sonnenschein:2024rzw} I & $\cdots$ & $4554\pm55$ & $\cdots$ & $4552\pm55$ & $\cdots$ \\
\midrule[1.0pt]
$3S$ & $\cdots$ & $0(1^+)$ & $1(0^+)$ & $1(1^+)$ & $1(2^+)$ & $\cdots$ & $3S$ & $\cdots$ & $0(1^+)$ & $1(0^+)$ & $1(1^+)$ & $1(2^+)$ \\
\midrule[1.0pt]
Regge \cite{Chen:2023web} I & $\cdots$ & $4611$ & $\cdots$ & $4811$ & $\cdots$ & $\cdots$ & HQCD \cite{Sonnenschein:2024rzw} I & $\cdots$ & $4867.5\pm49.5$ & $\cdots$ & $4865.5\pm49.5$ & $\cdots$ \\
HQCD \cite{Sonnenschein:2024rzw} II & $\cdots$ & $4615$ & $\cdots$ & $4657\pm55$ & $\cdots$ & $\cdots$ & MGI ($\mu=30$) & $\cdots$ & $4868$ & $4912$ & $4937$ & $4988$ \\
Regge \cite{Chen:2023web} II & $\cdots$ & $4615$ & $\cdots$ & $4809$ & $\cdots$ & $\cdots$ & NR-G & $\cdots$ & $4913$ & $5027$ & $5046$ & $5085$ \\
MGI ($\mu=70$) & $\cdots$ & $4750$ & $4795$ & $4817$ & $4862$ & $\cdots$ & GI & $\cdots$ & $4956$ & $4999$ & $5026$ & $5082$ \\
NR-Y & $\cdots$ & $4801$ & $4911$ & $4923$ & $4946$ & $\cdots$ & CQM \cite{Meng:2024yhu} & $\cdots$ & $4986$ & $4958$ & $4958$ & $4956$ \\
MGI ($\mu=50$) & $\cdots$ & $4809$ & $4853$ & $4877$ & $4925$ & $\cdots$ & $\cdots$ & $\cdots$ & $\cdots$ & $\cdots$ & $\cdots$ & $\cdots$ \\
\midrule[1.0pt]
$4S$ & $\cdots$ & $0(1^+)$ & $1(0^+)$ & $1(1^+)$ & $1(2^+)$ & $\cdots$ & $4S$ & $\cdots$ & $0(1^+)$ & $1(0^+)$ & $1(1^+)$ & $1(2^+)$ \\
\midrule[1.0pt]
Regge \cite{Chen:2023web} I & $\cdots$ & $4808$ & $\cdots$ & $5009$ & $\cdots$ & $\cdots$ & HQCD \cite{Sonnenschein:2024rzw} I & $\cdots$ & $5152.5\pm45.5$ & $\cdots$ & $5150.5\pm45.5$ & $\cdots$ \\
Regge \cite{Chen:2023web} II & $\cdots$ & $4818$ & $\cdots$ & $5008$ & $\cdots$ & $\cdots$ & NR-Y & $\cdots$ & $5164$ & $5250$ & $5261$ & $5280$ \\
HQCD \cite{Sonnenschein:2024rzw} II & $\cdots$ & $4922$ & $\cdots$ & $4960\pm50$ & $\cdots$ & $\cdots$ & MGI ($\mu=30$) & $\cdots$ & $5180$ & $5234$ & $5252$ & $5292$ \\
CQM \cite{Meng:2024yhu} & $\cdots$ & $\cdots$ & $\cdots$ & $4985$ & $5027$ & $\cdots$ & NR-G & $\cdots$ & $5270$ & $5368$ & $5382$ & $5414$ \\
MGI ($\mu=70$) & $\cdots$ & $5012$ & $5069$ & $5084$ & $5118$ & $\cdots$ & GI & $\cdots$ & $5305$ & $5357$ & $5377$ & $5421$ \\
MGI ($\mu=50$) & $\cdots$ & $5096$ & $5151$ & $5169$ & $5205$ & $\cdots$ & $\cdots$ & $\cdots$ & $\cdots$ & $\cdots$ & $\cdots$ & $\cdots$ \\
\midrule[1.0pt]
$1D$ & $0(1^+)$ & $0(2^+)$ & $0(3^+)$ & $1(0^+)$ & $1(1^+)$ & $1(1^+)$ & $1(2^+)$ & $1(2^+)$ & $1(2^+)$ & $1(3^+)$ & $1(3^+)$ & $1(4^+)$ \\
\midrule[1.0pt]
HQCD \cite{Sonnenschein:2024rzw} II & $\cdots$ & $\cdots$ & $4409$ & $\cdots$ & $\cdots$ & $\cdots$ & $\cdots$ & $\cdots$ & $\cdots$ & $4427\pm60$ & $\cdots$ & $\cdots$ \\
Regge \cite{Chen:2023web} I & $\cdots$ & $\cdots$ & $4446$ & $\cdots$ & $\cdots$ & $\cdots$ & $\cdots$ & $\cdots$ & $\cdots$ & $4615$ & $\cdots$ & $\cdots$ \\
Regge \cite{Chen:2023web} II & $\cdots$ & $\cdots$ & $4447$ & $\cdots$ & $\cdots$ & $\cdots$ & $\cdots$ & $\cdots$ & $\cdots$ & $4613$ & $\cdots$ & $\cdots$ \\
NR-Y & $4428$ & $4440$ & $4457$ & $4549$ & $4575$ & $4558$ & $4616$ & $4588$ & $4575$ & $4628$ & $4604$ & $4644$ \\
HQCD \cite{Sonnenschein:2024rzw} I & $\cdots$ & $\cdots$ & $4554\pm55$ & $\cdots$ & $\cdots$ & $\cdots$ & $\cdots$ & $\cdots$ & $\cdots$ & $4552\pm55$ & $\cdots$ & $\cdots$ \\
MGI ($\mu=70$) & $4554$ & $4564$ & $4579$ & $4628$ & $4643$ & $4636$ & $4646$ & $4668$ & $4650$ & $4654$ & $4669$ & $4664$ \\
MGI ($\mu=50$) & $4592$ & $4603$ & $4619$ & $4670$ & $4686$ & $4678$ & $4688$ & $4712$ & $4693$ & $4696$ & $4713$ & $4707$ \\
NR-G & $4607$ & $4616$ & $4631$ & $4790$ & $4789$ & $4803$ & $4790$ & $4828$ & $4801$ & $4797$ & $4820$ & $4805$ \\
MGI ($\mu=30$) & $4631$ & $4643$ & $4659$ & $4712$ & $4729$ & $4720$ & $4730$ & $4756$ & $4735$ & $4739$ & $4756$ & $4750$ \\
GI & $4689$ & $4701$ & $4718$ & $4776$ & $4793$ & $4783$ & $4792$ & $4822$ & $4799$ & $4802$ & $4821$ & $4813$ \\
\bottomrule[1.0pt]\bottomrule[1.0pt]
\end{tabular}
\end{table*}

\renewcommand\tabcolsep{0.21cm}
\renewcommand{\arraystretch}{1.5}
\begin{table*}[!htbp]
\caption{A comparison of the $1P$- and $1F$-wave doubly charmed tetraquark masses from this work (GI, MGI, and NR models) and other phenomenological approaches, categorized by the $I(J^P)$ feature (in unit of MeV).}\label{Tcc4}
\begin{tabular}{ccccccccccccc}
\toprule[1.0pt]\toprule[1.0pt]
$1P$ & $\cdots$ & $0(0^-)$ & $0(1^-)$ & $0(2^-)$ & $1(0^-)$ & $1(1^-)$ & $1(1^-)$ & $1(1^-)$ & $1(2^-)$ & $1(2^-)$ & $1(3^-)$ & $\cdots$ \\
\midrule[1.0pt]
CQM \cite{Yan:2018gik} & $\cdots$ & $\cdots$ & $3920$ & $\cdots$ & $\cdots$ & $4410$ & $4390$ & $4340$ & $\cdots$ & $\cdots$ & $\cdots$ & $\cdots$ \\
CQM \cite{Vijande:2009kj} IV & $\cdots$ & $\cdots$ & $\cdots$ & $\cdots$ & $\cdots$ & $4380$ & $\cdots$ & $\cdots$ & $\cdots$ & $\cdots$ & $4502$ & $\cdots$ \\
CQM \cite{Vijande:2009kj} VI & $\cdots$ & $\cdots$ & $\cdots$ & $3927$ & $\cdots$ & $4420$ & $\cdots$ & $\cdots$ & $3918$ & $\cdots$ & $4461$ & $\cdots$ \\
CQM \cite{Vijande:2009kj} V & $\cdots$ & $\cdots$ & $\cdots$ & $3996$ & $\cdots$ & $4426$ & $\cdots$ & $\cdots$ & $4004$ & $\cdots$ & $4461$ & $\cdots$ \\
CQM \cite{Semay:1994ht} II & $\cdots$ & $\cdots$ & $\cdots$ & $\cdots$ & $\cdots$ & $\cdots$ & $\cdots$ & $\cdots$ & $\cdots$ & $\cdots$ & $4464$ & $\cdots$ \\
CQM \cite{Semay:1994ht} III & $\cdots$ & $\cdots$ & $\cdots$ & $\cdots$ & $\cdots$ & $\cdots$ & $\cdots$ & $\cdots$ & $\cdots$ & $\cdots$ & $4477$ & $\cdots$ \\
CQM \cite{Semay:1994ht} IV & $\cdots$ & $\cdots$ & $\cdots$ & $\cdots$ & $\cdots$ & $\cdots$ & $\cdots$ & $\cdots$ & $\cdots$ & $\cdots$ & $4484$ & $\cdots$ \\
CQM \cite{Semay:1994ht} V & $\cdots$ & $\cdots$ & $\cdots$ & $\cdots$ & $\cdots$ & $\cdots$ & $\cdots$ & $\cdots$ & $\cdots$ & $\cdots$ & $4504$ & $\cdots$ \\
CQM \cite{Semay:1994ht} I & $\cdots$ & $\cdots$ & $\cdots$ & $\cdots$ & $\cdots$ & $\cdots$ & $\cdots$ & $\cdots$ & $\cdots$ & $\cdots$ & $4512$ & $\cdots$ \\
HQCD \cite{Sonnenschein:2024rzw} II & $\cdots$ & $\cdots$ & $\cdots$ & $4155$ & $\cdots$ & $\cdots$ & $\cdots$ & $\cdots$ & $4175\pm66$ & $\cdots$ & $\cdots$ & $\cdots$ \\
CQM \cite{Wang:2024vjc} & $\cdots$ & $4150$ & $4159$ & $4193$ & $4393$ & $4292$ & $4388$ & $\cdots$ & $4309$ & $4420$ & $4459$ & $\cdots$ \\
NR-Y & $\cdots$ & $4161$ & $4174$ & $4199$ & $4310$ & $4370$ & $4331$ & $4308$ & $4386$ & $4354$ & $4409$ & $\cdots$ \\
CQM \cite{Kim:2022mpa} & $\cdots$ & $\cdots$ & $\cdots$ & $4253$ & $\cdots$ & $4423$ & $\cdots$ & $\cdots$ & $4430$ & $\cdots$ & $4442$ & $\cdots$ \\
Regge \cite{Song:2023izj} & $\cdots$ & $4253$ & $4268$ & $4298$ & $4429$ & $4446$ & $4447$ & $4487$ & $4484$ & $4499$ & $4517$ & $\cdots$ \\
MGI ($\mu=70$) & $\cdots$ & $4252$ & $4271$ & $4305$ & $4295$ & $4369$ & $4361$ & $4300$ & $4388$ & $4385$ & $4412$ & $\cdots$ \\
HQCD \cite{Sonnenschein:2024rzw} I & $\cdots$ & $\cdots$ & $\cdots$ & $4315\pm60$ & $\cdots$ & $\cdots$ & $\cdots$ & $\cdots$ & $4313\pm60$ & $\cdots$ & $\cdots$ & $\cdots$ \\
MGI ($\mu=50$) & $\cdots$ & $4279$ & $4299$ & $4335$ & $4326$ & $4400$ & $4394$ & $4331$ & $4419$ & $4420$ & $4446$ & $\cdots$ \\
NR-G & $\cdots$ & $4289$ & $4311$ & $4356$ & $4408$ & $4534$ & $4501$ & $4425$ & $4554$ & $4535$ & $4577$ & $\cdots$ \\
MGI ($\mu=30$) & $\cdots$ & $4307$ & $4327$ & $4364$ & $4357$ & $4432$ & $4427$ & $4362$ & $4452$ & $4453$ & $4479$ & $\cdots$ \\
GI & $\cdots$ & $4349$ & $4370$ & $4408$ & $4404$ & $4477$ & $4480$ & $4410$ & $4501$ & $4503$ & $4529$ & $\cdots$ \\
\midrule[1.0pt]
$1F$ & $0(2^-)$ & $0(3^-)$ & $0(4^-)$ & $1(1^-)$ & $1(2^-)$ & $1(2^-)$ & $1(3^-)$ & $1(3^-)$ & $1(3^-)$ & $1(4^-)$ & $1(4^-)$ & $1(5^-)$ \\
\midrule[1.0pt]
Regge \cite{Chen:2023web} I & $\cdots$ & $\cdots$ & $4601$ & $\cdots$ & $\cdots$ & $\cdots$ & $\cdots$ & $\cdots$ & $\cdots$ & $4763$ & $\cdots$ & $\cdots$ \\
Regge \cite{Chen:2023web} II & $\cdots$ & $\cdots$ & $4604$ & $\cdots$ & $\cdots$ & $\cdots$ & $\cdots$ & $\cdots$ & $\cdots$ & $4761$ & $\cdots$ & $\cdots$ \\
HQCD \cite{Sonnenschein:2024rzw} II & $\cdots$ & $\cdots$ & $4640$ & $\cdots$ & $\cdots$ & $\cdots$ & $\cdots$ & $\cdots$ & $\cdots$ & $4657\pm55$ & $\cdots$ & $\cdots$ \\
NR-Y & $4664$ & $4674$ & $4687$ & $4777$ & $4800$ & $4785$ & $4830$ & $4809$ & $4798$ & $4840$ & $4822$ & $4852$ \\
HQCD \cite{Sonnenschein:2024rzw} I & $\cdots$ & $\cdots$ & $4774\pm51$ & $\cdots$ & $\cdots$ & $\cdots$ & $\cdots$ & $\cdots$ & $\cdots$ & $4772\pm51$ & $\cdots$ & $\cdots$ \\
MGI ($\mu=70$) & $4780$ & $4785$ & $4792$ & $4874$ & $4868$ & $4881$ & $4858$ & $4893$ & $4874$ & $4862$ & $4882$ & $4867$ \\
MGI ($\mu=50$) & $4831$ & $4837$ & $4845$ & $4929$ & $4922$ & $4937$ & $4912$ & $4949$ & $4929$ & $4916$ & $4938$ & $4921$ \\
NR-G & $4852$ & $4857$ & $4864$ & $5049$ & $5029$ & $5056$ & $5003$ & $5066$ & $5034$ & $5005$ & $5042$ & $5009$ \\
MGI ($\mu=30$) & $4882$ & $4888$ & $4896$ & $4984$ & $4976$ & $4992$ & $4964$ & $5005$ & $4983$ & $4969$ & $4993$ & $4974$ \\
GI & $4958$ & $4965$ & $4973$ & $5065$ & $5056$ & $5074$ & $5043$ & $5088$ & $5064$ & $5047$ & $5074$ & $5053$ \\
\bottomrule[1.0pt]\bottomrule[1.0pt]
\end{tabular}
\end{table*}

\subsection{Root-mean square distance}\label{subsec41}

As mentioned in Section \ref{sec1}, the $T_{cc}(3875)^+$ state was identified as the hadronic molecule by quite a lot of studies \cite{Feijoo:2021ppq,Chen:2021vhg,Huang:2021urd,Fleming:2021wmk,Meng:2021jnw,Hu:2021gdg,Dong:2021bvy,Xin:2021wcr,Ling:2021bir,Du:2021zzh,Chen:2021tnn,Zhao:2021cvg,
Ke:2021rxd,Chen:2021cfl,Deng:2021gnb,Albaladejo:2021vln,Braaten:2022elw,Deng:2022cld,Ortega:2022efc,Wang:2022jop,Dai:2023mxm,Meng:2023jqk,Asanuma:2023atv,
Sakai:2023syt,Qiu:2023uno,Ma:2023int}. For instance, by taking into account the one-boson exchange potential (OBEP) as the dominant interaction within the hadronic molecule, Ref. \cite{Sakai:2023syt} acquired the $T_{cc}$ bound state with $I(J^P)=0(1^+)$, whose binding energy $0.273$ MeV is equal to the one observed by the LHCb Collaboration. In spite of this, the ground state isoscalar doubly charmed tetraquark masses predicted by Refs. \cite{Maiani:2022qze,Noh:2023zoq,Noh:2021lqs,Wu:2022gie,Janc:2004qn,Karliner:2017qjm,Li:2023wug}, MGI ($\mu=70$), and NR-Y diquark scenarios in this work are somewhat close to the experimental value $3874.83\pm0.11$ MeV of the $T_{cc}(3875)^+$ mass. Therefore, it is requisite to make use of other physical indications to discern the diquark-antidiquark mechanism from the hadronic molecular picture. Generally, the multiquarks and hadronic molecules are defined as the compact multiquarks and loosely bound hadronic molecules, respectively, based upon the root-mean square distance. In order to clarify the validity of the diquark models, the root-mean square distance of the $1S$-wave $T_{cc}$ states from the GI, MGI, NR-G, and NR-Y scenarios is presented in Table \ref{rms}. Thereinto, the root-mean square distance of the isoscalar $T_{cc}$ state possesses the minimum value of $0.36$ fm from the GI diquark model and the maximum value of $0.56$ fm from the NR-Y diquark model. Besides, the root-mean square distance of three isovector $T_{cc}$ states possesses the minimum value of $0.28$ fm from the GI diquark model and the maximum value of $0.54$ fm from the NR-Y diquark model. However, in the framework of the hadronic molecular picture, Ref. \cite{Sakai:2023syt} unravels that the root-mean square distance of the $T_{cc}$ state with $I(J^P)=0(1^+)$ is $6.43$ fm. It substantiates that the loosely bound hadronic molecules indeed have a spatial size much larger than the one of compact multiquarks. Hence, the root-mean square distance offers a crucial clue to ascertain the nature of $T_{cc}$ states.

\subsection{Regge trajectories}\label{subsec42}

In consideration of the miscellaneous radial and orbital excitations in Tables \ref{ccud1}-\ref{ccud3}, it is obligatory to examine the eligibility of the excited $T_{cc}$ states with higher orbital and radial numbers. For this purpose, the linear Regge trajectories are employed to reveal the spectroscopic behaviors of the highly excited hadron states. As a successful phenomenological approach with the universality, Regge trajectories have been popularly adopted in the mass spectra analyses of numerous heavy-light hadrons, involving singly heavy mesons \cite{Chen:2023web,Afonin:2014nya,Jia:2018vwl,Pan:2022egn,Chen:2017fcs,Jia:2019see}, singly heavy baryons \cite{Chen:2023web,Chen:2017fcs,Jia:2019see,Chen:2014nyo,Jia:2019bkr,Jia:2020vek,Jakhad:2023mni,Jakhad:2024fin,Pan:2023hwt}, doubly heavy baryons \cite{Chen:2023web,Song:2022csw}, doubly heavy tetraquarks \cite{Song:2023izj,Chen:2023web}, and heavy-light diquarks \cite{Chen:2023web,Chen:2023cws}. Accordingly, following the steps in Ref. \cite{Chen:2014nyo}, the generalized form of the Regge trajectories with regard to the heavy-light hadrons is expressed as
\begin{eqnarray}
(\bar M-m_h)^2&=&\beta_nn+\beta_lL+\beta_0,\label{eq40}
\end{eqnarray}
where $\bar M$ is the spin-averaged mass of the heavy-light hadron, $m_h$ is the mass of the heavy flavored constituent in the heavy-light hadron, $\beta_n$ is the radial slope of Regge trajectory, $\beta_l$ is the orbital slope of Regge trajectory, and $\beta_0$ is the intercept of Regge trajectory. In order to check the reliability of the highly excited $T_{cc}$ states predicted by this work, Regge trajectory for each diquark scenario is carried out by individually fitting the spectroscopic outcomes in Tables \ref{ccud1}-\ref{ccud3}. Thereinto, the fitting input values of $\bar M$ and $m_h$ in here are the theoretically predicted spin-averaged $T_{cc}$ mass $m_\text{th}$ and the doubly charmed diquark mass $m_{\{cc\}}$, respectively. Subsequently, a series of Regge trajectory coefficients $\beta_n$, $\beta_l$, and $\beta_0$ are determined. Furthermore, for the sake of estimating the uncertainties of the highly excited $T_{cc}$ states, the spin-averaged $T_{cc}$ mass $m_\text{RT}$ obtained by corresponding Regge trajectory and the mass difference between $m_\text{RT}$ and $m_\text{th}$ are unveiled in Table \ref{Regge}. It can be found that the mass difference $\left|m_\text{RT}-m_\text{th}\right|$ for most highly excited states is less than $20$ MeV except the NR-Y model. As a matter of fact, Regge trajectory is not very accurate for the ground and lowly excited states \cite{Chen:2017fcs}, so it is normal that there is mass discrepancy for the $1S$-, $1P$-, and $2S$-wave $T_{cc}$ states. If setting the maximum $\left|m_\text{RT}-m_\text{th}\right|$ value among all of the $I(nL)$ states within a single theoretical model as the assessment criteria, the MGI ($\mu=30$) scenario with the maximum $\left|m_\text{RT}-m_\text{th}\right|$ value of $13$ MeV will be the optimal one. Consequently, the following spectroscopic discussion is going to focus mostly on the predicted outcomes of the MGI ($\mu=30$) diquark model.

\subsection{Spectroscopic comparison}\label{subsec43}

Regarding the mass spectra of the $T_{cc}$ family, this subsection is aimed at the phenomenological comparison between this work and other theoretical approaches.

\subsubsection{$cc$ and $ud$ diquarks}\label{subsec431}

Admittedly, the masses of the doubly charmed and light diquarks are imperative for the $T_{cc}$ spectroscopy in the diquark-antidiquark picture. Currently, various results for the masses of the $cc$ and $ud$ diquarks have been rendered by a lot of Refs. \cite{Ebert:2007rn,Feng:2013kea,Yan:2018gik,Wallbott:2020jzh,Noh:2021lqs,Braaten:2020nwp,Weng:2021hje,Guo:2021yws,Song:2023izj,Wu:2022gie,Kucab:2024nkv,Mutuk:2024vzv,
Lebed:2024zrp}. A broad range of the diquark $cc$ masses between 2865.5 \cite{Song:2023izj} and 3609 MeV \cite{Noh:2021lqs} is clearly displayed in Table \ref{diquark}. Thereinto, the theoretical mass 3320 MeV of the diquark $cc$ acquired by the MGI ($\mu=30$) model is close to the predicted values 3340 and 3300.8 MeV offered by Refs. \cite{Yan:2018gik,Wu:2022gie}. When it comes to the scalar light diquark, the masses are spread across the energy interval between 395 \cite{Yan:2018gik} and 802 MeV \cite{Wallbott:2020jzh}. The scalar diquark $ud$ mass 673 MeV in the MGI ($\mu=30$) model is in the vicinity of the theoretical value 650 MeV reaped by Ref. \cite{Feng:2013kea}. In terms of the axial-vector light diquark, the masses lie on the range between 395 \cite{Yan:2018gik} and 1019 MeV \cite{Kim:2021ywp}. Here, the MGI ($\mu=30$) model mass 814 MeV is in consonance with the axial-vector diquark $ud$ mass 810 MeV utilized by Ref. \cite{Mutuk:2024vzv}. Additionally, the mass value 814 MeV is also adjacent to the axial-vector light diquark mass 835.7 MeV derived by Ref. \cite{Braaten:2020nwp}. It is noteworthy that the diquark masses in the GI (MGI) model are calculated on the basis of the universal parameters in the mesons \cite{Godfrey:1985xj}. In order to keep the parameter universality among the mesons, diquarks, and tetraquarks, the mass distribution in Table \ref{diquark} cannot be deemed as the systematical uncertainties.

\subsubsection{$1S$-wave $T_{cc}$ states}\label{subsec432}

Currently, the phenomenological explorations of the $T_{cc}$ spectroscopy are mainly focused on the $1S$-wave isoscalar state. In spite of this, a sizable discrepancy between the minimum theoretical value 3580 MeV \cite{Pepin:1996id} and the maximum theoretical value 4101 MeV \cite{Vijande:2003ki} with regard to the ground state isoscalar $cc\bar u\bar d$ tetraquark mass is listed in Tables \ref{Tcc1}-\ref{Tcc2}. Thereinto, most of the theoretical outcomes for the $1S$-wave isoscalar $cc\bar u\bar d$ tetraquark mass procured by this work (GI, MGI, NR-G, and NR-Y scenarios) are in the energy range between 3876 and 3948 MeV, somewhat higher than the observed value $3874.83\pm0.11$ MeV of the $T_{cc}(3875)^+$ structure. In addition, the predicted result 3917 MeV of the isoscalar $T_{cc}$ mass acquired by the MGI ($\mu=30$) scenario is in accord with the outcomes garnered by the MIT bag model \cite{Zhang:2021yul} and the constituent quark model \cite{Semay:1994ht,Vijande:2009kj,Noh:2021lqs,Ma:2023int}. As the isospin siblings, the $1S$-wave isovector $T_{cc}$ states are significant in the current $cc\bar u\bar d$ tetraquark issues. As enumerated in Tables \ref{Tcc1}-\ref{Tcc2}, this work (GI, MGI, NR-G, and NR-Y scenarios) exhibits the prediction ambit of the ground state isovector $T_{cc}$ masses between 3766 and 4135 MeV. As far as the $1S$-wave $cc\bar u\bar d$ state with $I(J^P)=1(0^+)$ is concerned, the predicted mass is lower than the one of the isoscalar state in the GI and MGI models. Concretely, the mass obtained by the MGI ($\mu=30$) scenario is $3809$ MeV, lower than the value $3917$ MeV of the isoscalar state. Moreover, the analogous case is espoused by the QCD sum rules \cite{Wang:2017uld,Wang:2017dtg,Agaev:2019qqn,Albuquerque:2022weq,Agaev:2021vur}, the BO approximation \cite{Maiani:2022qze}, the BS equation \cite{Wallbott:2020jzh}, and the lattice QCD \cite{Junnarkar:2018twb}. However, the NR-G and NR-Y models predict that the isovector scalar $T_{cc}$ mass is higher than the isoscalar axial-vector $T_{cc}$ mass. Hence, the mass gap between these two states necessitates the further quest of the experiments. Besides, the mass $4083$ MeV of the isovector tensor state is predicted by the MGI ($\mu=30$) model.

\subsubsection{$2S$-, $3S$-, and $4S$-wave $T_{cc}$ states}\label{subsec433}

Nowadays, the $cc\bar u\bar d$ tetraquark states with radial excitations are rarely probed by phenomenological theories \cite{Yang:2019itm,Kim:2022mpa,Chen:2021tnn,Chen:2023web,Meng:2024yhu,Sonnenschein:2024rzw}. As unveiled in Table \ref{Tcc3}, the minimum and maximum theoretical masses of the $2S$-wave doubly charmed tetraquark states predicted by this work (GI, MGI, NR-G, and NR-Y scenarios) are 4388 and 4694 MeV, respectively. The masses 4478, 4497, 4537, and 4612 MeV of the $2S$-wave $cc\bar u\bar d$ tetraquarks are predicted by the MGI ($\mu=30$) model. Thereinto, the isoscalar value 4478 MeV is higher than the one acquired by the constituent quark model \cite{Yang:2019itm,Kim:2022mpa,Meng:2024yhu,Chen:2021tnn}. When it comes to the $3S$-wave $cc\bar u\bar d$ tetraquarks, the predicted masses offered by this work are in the realm between 4750 and 5085 MeV, approaching to the consequences garnered by Refs. \cite{Meng:2024yhu,Sonnenschein:2024rzw}. Concerning the masses of the $4S$-wave doubly charmed tetraquark states, the theoretical values achieved by this work are in the range between 5012 and 5421 MeV, solely comporting with the outcomes reaped by Ref. \cite{Sonnenschein:2024rzw}. Based on the $1S$- and $2S$-wave $T_{cc}$ masses predicted by Ref. \cite{Kim:2022mpa}, the Regge trajectory relation is exploited by Ref. \cite{Chen:2023web} to reveal the $3S$- and $4S$-wave $cc\bar u\bar d$ tetraquark masses which are lower than the results of all the scenarios in this work.

\subsubsection{$1P$-, $1D$-, and $1F$-wave $T_{cc}$ states}\label{subsec434}

So far, the spectroscopic properties of the orbitally excited doubly charmed tetraquark states are merely surveyed by a minority of theoretical prescriptions, including the constituent quark model \cite{Semay:1994ht,Vijande:2009kj,Yan:2018gik,Kim:2022mpa,Wang:2024vjc}, the Regge trajectory relation \cite{Song:2023izj,Chen:2023web}, and the holographic QCD \cite{Sonnenschein:2024rzw}. The masses of the $1P$-wave isoscalar and isovector doubly charmed tetraquark states predicted by this work lie on the range between 4161 and 4408 MeV and the extent between 4295 and 4577 MeV, respectively. As Table \ref{Tcc4} demonstrates, the $1P$-wave isoscalar $cc\bar u\bar d$ tetraquark masses predicted by Refs. \cite{Vijande:2009kj,Yan:2018gik} are manifestly lower than the theoretical outcomes of the MGI ($\mu=30$) scenario. Nevertheless, the predicted masses of the $1P$-wave isovector $T_{cc}$ states from the MGI ($\mu=30$) model are close to the values derived by the constituent quark model \cite{Semay:1994ht,Yan:2018gik,Kim:2022mpa,Wang:2024vjc} and the Regge trajectory relation \cite{Song:2023izj}. According to Tables \ref{Tcc3}-\ref{Tcc4}, the holographic QCD \cite{Sonnenschein:2024rzw} and the Regge trajectory relation \cite{Chen:2023web} are the few available theoretical recipes with respect to the mass spectra of the $1D$- and $1F$-wave $T_{cc}$ tetraquarks except this work. The masses of the $1D$-wave isoscalar and isovector $cc\bar u\bar d$ states in this work are in the range between 4428 and 4718 MeV and the scope between 4549 and 4828 MeV, respectively. Apart from that, the $1F$-wave isoscalar and isovector $T_{cc}$ masses in this work are in the interval between 4664 and 4973 MeV and the realm between 4777 and 5088 MeV, respectively. In light of the $1S$- and $1P$-wave $T_{cc}$ masses reaped by the constituent quark model \cite{Kim:2022mpa}, the Regge trajectory relation \cite{Chen:2023web} predicts the $1D$- and $1F$-wave $cc\bar u\bar d$ tetraquark masses which are lower than the outcomes of all the models in this work. Owing to the sheer paucity of the spectroscopic information about the $cc\bar u\bar d$ states with the joint radial and orbital excitations, this work investigates the mass spectra of the $2P$-, $3P$-, and $2D$-wave $T_{cc}$ tetraquarks for the first time. Thus it can be seen that the further phenomenological exploration and experimental probe are requisite in order to shed light on the spectroscopic properties of the low-lying excited doubly charmed tetraquark states.

\subsection{Mixing angles}\label{subsec44}

\renewcommand\tabcolsep{0.24cm}
\renewcommand{\arraystretch}{1.5}
\begin{table*}[!htbp]
\caption{The mixing angles of the isovector $T_{cc}$ states and the magic mixing angles in the ideal heavy quark limit achieved by this work.}\label{angle}
\begin{tabular}{p{1.77cm}<{\centering}p{1.77cm}<{\centering}p{1.77cm}<{\centering}p{1.77cm}<{\centering}p{1.77cm}<{\centering}p{1.77cm}<{\centering}
p{1.77cm}<{\centering}p{1.77cm}<{\centering}}
\toprule[1.0pt]\toprule[1.0pt]
Mixing & GI & MGI ($\mu=30$) & MGI ($\mu=50$) & MGI ($\mu=70$) & NR-G & NR-Y & HQL \\
\midrule[1.0pt]
$1^1P_1$$\leftrightarrow$$1^3P_1$ & $-44.4^\circ$ & $44.5^\circ$ & $43.6^\circ$ & $42.9^\circ$ & $41.9^\circ$ & $40.5^\circ$ & $40.9^\circ$ \\
$1^3P_1$$\leftrightarrow$$1^5P_1$ & $9.4^\circ$ & $-7.4^\circ$ & $-7.0^\circ$ & $-6.6^\circ$ & $-7.7^\circ$ & $12.4^\circ$ & $40.9^\circ$ \\
$1^1P_1$$\leftrightarrow$$1^5P_1$ & $8.1^\circ$ & $9.6^\circ$ & $9.7^\circ$ & $9.8^\circ$ & $10.8^\circ$ & $9.5^\circ$ & $9.6^\circ$ \\
$1^3P_2$$\leftrightarrow$$1^5P_2$ & $7.4^\circ$ & $-20.1^\circ$ & $-41.2^\circ$ & $39.7^\circ$ & $28.3^\circ$ & $28.7^\circ$ & $30.0^\circ$ \\
$2^1P_1$$\leftrightarrow$$2^3P_1$ & $37.7^\circ$ & $37.2^\circ$ & $36.6^\circ$ & $35.6^\circ$ & $42.3^\circ$ & $40.6^\circ$ & $40.9^\circ$ \\
$2^3P_1$$\leftrightarrow$$2^5P_1$ & $-11.3^\circ$ & $-10.6^\circ$ & $-10.1^\circ$ & $-9.4^\circ$ & $-7.1^\circ$ & $12.3^\circ$ & $40.9^\circ$ \\
$2^1P_1$$\leftrightarrow$$2^5P_1$ & $11.5^\circ$ & $11.4^\circ$ & $11.3^\circ$ & $11.3^\circ$ & $11.1^\circ$ & $9.6^\circ$ & $9.6^\circ$ \\
$2^3P_2$$\leftrightarrow$$2^5P_2$ & $23.2^\circ$ & $22.7^\circ$ & $22.1^\circ$ & $21.1^\circ$ & $30.5^\circ$ & $29.1^\circ$ & $30.0^\circ$ \\
$3^1P_1$$\leftrightarrow$$3^3P_1$ & $38.1^\circ$ & $37.4^\circ$ & $36.8^\circ$ & $35.9^\circ$ & $42.3^\circ$ & $40.7^\circ$ & $40.9^\circ$ \\
$3^3P_1$$\leftrightarrow$$3^5P_1$ & $-14.1^\circ$ & $-12.9^\circ$ & $-12.1^\circ$ & $-11.2^\circ$ & $-6.1^\circ$ & $12.3^\circ$ & $40.9^\circ$ \\
$3^1P_1$$\leftrightarrow$$3^5P_1$ & $13.8^\circ$ & $13.1^\circ$ & $12.7^\circ$ & $12.4^\circ$ & $10.9^\circ$ & $9.6^\circ$ & $9.6^\circ$ \\
$3^3P_2$$\leftrightarrow$$3^5P_2$ & $21.8^\circ$ & $21.2^\circ$ & $20.7^\circ$ & $19.7^\circ$ & $31.1^\circ$ & $29.4^\circ$ & $30.0^\circ$ \\
$1^3D_1$$\leftrightarrow$$1^5D_1$ & $-35.8^\circ$ & $-33.9^\circ$ & $-32.4^\circ$ & $-30.8^\circ$ & $44.5^\circ$ & $23.4^\circ$ & $30.0^\circ$ \\
$1^1D_2$$\leftrightarrow$$1^3D_2$ & $43.7^\circ$ & $44.0^\circ$ & $44.3^\circ$ & $44.7^\circ$ & $45.3^\circ$ & $43.8^\circ$ & $45.0^\circ$ \\
$1^3D_2$$\leftrightarrow$$1^5D_2$ & $-25.6^\circ$ & $-25.0^\circ$ & $-24.7^\circ$ & $-24.3^\circ$ & $-27.4^\circ$ & $38.5^\circ$ & $52.2^\circ$ \\
$1^1D_2$$\leftrightarrow$$1^5D_2$ & $12.5^\circ$ & $13.1^\circ$ & $13.6^\circ$ & $14.3^\circ$ & $15.3^\circ$ & $13.7^\circ$ & $15.0^\circ$ \\
$1^3D_3$$\leftrightarrow$$1^5D_3$ & $34.3^\circ$ & $34.7^\circ$ & $35.0^\circ$ & $35.3^\circ$ & $35.9^\circ$ & $32.4^\circ$ & $35.3^\circ$ \\
$2^3D_1$$\leftrightarrow$$2^5D_1$ & $-42.7^\circ$ & $-40.7^\circ$ & $-39.2^\circ$ & $-37.3^\circ$ & $-43.0^\circ$ & $23.8^\circ$ & $30.0^\circ$ \\
$2^1D_2$$\leftrightarrow$$2^3D_2$ & $44.0^\circ$ & $44.3^\circ$ & $44.5^\circ$ & $44.7^\circ$ & $44.7^\circ$ & $44.0^\circ$ & $45.0^\circ$ \\
$2^3D_2$$\leftrightarrow$$2^5D_2$ & $-27.2^\circ$ & $-26.6^\circ$ & $-26.1^\circ$ & $-25.6^\circ$ & $-26.3^\circ$ & $39.5^\circ$ & $52.2^\circ$ \\
$2^1D_2$$\leftrightarrow$$2^5D_2$ & $13.3^\circ$ & $13.7^\circ$ & $14.0^\circ$ & $14.4^\circ$ & $14.1^\circ$ & $13.9^\circ$ & $15.0^\circ$ \\
$2^3D_3$$\leftrightarrow$$2^5D_3$ & $34.2^\circ$ & $34.5^\circ$ & $34.7^\circ$ & $34.9^\circ$ & $35.9^\circ$ & $32.9^\circ$ & $35.3^\circ$ \\
$1^3F_2$$\leftrightarrow$$1^5F_2$ & $38.0^\circ$ & $38.4^\circ$ & $38.8^\circ$ & $39.1^\circ$ & $37.1^\circ$ & $29.6^\circ$ & $35.3^\circ$ \\
$1^1F_3$$\leftrightarrow$$1^3F_3$ & $46.0^\circ$ & $46.2^\circ$ & $46.3^\circ$ & $46.4^\circ$ & $46.5^\circ$ & $45.2^\circ$ & $46.7^\circ$ \\
$1^3F_3$$\leftrightarrow$$1^5F_3$ & $-31.5^\circ$ & $-31.1^\circ$ & $-30.9^\circ$ & $-30.7^\circ$ & $-32.1^\circ$ & $46.9^\circ$ & $55.9^\circ$ \\
$1^1F_3$$\leftrightarrow$$1^5F_3$ & $16.3^\circ$ & $16.6^\circ$ & $16.8^\circ$ & $17.0^\circ$ & $17.1^\circ$ & $15.5^\circ$ & $17.4^\circ$ \\
$1^3F_4$$\leftrightarrow$$1^5F_4$ & $36.7^\circ$ & $37.0^\circ$ & $37.2^\circ$ & $37.4^\circ$ & $37.4^\circ$ & $34.4^\circ$ & $37.8^\circ$ \\
\bottomrule[1.0pt]\bottomrule[1.0pt]
\end{tabular}
\end{table*}

Conventionally, the meson with the certain orbital angular momentum may exist as the states that possess the identical total angular momentum and the discrepant spin quantum number, e.g., the $^1P_1$ and $^3P_1$ states. For the case of the hidden-flavor quarkonium, the $^1P_1$ and $^3P_1$ states cannot mix since the equal mass of the quark and antiquark engenders the diagonal spin-orbit interaction \cite{Barnes:2005pb,Godfrey:2015dia}. On the contrary, the heavy-light meson constituted of a quark and antiquark with unequal mass may manifest as an admixture of $^1P_1$ and $^3P_1$ states due to the advent of the off-diagonal term in the spin-orbit interaction \cite{Godfrey:2016nwn,Godfrey:2015dva}. For instance, the physical $P_1$ states can be expressed as the linear combinations of the unmixing $^1P_1$ and $^3P_1$ states \cite{Godfrey:1985xj,Godfrey:2016nwn,Godfrey:2015dva,Godfrey:2004ya}, i.e.,
\begin{eqnarray}
P_1&=&{^1}P_1\cos\theta_{nP_1}+{^3}P_1\sin\theta_{nP_1},\nonumber\\
P_1^\prime&=&-{^1}P_1\sin\theta_{nP_1}+{^3}P_1\cos\theta_{nP_1}.\label{eq41}
\end{eqnarray}

Unlike a mixture of two states in the heavy-light meson, the isovector doubly charmed tetraquark with the diquark-antidiquark configuration not only involves the mixing between two states, but also touches upon the mixing among three states, e.g., the $^1P_1$, $^3P_1$, and $^5P_1$ states. Concretely, the mixing of two states in the isovector doubly charmed tetraquark holds the generic form of
\begin{eqnarray}
\left(\begin{array}{c}
nL_J\\
nL_J^\prime\\
\end{array}\right)&=&
\left(\begin{array}{cc}
\cos\theta_{nL_J}&\sin\theta_{nL_J}\\
-\sin\theta_{nL_J}&\cos\theta_{nL_J}\\
\end{array}\right)
\left(\begin{array}{c}
n^3L_J\\
n^5L_J\\
\end{array}\right).\label{eq42}
\end{eqnarray}
Akin to the cause of mixing in the heavy-light meson, the mixture between $n^3L_J$ and $n^5L_J$ states is solely generated by the off-diagonal term in the spin-orbit interaction. An intriguing fingerprint of the orbitally excited isovector $T_{cc}$ states is the admixture among three states, expressed as a $3\times3$ unitary matrix by mimicking the prestigious Cabibbo-Kobayashi-Maskawa (CKM) matrix \cite{ParticleDataGroup:2024cfk}, i.e.,
\begin{eqnarray}
\left(\begin{array}{c}
nL_J\\
nL_J^\prime\\
nL_J^{\prime\prime}\\
\end{array}\right)&=&V_{123}
\left(\begin{array}{c}
n^1L_J\\
n^3L_J\\
n^5L_J\\
\end{array}\right),\label{eq43}
\end{eqnarray}
with
\begin{eqnarray}
V_{123}&=&\left(\begin{array}{ccc}
1&0&0\\
0&c_{23}&s_{23}\\
0&-s_{23}&c_{23}\\
\end{array}\right)
\left(\begin{array}{ccc}
c_{13}&0&s_{13}\\
0&1&0\\
-s_{13}&0&c_{13}\\
\end{array}\right)
\left(\begin{array}{ccc}
c_{12}&s_{12}&0\\
-s_{12}&c_{12}&0\\
0&0&1\\
\end{array}\right)\nonumber\\
&=&\left(\begin{array}{ccc}
c_{12}c_{13}&s_{12}c_{13}&s_{13}\\
-s_{12}c_{23}-c_{12}s_{23}s_{13}&c_{12}c_{23}-s_{12}s_{23}s_{13}&s_{23}c_{13}\\
s_{12}s_{23}-c_{12}c_{23}s_{13}&-c_{12}s_{23}-s_{12}c_{23}s_{13}&c_{23}c_{13}\\
\end{array}\right).\nonumber
\end{eqnarray}
Here, the $\cos\theta_{nL_J}$ and $\sin\theta_{nL_J}$ are abbreviated as $c$ and $s$, respectively, for the sake of brevity. Apparently, any two of the $n^1L_J$, $n^3L_J$, and $n^5L_J$ states are able to mix each other when the total angular momentum is equal to the orbital angular momentum, i.e., $J=L$. Thereinto, two sorts of mixings ($^1L_J$$\leftrightarrow$$^3L_J$ and $^3L_J$$\leftrightarrow$$^5L_J$) are derived from the off-diagonal terms in the spin-orbit interaction. Moreover, the $^1L_J$$\leftrightarrow$$^5L_J$ mixing is caused by the off-diagonal term in the tensor interaction, distinct from the conventional heavy-light meson. In this work, the mixed states in Eq. (\ref{eq42}) whose major ingredients are the $n^3L_J$ and $n^5L_J$ states are assigned as the $nL_J$ and $nL_J^\prime$ states, respectively. Analogously, the mixed states in Eq. (\ref{eq43}) whose major ingredients are the $n^1L_J$, $n^3L_J$, and $n^5L_J$ states are assigned as the $nL_J$, $nL_J^\prime$, and $nL_J^{\prime\prime}$ states, respectively. As uncovered in Table \ref{angle}, the sign of the mixing angle is contingent on two facets: the sign of the off-diagonal matrix element and the order of the $nL_J$, $nL_J^\prime$, and $nL_J^{\prime\prime}$ masses. Taking the $1^3P_2$$\leftrightarrow$$1^5P_2$ mixing as an illustration, although the signs of the off-diagonal matrix elements reaped by the GI and NR-G models are inverse, the signs of the mixing angles exported by these two models are fortuitously same inasmuch as the orders of the $1P_2$ and $1P_2^\prime$ masses stemming from the GI and NR-G models are also inverse.

In the ideal heavy quark limit (HQL), the total angular momentum $J$ of the double-heavy tetraquark is limned by the spin quantum number $S_{QQ}$ of the infinitely heavy diquark and the total angular momentum $J_l$ of the light diquark degrees of freedom, on the basis of the $j-j$ coupling scheme. According to the explicit relations between the $j-j$ and $L-S$ coupling schemes with respect to the heavy-light hadrons in the Appendix \ref{app}, the magic mixing angles of the orbitally excited double-heavy tetraquarks are systematically enumerated in Table \ref{angle}. In the $m_Q\rightarrow\infty$ limit, the $nL_{L-1}$, $nL_{L-1}^\prime$, $nL_{L+1}$, and $nL_{L+1}^\prime$ states garnered by Eq. (\ref{eq42}) automatically degenerate into the $nL_{L-1}(J_l=L)$, $nL_{L-1}(J_l=L-1)$, $nL_{L+1}(J_l=L+1)$, and $nL_{L+1}(J_l=L)$ states, respectively. In a like manner, the $nL_L$, $nL_L^\prime$, and $nL_L^{\prime\prime}$ states in the HQL derived by Eq. (\ref{eq43}) are prone to turn into the $nL_L(J_l=L+1)$, $nL_L(J_l=L)$, and $nL_L(J_l=L-1)$ states, respectively. Conspicuously, there are two fascinating equalities concerning the magic mixing angles of the doubly heavy tetraquarks, i.e., the equality between the $^1P_1$$\leftrightarrow$$^3P_1$ and $^3P_1$$\leftrightarrow$$^5P_1$ mixing angles and the equality between the $^3L_{L+1}$$\leftrightarrow$$^5L_{L+1}$ and $^3(L+1)_L$$\leftrightarrow$$^5(L+1)_L$ mixing angles. Apart from that, the $^3L_{L-1}$$\leftrightarrow$$^5L_{L-1}$ mixing is feasible only when the orbital excitation of the double-heavy tetraquark is higher than $P$-wave ($L>1$), owing to the fact that the existence of the $^5P_0$ state is forbidden. One thing to point out here is that the definition of the mixing angle is replete with ambiguities, e.g., the flip-flop sign of the mixing angle induced by the charge conjugation. In order to refrain from the latent perplexity, this work (GI, MGI, NR-G, NR-Y, and HQL scenarios) has uniformly employed a phase convention corresponding to the order of coupling $\vec{L}\times\vec{S}_{\bar{q}\bar{q}}\times\vec{S}_{QQ}$.

\section{Summary}\label{sec5}

Accompanied by the precision enhancement of the experimental detection, a zoo of exotic hadrons is in the process of gradual establishment. It is certain that the phenomenological explorations for the spectroscopy of multiquark states are imperative to demystify the nature of these novel hadrons. Lately, the first $T_{cc}$ state observed by the LHCb Collaboration offers a fantastic opportunity to the spectroscopists, thanks to its narrow width and definite signal \cite{LHCb:2021vvq,LHCb:2021auc}. By taking advantage of relativized and nonrelativistic diquark-antidiquark scenarios, this work aspires to pin down the mass spectrum of the prospective double-charm tetraquark family. Specifically, these several sorts of diquark formalisms cover the Godfrey-Isgur (GI) relativized diquark model, the modified Godfrey-Isgur (MGI) relativized diquark model incorporating the color screening effects, the nonrelativistic (NR-G) diquark model with the Gaussian type hyperfine potential, and the nonrelativistic (NR-Y) diquark model with the Yukawa type hyperfine potential.

To sum up, for the sake of shedding light on the low-lying isoscalar and isovector members of the entire $T_{cc}$ family, this work comprehensively investigates the spectroscopic properties of the doubly charmed tetraquarks with the diverse orbital and radial excitations, comprising the $1S$-, $2S$-, $3S$-, $4S$-, $1P$-, $2P$-, $3P$-, $1D$-, $2D$-, and $1F$-wave states. Thereinto, with regard to the mass of the $1S$-wave isoscalar double-charm tetraquark, the predicted values from most diquark-antidiquark scenarios are partly higher than the experimental mass of the $T_{cc}(3875)^+$ state. Furthermore, in terms of the mass spectra of the $1S$-wave isovector $T_{cc}$ tetraquark states, the theoretical outcomes 3809, 3925, and 4083 MeV performed by the MGI ($\mu=30$) diquark approach deliver the available hints to the latent experimental observation. In comparison with the existing theoretical scenarios, the spectroscopy of the $2P$-, $3P$-, and $2D$-wave doubly charmed tetraquarks is surveyed by this work for the first time. Moreover, this work carries out the first theoretical inquiry into the mixing angles of the orbitally excited isovector $T_{cc}$ states and the magic mixing angles of the ideal heavy-light tetraquarks in the heavy quark limit. As a consequence, the spectroscopic predictions rendered by this work are not only capable of facilitating the phenomenological construction of the complete double-charm tetraquark family, but also capable of expediting the experimental quest for the promising low-lying excited $T_{cc}$ states.

\section*{ACKNOWLEDGMENTS}

W. C. Dong would like to thank Duojie Jia for helpful discussions and valuable comments. This work is supported by the National Natural Science Foundation of China (Grant No. 12175068) and the Hebei Natural Science Foundation (Grant No. A2024502002).

\appendix
\renewcommand{\theequation}{A\arabic{equation}}
\section{Coupling relations}\label{app}

As far as the heavy-light hadrons ($Q\bar q$ mesons, $Qqq$ baryons, $QQq$ baryons, and $QQ\bar q\bar q$ tetraquarks) are concerned, the relations between the $j-j$ and $L-S$ coupling schemes are unraveled in this appendix. Additionally, the coupling relations of the singly heavy mesons and doubly heavy baryons have been explored by Refs. \cite{Cahn:2003cw} and \cite{Matsuki:2021eck}, respectively.

\subsection{Singly heavy meson}\label{subapp1}

For the case of the $Q\bar q$ meson, there is only one sort of coupling relation corresponding to $J=L$, i.e.,
\begin{gather}
\left(\begin{array}{cc}
\left|J=L,J_l=L+1/2\right\rangle\\
\left|J=L,J_l=L-1/2\right\rangle
\end{array}\right)\nonumber\\
=\frac{1}{\sqrt{2L+1}}\left(\begin{array}{cc}
\sqrt{L+1}&\sqrt{L}\\
-\sqrt{L}&\sqrt{L+1}
\end{array}\right)\nonumber\\
\times\left(\begin{array}{cc}
\left|J=L,S=0\right\rangle\\
\left|J=L,S=1\right\rangle
\end{array}\right).\label{eqa1}
\end{gather}
Thereupon, the magic mixing angle of the ideal $Q\bar q$ meson is $\theta=\arctan(\sqrt{L}/\sqrt{L+1})$ for $J=L$.

\subsection{Singly heavy baryon}\label{subapp2}

In regard to the $Qqq$ baryon, there are two sorts of coupling relations corresponding to $J=L-1/2$ and $J=L+1/2$, i.e.,
\begin{gather}
\left(\begin{array}{cc}
\left|J=L-1/2,J_l=L\right\rangle\\
\left|J=L-1/2,J_l=L-1\right\rangle
\end{array}\right)\nonumber\\
=\frac{1}{\sqrt{3L}}\left(\begin{array}{cc}
\sqrt{L+1}&\sqrt{2L-1}\\
-\sqrt{2L-1}&\sqrt{L+1}
\end{array}\right)\nonumber\\
\times\left(\begin{array}{cc}
\left|J=L-1/2,S=1/2\right\rangle\\
\left|J=L-1/2,S=3/2\right\rangle
\end{array}\right),\label{eqa2}\\
\left(\begin{array}{cc}
\left|J=L+1/2,J_l=L+1\right\rangle\\
\left|J=L+1/2,J_l=L\right\rangle
\end{array}\right)\nonumber\\
=\frac{1}{\sqrt{3(L+1)}}\left(\begin{array}{cc}
\sqrt{2L+3}&\sqrt{L}\\
-\sqrt{L}&\sqrt{2L+3}
\end{array}\right)\nonumber\\
\times\left(\begin{array}{cc}
\left|J=L+1/2,S=1/2\right\rangle\\
\left|J=L+1/2,S=3/2\right\rangle
\end{array}\right).\label{eqa3}
\end{gather}
Consequently, the magic mixing angles of the ideal $Qqq$ baryon are $\theta=\arctan(\sqrt{2L-1}/\sqrt{L+1})$ for $J=L-1/2$ and $\theta=\arctan(\sqrt{L}/\sqrt{2L+3})$ for $J=L+1/2$.

\subsection{Doubly heavy baryon}\label{subapp3}

In terms of the $QQq$ baryon, there are two sorts of coupling relations corresponding to $J=L-1/2$ and $J=L+1/2$, i.e.,
\begin{gather}
\left(\begin{array}{cc}
\left|J=L-1/2,J_l=L+1/2\right\rangle\\
\left|J=L-1/2,J_l=L-1/2\right\rangle
\end{array}\right)\nonumber\\
=\frac{1}{\sqrt{3(2L+1)}}\left(\begin{array}{cc}
2\sqrt{L+1}&\sqrt{2L-1}\\
-\sqrt{2L-1}&2\sqrt{L+1}
\end{array}\right)\nonumber\\
\times\left(\begin{array}{cc}
\left|J=L-1/2,S=1/2\right\rangle\\
\left|J=L-1/2,S=3/2\right\rangle
\end{array}\right),\label{eqa4}\\
\left(\begin{array}{cc}
\left|J=L+1/2,J_l=L+1/2\right\rangle\\
\left|J=L+1/2,J_l=L-1/2\right\rangle
\end{array}\right)\nonumber\\
=\frac{1}{\sqrt{3(2L+1)}}\left(\begin{array}{cc}
\sqrt{2L+3}&2\sqrt{L}\\
-2\sqrt{L}&\sqrt{2L+3}
\end{array}\right)\nonumber\\
\times\left(\begin{array}{cc}
\left|J=L+1/2,S=1/2\right\rangle\\
\left|J=L+1/2,S=3/2\right\rangle
\end{array}\right).\label{eqa5}
\end{gather}
Accordingly, the magic mixing angles of the ideal $QQq$ baryon are $\theta=\arctan(\sqrt{2L-1}/(2\sqrt{L+1}))$ for $J=L-1/2$ and $\theta=\arctan(2\sqrt{L}/\sqrt{2L+3})$ for $J=L+1/2$.

\subsection{Doubly heavy tetraquark}\label{subapp4}

With respect to the $QQ\bar q\bar q$ tetraquark, there are three sorts of coupling relations corresponding to $J=L-1$, $J=L$, and $J=L+1$, i.e.,
\begin{gather}
\left(\begin{array}{cc}
\left|J=L-1,J_l=L\right\rangle\\
\left|J=L-1,J_l=L-1\right\rangle
\end{array}\right)\nonumber\\
=\frac{1}{\sqrt{2L}}\left(\begin{array}{cc}
\sqrt{L+1}&\sqrt{L-1}\\
-\sqrt{L-1}&\sqrt{L+1}
\end{array}\right)\nonumber\\
\times\left(\begin{array}{cc}
\left|J=L-1,S=1\right\rangle\\
\left|J=L-1,S=2\right\rangle
\end{array}\right),\label{eqa6}\\
\left(\begin{array}{ccc}
\left|J=L,J_l=L+1\right\rangle\\
\left|J=L,J_l=L\right\rangle\\
\left|J=L,J_l=L-1\right\rangle
\end{array}\right)\nonumber\\
=\left(\begin{array}{ccc}
\displaystyle\sqrt{\frac{2L+3}{3(2L+1)}}&\displaystyle\sqrt{\frac{L(2L+3)}{2(L+1)(2L+1)}}&\displaystyle\sqrt{\frac{L(2L-1)}{6(L+1)(2L+1)}}\\
\displaystyle-\sqrt{\frac{1}{3}}&\displaystyle\sqrt{\frac{1}{2L(L+1)}}&\displaystyle\sqrt{\frac{(2L-1)(2L+3)}{6L(L+1)}}\\
\displaystyle\sqrt{\frac{2L-1}{3(2L+1)}}&\displaystyle-\sqrt{\frac{(L+1)(2L-1)}{2L(2L+1)}}&\displaystyle\sqrt{\frac{(L+1)(2L+3)}{6L(2L+1)}}\\
\end{array}\right)\nonumber\\
\times\left(\begin{array}{ccc}
\left|J=L,S=0\right\rangle\\
\left|J=L,S=1\right\rangle\\
\left|J=L,S=2\right\rangle
\end{array}\right),\label{eqa7}\\
\left(\begin{array}{cc}
\left|J=L+1,J_l=L+1\right\rangle\\
\left|J=L+1,J_l=L\right\rangle
\end{array}\right)\nonumber\\
=\frac{1}{\sqrt{2(L+1)}}\left(\begin{array}{cc}
\sqrt{L+2}&\sqrt{L}\\
-\sqrt{L}&\sqrt{L+2}
\end{array}\right)\nonumber\\
\times\left(\begin{array}{cc}
\left|J=L+1,S=1\right\rangle\\
\left|J=L+1,S=2\right\rangle
\end{array}\right).\label{eqa8}
\end{gather}
Hence, the magic mixing angles of the ideal $QQ\bar q\bar q$ tetraquark are $\theta=\arctan(\sqrt{L-1}/\sqrt{L+1})$ for $J=L-1$ and $\theta=\arctan(\sqrt{L}/\sqrt{L+2})$ for $J=L+1$. When it comes to $J=L$, the magic mixing angles are enumerated in Table \ref{angle}.

\section{Uncertainty analyses}\label{un}

With regard to parameters of all types of models, this work makes no attempt to alter their values from original references. In the original Refs. \cite{Song:2015nia,Song:2015fha} of the MGI model, several $\mu$ values are taken to show $\mu$ dependence of the MGI model. Then the optimal $\mu$ value for charmed mesons is determined by fitting their observed masses \cite{Song:2015nia,Song:2015fha}. Regrettably, the nature of the $T_{cc}(3875)^+$ structure is a moot point at present. If its observed mass is the input of fitting, the output of $\mu$ value will be $0.07$ GeV. As shown in Refs. \cite{Song:2015nia,Song:2015fha}, the predicted outcomes of the GI model are obviously higher than the experimental values. The MGI model with the optimal $\mu$ value is successful for depicting the global (ground and excited) charmed meson spectrum, while the predicted spin-averaged ground state mass is a little higher than the experimental data. Considering the performances of GI and MGI models in charmed meson spectroscopy, for the MGI model uncertainty of this work, results of the GI model and the MGI model with $\mu=0.07$ GeV can be adopted as the upper and lower limits, respectively. Following the model uncertainty formula given by Ref. \cite{Wu:2022gie}, the mass difference between experimental data and three types of theoretical predictions for conventional heavy-light hadrons is presented in Table \ref{sa}. Differing from the light diquark masses of NR-G and NR-Y models obtained by chiral effective theory \cite{Kim:2021ywp}, the heavy diquark mass of the NR-Y model is acquired by Eqs. (\ref{eq31})-(\ref{eq33}) (heavy quark symmetry) and corresponding experimental data ($D$, $\Lambda_c$, $\Xi_{cc}$, and $T_{cc}$). According to the observed masses and errors of these hadrons \cite{ParticleDataGroup:2024cfk}, the NR-Y model diquark $cc$ mass with uncertainty is $3369.31\pm0.35$ MeV. Taking five $T_{cc}$ tetraquarks predicted by the NR-Y model for example, the mass variation caused by the experimental error ($\pm0.35$ MeV) is discussed in Table \ref{sb}. It shows that the mass uncertainty of the NR-Y model should include $\pm0.35$ MeV.

\renewcommand\tabcolsep{0.22cm}
\renewcommand{\arraystretch}{1.5}
\begin{table*}[!htbp]
\caption{Spin-averaged masses of ground state heavy-light hadrons obtained by experimental and theoretical approaches (in unit of MeV).}\label{sa}
\begin{tabular}{cccccccccc}
\toprule[1.0pt]\toprule[1.0pt]
State & PDG \cite{ParticleDataGroup:2024cfk} & GI \cite{Godfrey:1985xj} & $\left|m_\text{GI}-m_\text{exp}\right|$ & State & PDG \cite{ParticleDataGroup:2024cfk} & NR-G \cite{Kim:2020imk} & $\left|m_\text{NR-G}-m_\text{exp}\right|$ & NR-Y \cite{Kim:2020imk} & $\left|m_\text{NR-Y}-m_\text{exp}\right|$ \\
\midrule[1.0pt]
$K$ & $794.12$ & $792.40$ & $1.72$ & $\Lambda_c$ & $2286.46$ & $2285.71$ & $0.75$ & $2286.17$ & $0.29$ \\
$D$ & $1973.23$ & $1996.95$ & $23.72$ & $\Sigma_c$ & $2496.55$ & $2496.36$ & $0.19$ & $2496.41$ & $0.14$ \\
$D_s$ & $2076.24$ & $2087.11$ & $10.87$ & $\Xi_c$ & $2469.08$ & $2468.78$ & $0.29$ & $2469.29$ & $0.21$ \\
$B$ & $5313.45$ & $5354.49$ & $41.04$ & $\Xi_c^\prime$ & $2623.24$ & $2623.16$ & $0.08$ & $2623.70$ & $0.46$ \\
$B_s$ & $5403.28$ & $5433.29$ & $30.01$ & $\Omega_c$ & $2742.33$ & $2739.52$ & $2.81$ & $2738.25$ & $4.09$ \\
\multicolumn{4}{c}{$\displaystyle\chi_\text{GI}=\sqrt{\sum_{i=1}^5(m_\text{GI}-m_\text{exp})^2/5}=25.57$} & \multicolumn{3}{c}{$\displaystyle\chi_\text{NR-G}=\sqrt{\sum_{i=1}^5(m_\text{NR-G}-m_\text{exp})^2/5}=1.31$} & \multicolumn{3}{c}{$\displaystyle\chi_\text{NR-Y}=\sqrt{\sum_{i=1}^5(m_\text{NR-Y}-m_\text{exp})^2/5}=1.85$} \\
\bottomrule[1.0pt]\bottomrule[1.0pt]
\end{tabular}
\end{table*}

\renewcommand\tabcolsep{0.16cm}
\renewcommand{\arraystretch}{1.5}
\begin{table*}[!htbp]
\caption{The mass variation of several double-charm tetraquark states caused by the NR-Y model diquark $cc$ uncertainty (in unit of MeV).}\label{sb}
\begin{tabular}{ccccccccccc}
\toprule[1.0pt]\toprule[1.0pt]
$nL$, $I(J^P)$ & \multicolumn{2}{c}{$1S$, $0(1^+)$} & \multicolumn{2}{c}{$1S$, $1(2^+)$} & \multicolumn{2}{c}{$1P$, $0(0^-)$} & \multicolumn{2}{c}{$2S$, $1(1^+)$} & \multicolumn{2}{c}{$1D$, $0(2^+)$} \\
$m_{\{cc\}}$ & $m_{T_{cc}}$ & $m_{T_{cc}}-m_{T_{cc}}^\text{center}$ & $m_{T_{cc}}$ & $m_{T_{cc}}-m_{T_{cc}}^\text{center}$ & $m_{T_{cc}}$ & $m_{T_{cc}}-m_{T_{cc}}^\text{center}$ & $m_{T_{cc}}$ & $m_{T_{cc}}-m_{T_{cc}}^\text{center}$ & $m_{T_{cc}}$ & $m_{T_{cc}}-m_{T_{cc}}^\text{center}$ \\
\midrule[1.0pt]
$3368.96$ & $3876.12$ & $-0.35$ & $4104.23$ & $-0.34$ & $4161.14$ & $-0.35$ & $4538.41$ & $-0.34$ & $4439.21$ & $-0.34$ \\
$3369.31$ & $3876.46$ & $0$ & $4104.57$ & $0$ & $4161.49$ & $0$ & $4538.75$ & $0$ & $4439.56$ & $0$ \\
$3369.66$ & $3876.81$ & $+0.35$ & $4104.92$ & $+0.34$ & $4161.83$ & $+0.35$ & $4539.09$ & $+0.34$ & $4439.90$ & $+0.34$ \\
\bottomrule[1.0pt]\bottomrule[1.0pt]
\end{tabular}
\end{table*}


\begin{thebibliography}{9999}
%\cite{ParticleDataGroup:2024cfk}
\bibitem{ParticleDataGroup:2024cfk}
S.~Navas \textit{et al.} (Particle Data Group),
Review of particle physics,
\href{https://doi.org/10.1103/PhysRevD.110.030001}{Phys. Rev. D \textbf{110}, 030001 (2024)}.
%doi:10.1103/PhysRevD.110.030001
%391 citations counted in INSPIRE as of 19 Nov 2024

%\cite{Olsen:2017bmm}
\bibitem{Olsen:2017bmm}
S.~L.~Olsen, T.~Skwarnicki and D.~Zieminska,
Nonstandard heavy mesons and baryons: Experimental evidence,
\href{https://doi.org/10.1103/RevModPhys.90.015003}{Rev. Mod. Phys. \textbf{90}, 015003 (2018)}.
%doi:10.1103/RevModPhys.90.015003
%[arXiv:1708.04012 [hep-ph]].
%492 citations counted in INSPIRE as of 10 Aug 2022

%\cite{Hosaka:2016pey}
\bibitem{Hosaka:2016pey}
A.~Hosaka, T.~Iijima, K.~Miyabayashi, Y.~Sakai and S.~Yasui,
Exotic hadrons with heavy flavors: X, Y, Z, and related states,
\href{https://doi.org/10.1093/ptep/ptw045}{PTEP \textbf{2016}, 062C01 (2016)}.
%doi:10.1093/ptep/ptw045
%[arXiv:1603.09229 [hep-ph]].
%210 citations counted in INSPIRE as of 10 Aug 2022

%\cite{Ali:2017jda}
\bibitem{Ali:2017jda}
A.~Ali, J.~S.~Lange and S.~Stone,
Exotics: Heavy Pentaquarks and Tetraquarks,
\href{https://doi.org/10.1016/j.ppnp.2017.08.003}{Prog. Part. Nucl. Phys. \textbf{97}, 123-198 (2017)}.
%doi:10.1016/j.ppnp.2017.08.003
%[arXiv:1706.00610 [hep-ph]].
%374 citations counted in INSPIRE as of 10 Aug 2022

%\cite{Lebed:2016hpi}
\bibitem{Lebed:2016hpi}
R.~F.~Lebed, R.~E.~Mitchell and E.~S.~Swanson,
Heavy-Quark QCD Exotica,
\href{https://doi.org/10.1016/j.ppnp.2016.11.003}{Prog. Part. Nucl. Phys. \textbf{93}, 143-194 (2017)}.
%doi:10.1016/j.ppnp.2016.11.003
%[arXiv:1610.04528 [hep-ph]].
%447 citations counted in INSPIRE as of 10 Aug 2022

%\cite{Esposito:2016noz}
\bibitem{Esposito:2016noz}
A.~Esposito, A.~Pilloni and A.~D.~Polosa,
Multiquark Resonances,
\href{https://doi.org/10.1016/j.physrep.2016.11.002}{Phys. Rept. \textbf{668}, 1-97 (2017)}.
%doi:10.1016/j.physrep.2016.11.002
%[arXiv:1611.07920 [hep-ph]].
%494 citations counted in INSPIRE as of 10 Aug 2022

%\cite{Guo:2017jvc}
\bibitem{Guo:2017jvc}
F.~K.~Guo, C.~Hanhart, U.~G.~Mei\ss{}ner, Q.~Wang, Q.~Zhao and B.~S.~Zou,
Hadronic molecules,
\href{https://doi.org/10.1103/RevModPhys.90.015004}{Rev. Mod. Phys. \textbf{90}, 015004 (2018)}
[erratum: \href{https://doi.org/10.1103/RevModPhys.94.029901}{Rev. Mod. Phys. \textbf{94}, 029901 (2022)}].
%doi:10.1103/RevModPhys.90.015004
%[arXiv:1705.00141 [hep-ph]].
%783 citations counted in INSPIRE as of 10 Aug 2022

%\cite{Karliner:2017qhf}
\bibitem{Karliner:2017qhf}
M.~Karliner, J.~L.~Rosner and T.~Skwarnicki,
Multiquark States,
\href{https://doi.org/10.1146/annurev-nucl-101917-020902}{Ann. Rev. Nucl. Part. Sci. \textbf{68}, 17-44 (2018)}.
%doi:10.1146/annurev-nucl-101917-020902
%[arXiv:1711.10626 [hep-ph]].
%167 citations counted in INSPIRE as of 10 Aug 2022

%\cite{Brambilla:2019esw}
\bibitem{Brambilla:2019esw}
N.~Brambilla, S.~Eidelman, C.~Hanhart, A.~Nefediev, C.~P.~Shen, C.~E.~Thomas, A.~Vairo and C.~Z.~Yuan,
The $XYZ$ states: experimental and theoretical status and perspectives,
\href{https://doi.org/10.1016/j.physrep.2020.05.001}{Phys. Rept. \textbf{873}, 1-154 (2020)}.
%doi:10.1016/j.physrep.2020.05.001
%[arXiv:1907.07583 [hep-ex]].
%396 citations counted in INSPIRE as of 10 Aug 2022

%\cite{Chen:2022asf}
\bibitem{Chen:2022asf}
H.~X.~Chen, W.~Chen, X.~Liu, Y.~R.~Liu and S.~L.~Zhu,
An updated review of the new hadron states,
\href{https://doi.org/10.1088/1361-6633/aca3b6}{Rept. Prog. Phys. \textbf{86}, 026201 (2023)}.
%doi:10.1088/1361-6633/aca3b6
%[arXiv:2204.02649 [hep-ph]].
%102 citations counted in INSPIRE as of 14 Feb 2023

%\cite{LHCb:2021vvq}
\bibitem{LHCb:2021vvq}
R.~Aaij \textit{et al.} (LHCb Collaboration),
Observation of an exotic narrow doubly charmed tetraquark,
\href{https://doi.org/10.1038/s41567-022-01614-y}{Nature Phys. \textbf{18}, 751-754 (2022)}.
%doi:10.1038/s41567-022-01614-y
%[arXiv:2109.01038 [hep-ex]].
%307 citations counted in INSPIRE as of 29 Feb 2024

%\cite{LHCb:2021auc}
\bibitem{LHCb:2021auc}
R.~Aaij \textit{et al.} (LHCb Collaboration),
Study of the doubly charmed tetraquark $T_{cc}^{+}$,
\href{https://doi.org/10.1038/s41467-022-30206-w}{Nature Commun. \textbf{13}, 3351 (2022)}.
%doi:10.1038/s41467-022-30206-w
%[arXiv:2109.01056 [hep-ex]].
%273 citations counted in INSPIRE as of 29 Feb 2024

%\cite{Gell-Mann:1964ewy}
\bibitem{Gell-Mann:1964ewy}
M.~Gell-Mann,
A Schematic Model of Baryons and Mesons,
\href{https://doi.org/10.1016/S0031-9163(64)92001-3}{Phys. Lett. \textbf{8}, 214-215 (1964)}.
%doi:10.1016/S0031-9163(64)92001-3
%3750 citations counted in INSPIRE as of 22 Aug 2022

%\cite{Zweig:1964ruk}
\bibitem{Zweig:1964ruk}
G.~Zweig,
An SU$_3$ model for strong interaction symmetry and its breaking; Version 2,
\href{https://cds.cern.ch/record/570209}{CERN-TH-412}.
%619 citations counted in INSPIRE as of 22 Aug 2022

%\cite{Ader:1981db}
\bibitem{Ader:1981db}
J.~P.~Ader, J.~M.~Richard and P.~Taxil,
Do narrow heavy multiquark states exist?,
\href{https://doi.org/10.1103/PhysRevD.25.2370}{Phys. Rev. D \textbf{25}, 2370 (1982)}.
%doi:10.1103/PhysRevD.25.2370
%264 citations counted in INSPIRE as of 22 Aug 2022

%\cite{Ballot:1983iv}
\bibitem{Ballot:1983iv}
J.~l.~Ballot and J.~M.~Richard,
Four quark states in additive potentials,
\href{https://doi.org/10.1016/0370-2693(83)90991-7}{Phys. Lett. B \textbf{123}, 449-451 (1983)}.
%doi:10.1016/0370-2693(83)90991-7
%78 citations counted in INSPIRE as of 29 Feb 2024

%\cite{Zouzou:1986qh}
\bibitem{Zouzou:1986qh}
S.~Zouzou, B.~Silvestre-Brac, C.~Gignoux and J.~M.~Richard,
Four-quark bound states,
\href{https://doi.org/10.1007/BF01557611}{Z. Phys. C \textbf{30}, 457 (1986)}.
%doi:10.1007/BF01557611
%255 citations counted in INSPIRE as of 29 Feb 2024

%\cite{Manohar:1992nd}
\bibitem{Manohar:1992nd}
A.~V.~Manohar and M.~B.~Wise,
Exotic $QQ\bar{q}\bar{q}$ states in QCD,
\href{https://doi.org/10.1016/0550-3213(93)90614-U}{Nucl. Phys. B \textbf{399}, 17-33 (1993)}.
%doi:10.1016/0550-3213(93)90614-U
%[arXiv:hep-ph/9212236 [hep-ph]].
%257 citations counted in INSPIRE as of 29 Feb 2024

%\cite{Lipkin:1986dw}
\bibitem{Lipkin:1986dw}
H.~J.~Lipkin,
A model-independent approach to multiquark bound states,
\href{https://doi.org/10.1016/0370-2693(86)90843-9}{Phys. Lett. B \textbf{172}, 242-247 (1986)}.
%doi:10.1016/0370-2693(86)90843-9
%155 citations counted in INSPIRE as of 29 Feb 2024

%\cite{Carlson:1987hh}
\bibitem{Carlson:1987hh}
J.~Carlson, L.~Heller and J.~A.~Tjon,
Stability of Dimesons,
\href{https://doi.org/10.1103/PhysRevD.37.744}{Phys. Rev. D \textbf{37}, 744 (1988)}.
%doi:10.1103/PhysRevD.37.744
%163 citations counted in INSPIRE as of 29 Feb 2024

%\cite{Zhang:2021yul}
\bibitem{Zhang:2021yul}
W.~X.~Zhang, H.~Xu and D.~Jia,
Masses and magnetic moments of hadrons with one and two open heavy quarks: Heavy baryons and tetraquarks,
\href{https://doi.org/10.1103/PhysRevD.104.114011}{Phys. Rev. D \textbf{104}, 114011 (2021)}.
%doi:10.1103/PhysRevD.104.114011
%[arXiv:2109.07040 [hep-ph]].
%30 citations counted in INSPIRE as of 01 Mar 2024

%\cite{Semay:1994ht}
\bibitem{Semay:1994ht}
C.~Semay and B.~Silvestre-Brac,
Diquonia and potential models,
\href{https://doi.org/10.1007/BF01413104}{Z. Phys. C \textbf{61}, 271-275 (1994)}.
%doi:10.1007/BF01413104
%152 citations counted in INSPIRE as of 29 Feb 2024

%\cite{Pepin:1996id}
\bibitem{Pepin:1996id}
S.~Pepin, F.~Stancu, M.~Genovese and J.~M.~Richard,
Tetraquarks with colour-blind forces in chiral quark models,
\href{https://doi.org/10.1016/S0370-2693(96)01597-3}{Phys. Lett. B \textbf{393}, 119-123 (1997)}.
%doi:10.1016/S0370-2693(96)01597-3
%[arXiv:hep-ph/9609348 [hep-ph]].
%104 citations counted in INSPIRE as of 29 Feb 2024

%\cite{Gelman:2002wf}
\bibitem{Gelman:2002wf}
B.~A.~Gelman and S.~Nussinov,
Does a narrow tetraquark $cc\bar{u}\bar{d}$ state exist?,
\href{https://doi.org/10.1016/S0370-2693(02)03069-1}{Phys. Lett. B \textbf{551}, 296-304 (2003)}.
%doi:10.1016/S0370-2693(02)03069-1
%[arXiv:hep-ph/0209095 [hep-ph]].
%100 citations counted in INSPIRE as of 29 Feb 2024

%\cite{Vijande:2003ki}
\bibitem{Vijande:2003ki}
J.~Vijande, F.~Fernandez, A.~Valcarce and B.~Silvestre-Brac,
Tetraquarks in a chiral constituent-quark model,
\href{https://doi.org/10.1140/epja/i2003-10128-9}{Eur. Phys. J. A \textbf{19}, 383 (2004)}.
%doi:10.1140/epja/i2003-10128-9
%[arXiv:hep-ph/0310007 [hep-ph]].
%104 citations counted in INSPIRE as of 29 Feb 2024

%\cite{Janc:2004qn}
\bibitem{Janc:2004qn}
D.~Janc and M.~Rosina,
The $T_{cc}=DD^*$ molecular state,
\href{https://doi.org/10.1007/s00601-004-0068-9}{Few Body Syst. \textbf{35}, 175-196 (2004)}.
%doi:10.1007/s00601-004-0068-9
%[arXiv:hep-ph/0405208 [hep-ph]].
%163 citations counted in INSPIRE as of 29 Feb 2024

%\cite{Ebert:2007rn}
\bibitem{Ebert:2007rn}
D.~Ebert, R.~N.~Faustov, V.~O.~Galkin and W.~Lucha,
Masses of tetraquarks with two heavy quarks in the relativistic quark model,
\href{https://doi.org/10.1103/PhysRevD.76.114015}{Phys. Rev. D \textbf{76}, 114015 (2007)}.
%doi:10.1103/PhysRevD.76.114015
%[arXiv:0706.3853 [hep-ph]].
%162 citations counted in INSPIRE as of 29 Feb 2024

%\cite{Vijande:2009kj}
\bibitem{Vijande:2009kj}
J.~Vijande, A.~Valcarce and N.~Barnea,
Exotic meson-meson molecules and compact four-quark states,
\href{https://doi.org/10.1103/PhysRevD.79.074010}{Phys. Rev. D \textbf{79}, 074010 (2009)}.
%doi:10.1103/PhysRevD.79.074010
%[arXiv:0903.2949 [hep-ph]].
%118 citations counted in INSPIRE as of 29 Feb 2024

%\cite{Lee:2009rt}
\bibitem{Lee:2009rt}
S.~H.~Lee and S.~Yasui,
Stable multiquark states with heavy quarks in a diquark model,
\href{https://doi.org/10.1140/epjc/s10052-009-1140-x}{Eur. Phys. J. C \textbf{64}, 283-295 (2009)}.
%doi:10.1140/epjc/s10052-009-1140-x
%[arXiv:0901.2977 [hep-ph]].
%98 citations counted in INSPIRE as of 29 Feb 2024

%\cite{Yang:2009zzp}
\bibitem{Yang:2009zzp}
Y.~Yang, C.~Deng, J.~Ping and T.~Goldman,
$S$-wave $QQ\bar{q}\bar{q}$ state in the constituent quark model,
\href{https://doi.org/10.1103/PhysRevD.80.114023}{Phys. Rev. D \textbf{80}, 114023 (2009)}.
%doi:10.1103/PhysRevD.80.114023
%86 citations counted in INSPIRE as of 29 Feb 2024

%\cite{Luo:2017eub}
\bibitem{Luo:2017eub}
S.~Q.~Luo, K.~Chen, X.~Liu, Y.~R.~Liu and S.~L.~Zhu,
Exotic tetraquark states with the $qq\bar{Q}\bar{Q}$ configuration,
\href{https://doi.org/10.1140/epjc/s10052-017-5297-4}{Eur. Phys. J. C \textbf{77}, 709 (2017)}.
%doi:10.1140/epjc/s10052-017-5297-4
%[arXiv:1707.01180 [hep-ph]].
%117 citations counted in INSPIRE as of 01 Mar 2024

%\cite{Karliner:2017qjm}
\bibitem{Karliner:2017qjm}
M.~Karliner and J.~L.~Rosner,
Discovery of doubly-charmed $\Xi_{cc}$ baryon implies a stable $bb\bar{u}\bar{d}$ tetraquark,
\href{https://doi.org/10.1103/PhysRevLett.119.202001}{Phys. Rev. Lett. \textbf{119}, 202001 (2017)}.
%doi:10.1103/PhysRevLett.119.202001
%[arXiv:1707.07666 [hep-ph]].
%266 citations counted in INSPIRE as of 01 Mar 2024

%\cite{Hyodo:2017hue}
\bibitem{Hyodo:2017hue}
T.~Hyodo, Y.~R.~Liu, M.~Oka and S.~Yasui,
Spectroscopy and production of doubly charmed tetraquarks,
\href{https://arxiv.org/abs/1708.05169}{arXiv:1708.05169}.
%18 citations counted in INSPIRE as of 01 Mar 2024

%\cite{Yan:2018gik}
\bibitem{Yan:2018gik}
X.~Yan, B.~Zhong and R.~Zhu,
Doubly charmed tetraquarks in a diquark\textendash{}antidiquark model,
\href{https://doi.org/10.1142/S0217751X18500963}{Int. J. Mod. Phys. A \textbf{33}, 1850096 (2018)}.
%doi:10.1142/S0217751X18500963
%[arXiv:1804.06761 [hep-ph]].
%26 citations counted in INSPIRE as of 01 Mar 2024

%\cite{Xing:2018bqt}
\bibitem{Xing:2018bqt}
Y.~Xing and R.~Zhu,
Weak Decays of Stable Doubly Heavy Tetraquark States,
\href{https://doi.org/10.1103/PhysRevD.98.053005}{Phys. Rev. D \textbf{98}, 053005 (2018)}.
%doi:10.1103/PhysRevD.98.053005
%[arXiv:1806.01659 [hep-ph]].
%41 citations counted in INSPIRE as of 01 Mar 2024

%\cite{Zhu:2019iwm}
\bibitem{Zhu:2019iwm}
R.~Zhu, X.~Liu, H.~Huang and C.~F.~Qiao,
Analyzing doubly heavy tetra- and penta-quark states by variational method,
\href{https://doi.org/10.1016/j.physletb.2019.134869}{Phys. Lett. B \textbf{797}, 134869 (2019)}.
%doi:10.1016/j.physletb.2019.134869
%[arXiv:1904.10285 [hep-ph]].
%66 citations counted in INSPIRE as of 01 Mar 2024

%\cite{Park:2018wjk}
\bibitem{Park:2018wjk}
W.~Park, S.~Noh and S.~H.~Lee,
Masses of the doubly heavy tetraquarks in a constituent quark model,
\href{https://doi.org/10.1016/j.nuclphysa.2018.12.019}{Nucl. Phys. A \textbf{983}, 1-19 (2019)}.
%doi:10.1016/j.nuclphysa.2018.12.019
%[arXiv:1809.05257 [nucl-th]].
%55 citations counted in INSPIRE as of 01 Mar 2024

%\cite{Yang:2019itm}
\bibitem{Yang:2019itm}
G.~Yang, J.~Ping and J.~Segovia,
Double-heavy tetraquarks,
\href{https://doi.org/10.1103/PhysRevD.101.014001}{Phys. Rev. D \textbf{101}, 014001 (2020)}.
%doi:10.1103/PhysRevD.101.014001
%[arXiv:1911.00215 [hep-ph]].
%68 citations counted in INSPIRE as of 01 Mar 2024

%\cite{Lu:2020rog}
\bibitem{Lu:2020rog}
Q.~F.~L\"u, D.~Y.~Chen and Y.~B.~Dong,
Masses of doubly heavy tetraquarks $T_{QQ'}$ in a relativized quark model,
\href{https://doi.org/10.1103/PhysRevD.102.034012}{Phys. Rev. D \textbf{102}, 034012 (2020)}.
%doi:10.1103/PhysRevD.102.034012
%[arXiv:2006.08087 [hep-ph]].
%60 citations counted in INSPIRE as of 01 Mar 2024

%\cite{Deng:2018kly}
\bibitem{Deng:2018kly}
C.~Deng, H.~Chen and J.~Ping,
Systematical investigation on the stability of doubly heavy tetraquark states,
\href{https://doi.org/10.1140/epja/s10050-019-00012-y}{Eur. Phys. J. A \textbf{56}, 9 (2020)}.
%doi:10.1140/epja/s10050-019-00012-y
%[arXiv:1811.06462 [hep-ph]].
%56 citations counted in INSPIRE as of 01 Mar 2024

%\cite{Tan:2020ldi}
\bibitem{Tan:2020ldi}
Y.~Tan, W.~Lu and J.~Ping,
Systematics of $QQ{\bar{q}}{\bar{q}}$ in a chiral constituent quark model,
\href{https://doi.org/10.1140/epjp/s13360-020-00741-w}{Eur. Phys. J. Plus \textbf{135}, 716 (2020)}.
%doi:10.1140/epjp/s13360-020-00741-w
%[arXiv:2004.02106 [hep-ph]].
%46 citations counted in INSPIRE as of 01 Mar 2024

%\cite{Cheng:2020wxa}
\bibitem{Cheng:2020wxa}
J.~B.~Cheng, S.~Y.~Li, Y.~R.~Liu, Z.~G.~Si and T.~Yao,
Double-heavy tetraquark states with heavy diquark-antiquark symmetry,
\href{https://doi.org/10.1088/1674-1137/abde2f}{Chin. Phys. C \textbf{45}, 043102 (2021)}.
%doi:10.1088/1674-1137/abde2f
%[arXiv:2008.00737 [hep-ph]].
%63 citations counted in INSPIRE as of 01 Mar 2024

%\cite{Noh:2021lqs}
\bibitem{Noh:2021lqs}
S.~Noh, W.~Park and S.~H.~Lee,
Doubly heavy tetraquarks, $qq'\bar{Q}\bar{Q'}$, in a nonrelativistic quark model with a complete set of harmonic oscillator bases,
\href{https://doi.org/10.1103/PhysRevD.103.114009}{Phys. Rev. D \textbf{103}, 114009 (2021)}.
%doi:10.1103/PhysRevD.103.114009
%[arXiv:2102.09614 [hep-ph]].
%32 citations counted in INSPIRE as of 01 Mar 2024

%\cite{Meng:2020knc}
\bibitem{Meng:2020knc}
Q.~Meng, E.~Hiyama, A.~Hosaka, M.~Oka, P.~Gubler, K.~U.~Can, T.~T.~Takahashi and H.~S.~Zong,
Stable double-heavy tetraquarks: spectrum and structure,
\href{https://doi.org/10.1016/j.physletb.2021.136095}{Phys. Lett. B \textbf{814}, 136095 (2021)}.
%doi:10.1016/j.physletb.2021.136095
%[arXiv:2009.14493 [nucl-th]].
%47 citations counted in INSPIRE as of 01 Mar 2024

%\cite{Kim:2022mpa}
\bibitem{Kim:2022mpa}
Y.~Kim, M.~Oka and K.~Suzuki,
Doubly heavy tetraquarks in a chiral-diquark picture,
\href{https://doi.org/10.1103/PhysRevD.105.074021}{Phys. Rev. D \textbf{105}, 074021 (2022)}.
%doi:10.1103/PhysRevD.105.074021
%[arXiv:2202.06520 [hep-ph]].
%26 citations counted in INSPIRE as of 01 Mar 2024

%\cite{Wang:2022clw}
\bibitem{Wang:2022clw}
J.~B.~Wang, G.~Li, C.~S.~An, C.~R.~Deng and J.~J.~Xie,
The low-lying hidden- and double-charm tetraquark states in a constituent quark model with instanton-induced interaction,
\href{https://doi.org/10.1140/epjc/s10052-022-10673-7}{Eur. Phys. J. C \textbf{82}, 721 (2022)}.
%doi:10.1140/epjc/s10052-022-10673-7
%[arXiv:2204.13320 [hep-ph]].
%9 citations counted in INSPIRE as of 02 Mar 2024

%\cite{Meng:2024yhu}
\bibitem{Meng:2024yhu}
Q.~Meng, G.~J.~Wang and M.~Oka,
Mass Spectra of Full-Heavy and Double-Heavy Tetraquark States in the Conventional Quark Model,
\href{https://arxiv.org/abs/2404.01238}{arXiv:2404.01238}.
%0 citations counted in INSPIRE as of 03 Apr 2024

%\cite{Weng:2021hje}
\bibitem{Weng:2021hje}
X.~Z.~Weng, W.~Z.~Deng and S.~L.~Zhu,
Doubly heavy tetraquarks in an extended chromomagnetic model,
\href{https://doi.org/10.1088/1674-1137/ac2ed0}{Chin. Phys. C \textbf{46}, 013102 (2022)}.
%doi:10.1088/1674-1137/ac2ed0
%[arXiv:2108.07242 [hep-ph]].
%48 citations counted in INSPIRE as of 01 Mar 2024

%\cite{Guo:2021yws}
\bibitem{Guo:2021yws}
T.~Guo, J.~Li, J.~Zhao and L.~He,
Mass spectra of doubly heavy tetraquarks in an improved chromomagnetic interaction model,
\href{https://doi.org/10.1103/PhysRevD.105.014021}{Phys. Rev. D \textbf{105}, 014021 (2022)}.
%doi:10.1103/PhysRevD.105.014021
%[arXiv:2108.10462 [hep-ph]].
%34 citations counted in INSPIRE as of 02 Mar 2024

%\cite{Liu:2023vrk}
\bibitem{Liu:2023vrk}
X.~Y.~Liu, W.~X.~Zhang and D.~Jia,
Doubly heavy tetraquarks: Heavy quark bindings and chromomagnetically mixings,
\href{https://doi.org/10.1103/PhysRevD.108.054019}{Phys. Rev. D \textbf{108}, 054019 (2023)}.
%doi:10.1103/PhysRevD.108.054019
%[arXiv:2303.03923 [hep-ph]].
%2 citations counted in INSPIRE as of 02 Mar 2024

%\cite{Meng:2023for}
\bibitem{Meng:2023for}
Q.~Meng, E.~Hiyama, M.~Oka, A.~Hosaka and C.~Xu,
Doubly heavy tetraquarks including one-pion exchange potential,
\href{https://doi.org/10.1016/j.physletb.2023.138221}{Phys. Lett. B \textbf{846}, 138221 (2023)}.
%doi:10.1016/j.physletb.2023.138221
%[arXiv:2308.05466 [nucl-th]].
%5 citations counted in INSPIRE as of 02 Mar 2024

%\cite{Noh:2023zoq}
\bibitem{Noh:2023zoq}
S.~Noh and W.~Park,
Nonrelativistic quark model analysis of $T_{cc}$,
\href{https://doi.org/10.1103/PhysRevD.108.014004}{Phys. Rev. D \textbf{108}, 014004 (2023)}.
%doi:10.1103/PhysRevD.108.014004
%[arXiv:2303.03285 [hep-ph]].
%2 citations counted in INSPIRE as of 02 Mar 2024

%\cite{Wu:2022gie}
\bibitem{Wu:2022gie}
T.~W.~Wu and Y.~L.~Ma,
Doubly heavy tetraquark multiplets as heavy antiquark-diquark symmetry partners of heavy baryons,
\href{https://doi.org/10.1103/PhysRevD.107.L071501}{Phys. Rev. D \textbf{107}, L071501 (2023)}.
%doi:10.1103/PhysRevD.107.L071501
%[arXiv:2211.15094 [hep-ph]].
%13 citations counted in INSPIRE as of 02 Mar 2024

%\cite{Mutuk:2023oyz}
\bibitem{Mutuk:2023oyz}
H.~Mutuk,
Masses and magnetic moments of doubly heavy tetraquarks via diffusion Monte Carlo method,
\href{https://doi.org/10.1140/epjc/s10052-024-12736-3}{Eur. Phys. J. C \textbf{84}, 395 (2024)}.
%doi:10.1140/epjc/s10052-024-12736-3
%[arXiv:2312.13383 [hep-ph]].
%4 citations counted in INSPIRE as of 18 Apr 2024

%\cite{Wang:2024vjc}
\bibitem{Wang:2024vjc}
D.~Wang, K.~R.~Song, W.~L.~Wang and F.~Huang,
Spectrum of $S$- and $P$-wave $cc\bar{q}\bar{q}'$ $(\bar{q},\bar{q}' = \bar{u}, \bar{d}, \bar{s})$ systems in a chiral SU(3) quark model,
\href{https://doi.org/10.1103/PhysRevD.109.074026}{Phys. Rev. D \textbf{109}, 074026 (2024)}.
%doi:10.1103/PhysRevD.109.074026
%[arXiv:2403.15187 [hep-ph]].
%0 citations counted in INSPIRE as of 24 Apr 2024

%\cite{Park:2024cic}
\bibitem{Park:2024cic}
D.~Park and S.~H.~Lee,
An AI-Inspired Numerical Method in the Quark Model: Application to Finding the Wave Functions for Heavy Tetraquark States,
\href{https://arxiv.org/abs/2406.00756}{arXiv:2406.00756}.
%0 citations counted in INSPIRE as of 04 Jun 2024

%\cite{Li:2023wug}
\bibitem{Li:2023wug}
S.~Y.~Li, Y.~R.~Liu, Z.~L.~Man, Z.~G.~Si and J.~Wu,
Doubly heavy tetraquark states in a mass splitting model,
\href{https://doi.org/10.1103/PhysRevD.110.094044}{Phys. Rev. D \textbf{110}, no.9, 094044 (2024)}.
%doi:10.1103/PhysRevD.110.094044
%[arXiv:2401.00115 [hep-ph]].
%5 citations counted in INSPIRE as of 23 Dec 2024

%\cite{He:2023ucd}
\bibitem{He:2023ucd}
B.~R.~He, M.~Harada and B.~S.~Zou,
Quark model with hidden local symmetry and its application to $T_{cc}$,
\href{https://doi.org/10.1103/PhysRevD.108.054025}{Phys. Rev. D \textbf{108}, 054025 (2023)}.
%doi:10.1103/PhysRevD.108.054025
%[arXiv:2306.03526 [hep-ph]].
%5 citations counted in INSPIRE as of 02 Mar 2024

%\cite{Chen:2021tnn}
\bibitem{Chen:2021tnn}
X.~Chen and Y.~Yang,
Doubly-heavy tetraquark states $cc\bar{u}\bar{d}$ and $bb\bar{u}\bar{d}$,
\href{https://doi.org/10.1088/1674-1137/ac4ee8}{Chin. Phys. C \textbf{46}, 054103 (2022)}.
%doi:10.1088/1674-1137/ac4ee8
%[arXiv:2109.02828 [hep-ph]].
%27 citations counted in INSPIRE as of 01 Mar 2024

%\cite{Deng:2021gnb}
\bibitem{Deng:2021gnb}
C.~Deng and S.~L.~Zhu,
$T_{cc}^{+}$ and its partners,
\href{https://doi.org/10.1103/PhysRevD.105.054015}{Phys. Rev. D \textbf{105}, 054015 (2022)}.
%doi:10.1103/PhysRevD.105.054015
%[arXiv:2112.12472 [hep-ph]].
%55 citations counted in INSPIRE as of 02 Mar 2024

%\cite{Deng:2022cld}
\bibitem{Deng:2022cld}
C.~R.~Deng and S.~L.~Zhu,
Decoding the double heavy tetraquark state $T_{cc}^+$,
\href{https://doi.org/10.1016/j.scib.2022.06.016}{Sci. Bull. \textbf{67}, 1522 (2022)}.
%doi:10.1016/j.scib.2022.06.016
%[arXiv:2204.11079 [hep-ph]].
%13 citations counted in INSPIRE as of 02 Mar 2024

%\cite{Ortega:2022efc}
\bibitem{Ortega:2022efc}
P.~G.~Ortega, J.~Segovia, D.~R.~Entem and F.~Fernandez,
Nature of the doubly-charmed tetraquark $T_{cc}^{+}$ in a constituent quark model,
\href{https://doi.org/10.1016/j.physletb.2023.137918}{Phys. Lett. B \textbf{841}, 137918 (2023)}
[erratum: \href{https://doi.org/10.1016/j.physletb.2023.138308}{Phys. Lett. B \textbf{847}, 138308 (2023)}].
%doi:10.1016/j.physletb.2023.137918
%[arXiv:2211.06118 [hep-ph]].
%17 citations counted in INSPIRE as of 02 Mar 2024

%\cite{Meng:2023jqk}
\bibitem{Meng:2023jqk}
L.~Meng, Y.~K.~Chen, Y.~Ma and S.~L.~Zhu,
Tetraquark bound states in constituent quark models: Benchmark test calculations,
\href{https://doi.org/10.1103/PhysRevD.108.114016}{Phys. Rev. D \textbf{108}, 114016 (2023)}.
%doi:10.1103/PhysRevD.108.114016
%[arXiv:2310.13354 [hep-ph]].
%5 citations counted in INSPIRE as of 02 Mar 2024

%\cite{Ma:2023int}
\bibitem{Ma:2023int}
Y.~Ma, L.~Meng, Y.~K.~Chen and S.~L.~Zhu,
Doubly heavy tetraquark states in the constituent quark model using diffusion Monte~Carlo method,
\href{https://doi.org/10.1103/PhysRevD.109.074001}{Phys. Rev. D \textbf{109}, 074001 (2024)}.
%doi:10.1103/PhysRevD.109.074001
%[arXiv:2309.17068 [hep-ph]].
%6 citations counted in INSPIRE as of 06 Apr 2024

%\cite{Xin:2021wcr}
\bibitem{Xin:2021wcr}
Q.~Xin and Z.~G.~Wang,
Analysis of the doubly-charmed tetraquark molecular states with the QCD sum rules,
\href{https://doi.org/10.1140/epja/s10050-022-00752-4}{Eur. Phys. J. A \textbf{58}, 110 (2022)}.
%doi:10.1140/epja/s10050-022-00752-4
%[arXiv:2108.12597 [hep-ph]].
%59 citations counted in INSPIRE as of 01 Mar 2024

%\cite{Navarra:2007yw}
\bibitem{Navarra:2007yw}
F.~S.~Navarra, M.~Nielsen and S.~H.~Lee,
QCD sum rules study of $QQ$-$\bar{u}\bar{d}$ mesons,
\href{https://doi.org/10.1016/j.physletb.2007.04.010}{Phys. Lett. B \textbf{649}, 166-172 (2007)}.
%doi:10.1016/j.physletb.2007.04.010
%[arXiv:hep-ph/0703071 [hep-ph]].
%151 citations counted in INSPIRE as of 29 Feb 2024

%\cite{Du:2012wp}
\bibitem{Du:2012wp}
M.~L.~Du, W.~Chen, X.~L.~Chen and S.~L.~Zhu,
Exotic $QQ\bar{q}\bar{q}$, $QQ\bar{q}\bar{s}$, and $QQ\bar{s}\bar{s}$ states,
\href{https://doi.org/10.1103/PhysRevD.87.014003}{Phys. Rev. D \textbf{87}, 014003 (2013)}.
%doi:10.1103/PhysRevD.87.014003
%[arXiv:1209.5134 [hep-ph]].
%107 citations counted in INSPIRE as of 01 Mar 2024

%\cite{Tang:2019nwv}
\bibitem{Tang:2019nwv}
L.~Tang, B.~D.~Wan, K.~Maltman and C.~F.~Qiao,
Doubly Heavy Tetraquarks in QCD Sum Rules,
\href{https://doi.org/10.1103/PhysRevD.101.094032}{Phys. Rev. D \textbf{101}, 094032 (2020)}.
%doi:10.1103/PhysRevD.101.094032
%[arXiv:1911.10951 [hep-ph]].
%41 citations counted in INSPIRE as of 01 Mar 2024

%\cite{Azizi:2021aib}
\bibitem{Azizi:2021aib}
K.~Azizi and U.~\"Ozdem,
Magnetic dipole moments of the $T_{cc}^+$ and $Z_V^{++}$ tetraquark states,
\href{https://doi.org/10.1103/PhysRevD.104.114002}{Phys. Rev. D \textbf{104}, 114002 (2021)}.
%doi:10.1103/PhysRevD.104.114002
%[arXiv:2109.02390 [hep-ph]].
%48 citations counted in INSPIRE as of 19 Nov 2024

%\cite{Agaev:2022ast}
\bibitem{Agaev:2022ast}
S.~S.~Agaev, K.~Azizi and H.~Sundu,
Hadronic molecule model for the doubly charmed state $T_{cc}^{+}$,
\href{https://doi.org/10.1007/JHEP06(2022)057}{JHEP \textbf{06}, 057 (2022)}.
%doi:10.1007/JHEP06(2022)057
%[arXiv:2201.02788 [hep-ph]].
%31 citations counted in INSPIRE as of 02 Mar 2024

%\cite{Albuquerque:2023rrf}
\bibitem{Albuquerque:2023rrf}
R.~Albuquerque, S.~Narison and D.~Rabetiarivony,
Pseudoscalar and Vector $T_{QQ\bar q\bar q'}$ Spectra and Couplings from LSR at NLO,
\href{https://doi.org/10.1016/j.nuclphysa.2023.122637}{Nucl. Phys. A \textbf{1034}, 122637 (2023)}.
%doi:10.1016/j.nuclphysa.2023.122637
%[arXiv:2301.08199 [hep-ph]].
%7 citations counted in INSPIRE as of 02 Mar 2024

%\cite{Wang:2017uld}
\bibitem{Wang:2017uld}
Z.~G.~Wang,
Analysis of the axialvector doubly heavy tetraquark states with QCD sum rules,
\href{https://doi.org/10.5506/APhysPolB.49.1781}{Acta Phys. Polon. B \textbf{49}, 1781 (2018)}.
%doi:10.5506/APhysPolB.49.1781
%[arXiv:1708.04545 [hep-ph]].
%46 citations counted in INSPIRE as of 01 Mar 2024

%\cite{Wang:2017dtg}
\bibitem{Wang:2017dtg}
Z.~G.~Wang and Z.~H.~Yan,
Analysis of the scalar, axialvector, vector, tensor doubly charmed tetraquark states with QCD sum rules,
\href{https://doi.org/10.1140/epjc/s10052-017-5507-0}{Eur. Phys. J. C \textbf{78}, 19 (2018)}.
%doi:10.1140/epjc/s10052-017-5507-0
%[arXiv:1710.02810 [hep-ph]].
%46 citations counted in INSPIRE as of 01 Mar 2024

%\cite{Agaev:2019qqn}
\bibitem{Agaev:2019qqn}
S.~S.~Agaev, K.~Azizi and H.~Sundu,
Strong decays of double-charmed pseudoscalar and scalar $cc\bar{u}\bar{d}$ tetraquarks,
\href{https://doi.org/10.1103/PhysRevD.99.114016}{Phys. Rev. D \textbf{99}, 114016 (2019)}.
%doi:10.1103/PhysRevD.99.114016
%[arXiv:1903.11975 [hep-ph]].
%24 citations counted in INSPIRE as of 01 Mar 2024

%\cite{Agaev:2021vur}
\bibitem{Agaev:2021vur}
S.~S.~Agaev, K.~Azizi and H.~Sundu,
Newly observed exotic doubly charmed meson $T_{cc}^{+}$,
\href{https://doi.org/10.1016/j.nuclphysb.2022.115650}{Nucl. Phys. B \textbf{975}, 115650 (2022)}.
%doi:10.1016/j.nuclphysb.2022.115650
%[arXiv:2108.00188 [hep-ph]].
%58 citations counted in INSPIRE as of 02 Mar 2024

%\cite{Albuquerque:2022weq}
\bibitem{Albuquerque:2022weq}
R.~Albuquerque, S.~Narison and D.~Rabetiarivony,
Improved XTZ masses and mass ratios from Laplace sum rules at NLO,
\href{https://doi.org/10.1016/j.nuclphysa.2022.122451}{Nucl. Phys. A \textbf{1023}, 122451 (2022)}.
%doi:10.1016/j.nuclphysa.2022.122451
%[arXiv:2201.13449 [hep-ph]].
%15 citations counted in INSPIRE as of 02 Mar 2024

%\cite{Gao:2020bvl}
\bibitem{Gao:2020bvl}
D.~Gao, D.~Jia, Y.~J.~Sun, Z.~Zhang, W.~N.~Liu and Q.~Mei,
Masses of doubly heavy tetraquarks $QQ\bar q\bar n$ with $J^P=1^+$,
\href{https://doi.org/10.1142/S0217732322502236}{Mod. Phys. Lett. A \textbf{37}, 2250223 (2022)}.
%doi:10.1142/S0217732322502236
%[arXiv:2007.15213 [hep-ph]].
%24 citations counted in INSPIRE as of 02 Mar 2024

%\cite{Maiani:2022qze}
\bibitem{Maiani:2022qze}
L.~Maiani, A.~Pilloni, A.~D.~Polosa and V.~Riquer,
Doubly heavy tetraquarks in the Born-Oppenheimer approximation,
\href{https://doi.org/10.1016/j.physletb.2022.137624}{Phys. Lett. B \textbf{836}, 137624 (2023)}.
%doi:10.1016/j.physletb.2022.137624
%[arXiv:2208.02730 [hep-ph]].
%10 citations counted in INSPIRE as of 02 Mar 2024

%\cite{Maiani:2019lpu}
\bibitem{Maiani:2019lpu}
L.~Maiani, A.~D.~Polosa and V.~Riquer,
Hydrogen bond of QCD in doubly heavy baryons and tetraquarks,
\href{https://doi.org/10.1103/PhysRevD.100.074002}{Phys. Rev. D \textbf{100}, 074002 (2019)}.
%doi:10.1103/PhysRevD.100.074002
%[arXiv:1908.03244 [hep-ph]].
%34 citations counted in INSPIRE as of 01 Mar 2024

%\cite{Mutuk:2024vzv}
\bibitem{Mutuk:2024vzv}
H.~Mutuk,
Doubly-charged $T_{cc}^{++}$ states in the dynamical diquark model,
\href{https://doi.org/10.1103/PhysRevD.110.034025}{Phys. Rev. D \textbf{110}, 034025 (2024)}.
%doi:10.1103/PhysRevD.110.034025
%[arXiv:2401.02788 [hep-ph]].
%4 citations counted in INSPIRE as of 18 Nov 2024

%\cite{Lebed:2024zrp}
\bibitem{Lebed:2024zrp}
R.~F.~Lebed and S.~R.~Martinez,
$T_{cc}$ in the diabatic diquark model: Effects of $D^*D$ isospin,
\href{https://doi.org/10.1103/PhysRevD.110.034033}{Phys. Rev. D \textbf{110}, 034033 (2024)}.
%doi:10.1103/PhysRevD.110.034033
%[arXiv:2406.08690 [hep-ph]].
%5 citations counted in INSPIRE as of 18 Nov 2024

%\cite{Ohkoda:2012hv}
\bibitem{Ohkoda:2012hv}
S.~Ohkoda, Y.~Yamaguchi, S.~Yasui, K.~Sudoh and A.~Hosaka,
Exotic mesons with double charm and bottom flavor,
\href{https://doi.org/10.1103/PhysRevD.86.034019}{Phys. Rev. D \textbf{86}, 034019 (2012)}.
%doi:10.1103/PhysRevD.86.034019
%[arXiv:1202.0760 [hep-ph]].
%66 citations counted in INSPIRE as of 01 Mar 2024

%\cite{Li:2012ss}
\bibitem{Li:2012ss}
N.~Li, Z.~F.~Sun, X.~Liu and S.~L.~Zhu,
Coupled-channel analysis of the possible $D^{(*)}D^{(*)}$, $\bar{B}^{(*)}\bar{B}^{(*)}$ and $D^{(*)}\bar{B}^{(*)}$ molecular states,
\href{https://doi.org/10.1103/PhysRevD.88.114008}{Phys. Rev. D \textbf{88}, 114008 (2013)}.
%doi:10.1103/PhysRevD.88.114008
%[arXiv:1211.5007 [hep-ph]].
%102 citations counted in INSPIRE as of 01 Mar 2024

%\cite{Liu:2019stu}
\bibitem{Liu:2019stu}
M.~Z.~Liu, T.~W.~Wu, M.~Pavon Valderrama, J.~J.~Xie and L.~S.~Geng,
Heavy-quark spin and flavor symmetry partners of the $X(3872)$ revisited: What can we learn from the one boson exchange model?,
\href{https://doi.org/10.1103/PhysRevD.99.094018}{Phys. Rev. D \textbf{99}, 094018 (2019)}.
%doi:10.1103/PhysRevD.99.094018
%[arXiv:1902.03044 [hep-ph]].
%91 citations counted in INSPIRE as of 01 Mar 2024

%\cite{Wang:2021yld}
\bibitem{Wang:2021yld}
F.~L.~Wang and X.~Liu,
Investigating new type of doubly charmed molecular tetraquarks composed of charmed mesons in the $H$ and $T$ doublets,
\href{https://doi.org/10.1103/PhysRevD.104.094030}{Phys. Rev. D \textbf{104}, 094030 (2021)}.
%doi:10.1103/PhysRevD.104.094030
%[arXiv:2108.09925 [hep-ph]].
%22 citations counted in INSPIRE as of 01 Mar 2024

%\cite{Wang:2021ajy}
\bibitem{Wang:2021ajy}
F.~L.~Wang, R.~Chen and X.~Liu,
A new group of doubly charmed molecule with $T$-doublet charmed meson pair,
\href{https://doi.org/10.1016/j.physletb.2022.137502}{Phys. Lett. B \textbf{835}, 137502 (2022)}.
%doi:10.1016/j.physletb.2022.137502
%[arXiv:2111.00208 [hep-ph]].
%17 citations counted in INSPIRE as of 01 Mar 2024

%\cite{Chen:2021vhg}
\bibitem{Chen:2021vhg}
R.~Chen, Q.~Huang, X.~Liu and S.~L.~Zhu,
Predicting another doubly charmed molecular resonance $T_{cc}^{\prime+}(3876)$,
\href{https://doi.org/10.1103/PhysRevD.104.114042}{Phys. Rev. D \textbf{104}, 114042 (2021)}.
%doi:10.1103/PhysRevD.104.114042
%[arXiv:2108.01911 [hep-ph]].
%51 citations counted in INSPIRE as of 01 Mar 2024

%\cite{Dong:2021bvy}
\bibitem{Dong:2021bvy}
X.~K.~Dong, F.~K.~Guo and B.~S.~Zou,
A survey of heavy\textendash{}heavy hadronic molecules,
\href{https://doi.org/10.1088/1572-9494/ac27a2}{Commun. Theor. Phys. \textbf{73}, 125201 (2021)}.
%doi:10.1088/1572-9494/ac27a2
%[arXiv:2108.02673 [hep-ph]].
%131 citations counted in INSPIRE as of 01 Mar 2024

%\cite{Asanuma:2023atv}
\bibitem{Asanuma:2023atv}
T.~Asanuma, Y.~Yamaguchi and M.~Harada,
Analysis of $DD^*$ and $\bar{D}^{(*)}\Xi_{cc}^{(*)}$ molecule by one boson exchange model based on heavy quark symmetry,
\href{https://doi.org/10.1103/PhysRevD.110.074030}{Phys. Rev. D \textbf{110}, 074030 (2024)}.
%doi:10.1103/PhysRevD.110.074030
%[arXiv:2311.04695 [hep-ph]].
%7 citations counted in INSPIRE as of 18 Nov 2024

%\cite{Sakai:2023syt}
\bibitem{Sakai:2023syt}
M.~Sakai and Y.~Yamaguchi,
Analysis of $T_{cc}$ and $T_{bb}$ based on the hadronic molecular model and their spin multiplets,
\href{https://doi.org/10.1103/PhysRevD.109.054016}{Phys. Rev. D \textbf{109}, 054016 (2024)}.
%doi:10.1103/PhysRevD.109.054016
%[arXiv:2312.08663 [hep-ph]].
%2 citations counted in INSPIRE as of 12 Mar 2024

%\cite{Qiu:2023uno}
\bibitem{Qiu:2023uno}
L.~Qiu, C.~Gong and Q.~Zhao,
Coupled-channel description of charmed heavy hadronic molecules within the meson-exchange model and its implication,
\href{https://doi.org/10.1103/PhysRevD.109.076016}{Phys. Rev. D \textbf{109}, 076016 (2024)}.
%doi:10.1103/PhysRevD.109.076016
%[arXiv:2311.10067 [hep-ph]].
%5 citations counted in INSPIRE as of 18 Apr 2024

%\cite{Zhao:2021cvg}
\bibitem{Zhao:2021cvg}
M.~J.~Zhao, Z.~Y.~Wang, C.~Wang and X.~H.~Guo,
Investigation of the possible $D\bar{D}^*/B\bar{B}^*$ and $DD^*/\bar{B}\bar{B}^*$ bound states,
\href{https://doi.org/10.1103/PhysRevD.105.096016}{Phys. Rev. D \textbf{105}, 096016 (2022)}.
%doi:10.1103/PhysRevD.105.096016
%[arXiv:2112.12633 [hep-ph]].
%16 citations counted in INSPIRE as of 02 Mar 2024

%\cite{Ke:2021rxd}
\bibitem{Ke:2021rxd}
H.~W.~Ke, X.~H.~Liu and X.~Q.~Li,
Possible molecular states of $D^{(*)}D^{(*)}$ and $B^{(*)}B^{(*)}$ within the Bethe\textendash{}Salpeter framework,
\href{https://doi.org/10.1140/epjc/s10052-022-10092-8}{Eur. Phys. J. C \textbf{82}, 144 (2022)}.
%doi:10.1140/epjc/s10052-022-10092-8
%[arXiv:2112.14142 [hep-ph]].
%32 citations counted in INSPIRE as of 02 Mar 2024

%\cite{Feng:2013kea}
\bibitem{Feng:2013kea}
G.~Q.~Feng, X.~H.~Guo and B.~S.~Zou,
$QQ'\bar u\bar d$ bound state in the Bethe-Salpeter equation approach,
\href{https://arxiv.org/abs/1309.7813}{arXiv:1309.7813}.
%34 citations counted in INSPIRE as of 01 Mar 2024

%\cite{Ding:2020dio}
\bibitem{Ding:2020dio}
Z.~M.~Ding, H.~Y.~Jiang and J.~He,
Molecular states from $D^{(*)}\bar{D}^{(*)}/B^{(*)}\bar{B}^{(*)}$ and $D^{(*)}D^{(*)}/\bar{B}^{(*)}\bar{B}^{(*)}$ interactions,
\href{https://doi.org/10.1140/epjc/s10052-020-08754-6}{Eur. Phys. J. C \textbf{80}, 1179 (2020)}.
%doi:10.1140/epjc/s10052-020-08754-6
%[arXiv:2011.04980 [hep-ph]].
%37 citations counted in INSPIRE as of 01 Mar 2024

%\cite{Wallbott:2020jzh}
\bibitem{Wallbott:2020jzh}
P.~C.~Wallbott, G.~Eichmann and C.~S.~Fischer,
Disentangling different structures in heavy-light four-quark states,
\href{https://doi.org/10.1103/PhysRevD.102.051501}{Phys. Rev. D \textbf{102}, 051501 (2020)}.
%doi:10.1103/PhysRevD.102.051501
%[arXiv:2003.12407 [hep-ph]].
%13 citations counted in INSPIRE as of 01 Mar 2024

%\cite{Mehen:2017nrh}
\bibitem{Mehen:2017nrh}
T.~Mehen,
Implications of Heavy Quark-Diquark Symmetry for Excited Doubly Heavy Baryons and Tetraquarks,
\href{https://doi.org/10.1103/PhysRevD.96.094028}{Phys. Rev. D \textbf{96}, 094028 (2017)}.
%doi:10.1103/PhysRevD.96.094028
%[arXiv:1708.05020 [hep-ph]].
%63 citations counted in INSPIRE as of 01 Mar 2024

%\cite{Eichten:2017ffp}
\bibitem{Eichten:2017ffp}
E.~J.~Eichten and C.~Quigg,
Heavy-quark symmetry implies stable heavy tetraquark mesons $Q_iQ_j\bar q_k\bar q_l$,
\href{https://doi.org/10.1103/PhysRevLett.119.202002}{Phys. Rev. Lett. \textbf{119}, 202002 (2017)}.
%doi:10.1103/PhysRevLett.119.202002
%[arXiv:1707.09575 [hep-ph]].
%256 citations counted in INSPIRE as of 01 Mar 2024

%\cite{Braaten:2020nwp}
\bibitem{Braaten:2020nwp}
E.~Braaten, L.~P.~He and A.~Mohapatra,
Masses of doubly heavy tetraquarks with error bars,
\href{https://doi.org/10.1103/PhysRevD.103.016001}{Phys. Rev. D \textbf{103}, 016001 (2021)}.
%doi:10.1103/PhysRevD.103.016001
%[arXiv:2006.08650 [hep-ph]].
%50 citations counted in INSPIRE as of 01 Mar 2024

%\cite{Ikeda:2013vwa}
\bibitem{Ikeda:2013vwa}
Y.~Ikeda, B.~Charron, S.~Aoki, T.~Doi, T.~Hatsuda, T.~Inoue, N.~Ishii, K.~Murano, H.~Nemura and K.~Sasaki,
Charmed tetraquarks $T_{cc}$ and $T_{cs}$ from dynamical lattice QCD simulations,
\href{https://doi.org/10.1016/j.physletb.2014.01.002}{Phys. Lett. B \textbf{729}, 85-90 (2014)}.
%doi:10.1016/j.physletb.2014.01.002
%[arXiv:1311.6214 [hep-lat]].
%112 citations counted in INSPIRE as of 01 Mar 2024

%\cite{Junnarkar:2018twb}
\bibitem{Junnarkar:2018twb}
P.~Junnarkar, N.~Mathur and M.~Padmanath,
Study of doubly heavy tetraquarks in Lattice QCD,
\href{https://doi.org/10.1103/PhysRevD.99.034507}{Phys. Rev. D \textbf{99}, 034507 (2019)}.
%doi:10.1103/PhysRevD.99.034507
%[arXiv:1810.12285 [hep-lat]].
%155 citations counted in INSPIRE as of 01 Mar 2024

%\cite{Padmanath:2022cvl}
\bibitem{Padmanath:2022cvl}
M.~Padmanath and S.~Prelovsek,
Signature of a Doubly Charm Tetraquark Pole in $DD^*$ Scattering on the Lattice,
\href{https://doi.org/10.1103/PhysRevLett.129.032002}{Phys. Rev. Lett. \textbf{129}, 032002 (2022)}.
%doi:10.1103/PhysRevLett.129.032002
%[arXiv:2202.10110 [hep-lat]].
%56 citations counted in INSPIRE as of 02 Mar 2024

%\cite{Lyu:2023xro}
\bibitem{Lyu:2023xro}
Y.~Lyu, S.~Aoki, T.~Doi, T.~Hatsuda, Y.~Ikeda and J.~Meng,
Doubly Charmed Tetraquark $T_{cc}^{+}$ from Lattice QCD near Physical Point,
\href{https://doi.org/10.1103/PhysRevLett.131.161901}{Phys. Rev. Lett. \textbf{131}, 161901 (2023)}.
%doi:10.1103/PhysRevLett.131.161901
%[arXiv:2302.04505 [hep-lat]].
%30 citations counted in INSPIRE as of 02 Mar 2024

%\cite{Dai:2021wxi}
\bibitem{Dai:2021wxi}
L.~Y.~Dai, X.~Sun, X.~W.~Kang, A.~P.~Szczepaniak and J.~S.~Yu,
Pole analysis on the doubly charmed meson in $D^0D^0\pi^+$ mass spectrum,
\href{https://doi.org/10.1103/PhysRevD.105.L051507}{Phys. Rev. D \textbf{105}, L051507 (2022)}.
%doi:10.1103/PhysRevD.105.L051507
%[arXiv:2108.06002 [hep-ph]].
%25 citations counted in INSPIRE as of 02 Mar 2024

%\cite{Dai:2021vgf}
\bibitem{Dai:2021vgf}
L.~R.~Dai, R.~Molina and E.~Oset,
Prediction of new $T_{cc}$ states of $D^*D^*$ and $D_s^*D^*$ molecular nature,
\href{https://doi.org/10.1103/PhysRevD.105.016029}{Phys. Rev. D \textbf{105}, 016029 (2022)}
[erratum: \href{https://doi.org/10.1103/PhysRevD.106.099902}{Phys. Rev. D \textbf{106}, 099902 (2022)}].
%doi:10.1103/PhysRevD.105.016029
%[arXiv:2110.15270 [hep-ph]].
%34 citations counted in INSPIRE as of 02 Mar 2024

%\cite{Feijoo:2021ppq}
\bibitem{Feijoo:2021ppq}
A.~Feijoo, W.~H.~Liang and E.~Oset,
$D^0D^0\pi^+$ mass distribution in the production of the $T_{cc}$ exotic state,
\href{https://doi.org/10.1103/PhysRevD.104.114015}{Phys. Rev. D \textbf{104}, 114015 (2021)}.
%doi:10.1103/PhysRevD.104.114015
%[arXiv:2108.02730 [hep-ph]].
%90 citations counted in INSPIRE as of 01 Mar 2024

%\cite{Huang:2021urd}
\bibitem{Huang:2021urd}
Y.~Huang, H.~Q.~Zhu, L.~S.~Geng and R.~Wang,
Production of $T_{cc}^+$ exotic state in the $\gamma p\rightarrow D^+\bar{T}_{cc}^-\Lambda_c^+$ reaction,
\href{https://doi.org/10.1103/PhysRevD.104.116008}{Phys. Rev. D \textbf{104}, 116008 (2021)}.
%doi:10.1103/PhysRevD.104.116008
%[arXiv:2108.13028 [hep-ph]].
%33 citations counted in INSPIRE as of 01 Mar 2024

%\cite{Ling:2021bir}
\bibitem{Ling:2021bir}
X.~Z.~Ling, M.~Z.~Liu, L.~S.~Geng, E.~Wang and J.~J.~Xie,
Can we understand the decay width of the $T_{cc}^+$ state?,
\href{https://doi.org/10.1016/j.physletb.2022.136897}{Phys. Lett. B \textbf{826}, 136897 (2022)}.
%doi:10.1016/j.physletb.2022.136897
%[arXiv:2108.00947 [hep-ph]].
%72 citations counted in INSPIRE as of 01 Mar 2024

%\cite{Du:2021zzh}
\bibitem{Du:2021zzh}
M.~L.~Du, V.~Baru, X.~K.~Dong, A.~Filin, F.~K.~Guo, C.~Hanhart, A.~Nefediev, J.~Nieves and Q.~Wang,
Coupled-channel approach to $T_{cc}^+$ including three-body effects,
\href{https://doi.org/10.1103/PhysRevD.105.014024}{Phys. Rev. D \textbf{105}, 014024 (2022)}.
%doi:10.1103/PhysRevD.105.014024
%[arXiv:2110.13765 [hep-ph]].
%90 citations counted in INSPIRE as of 01 Mar 2024

%\cite{Albaladejo:2021vln}
\bibitem{Albaladejo:2021vln}
M.~Albaladejo,
$T_{cc}^{+}$ coupled channel analysis and predictions,
\href{https://doi.org/10.1016/j.physletb.2022.137052}{Phys. Lett. B \textbf{829}, 137052 (2022)}.
%doi:10.1016/j.physletb.2022.137052
%[arXiv:2110.02944 [hep-ph]].
%70 citations counted in INSPIRE as of 02 Mar 2024

%\cite{Fleming:2021wmk}
\bibitem{Fleming:2021wmk}
S.~Fleming, R.~Hodges and T.~Mehen,
$T_{cc}^+$ decays: Differential spectra and two-body final states,
\href{https://doi.org/10.1103/PhysRevD.104.116010}{Phys. Rev. D \textbf{104}, 116010 (2021)}.
%doi:10.1103/PhysRevD.104.116010
%[arXiv:2109.02188 [hep-ph]].
%53 citations counted in INSPIRE as of 01 Mar 2024

%\cite{Meng:2021jnw}
\bibitem{Meng:2021jnw}
L.~Meng, G.~J.~Wang, B.~Wang and S.~L.~Zhu,
Probing the long-range structure of the $T_{cc}^+$ with the strong and electromagnetic decays,
\href{https://doi.org/10.1103/PhysRevD.104.L051502}{Phys. Rev. D \textbf{104}, 051502 (2021)}.
%doi:10.1103/PhysRevD.104.L051502
%[arXiv:2107.14784 [hep-ph]].
%72 citations counted in INSPIRE as of 01 Mar 2024

%\cite{Chen:2021cfl}
\bibitem{Chen:2021cfl}
K.~Chen, R.~Chen, L.~Meng, B.~Wang and S.~L.~Zhu,
Systematics of the heavy flavor hadronic molecules,
\href{https://doi.org/10.1140/epjc/s10052-022-10540-5}{Eur. Phys. J. C \textbf{82}, 581 (2022)}.
%doi:10.1140/epjc/s10052-022-10540-5
%[arXiv:2109.13057 [hep-ph]].
%57 citations counted in INSPIRE as of 02 Mar 2024

%\cite{Braaten:2022elw}
\bibitem{Braaten:2022elw}
E.~Braaten, L.~P.~He, K.~Ingles and J.~Jiang,
Triangle singularity in the production of $T_{cc}^{+}(3875)$ and a soft pion,
\href{https://doi.org/10.1103/PhysRevD.106.034033}{Phys. Rev. D \textbf{106}, 034033 (2022)}.
%doi:10.1103/PhysRevD.106.034033
%[arXiv:2202.03900 [hep-ph]].
%19 citations counted in INSPIRE as of 02 Mar 2024

%\cite{Wang:2022jop}
\bibitem{Wang:2022jop}
B.~Wang and L.~Meng,
Revisiting the $DD^*$ chiral interactions with the local momentum-space regularization up to the third order and the nature of $T_{cc}^{+}$,
\href{https://doi.org/10.1103/PhysRevD.107.094002}{Phys. Rev. D \textbf{107}, 094002 (2023)}.
%doi:10.1103/PhysRevD.107.094002
%[arXiv:2212.08447 [hep-ph]].
%18 citations counted in INSPIRE as of 02 Mar 2024

%\cite{Dai:2023mxm}
\bibitem{Dai:2023mxm}
L.~Dai, S.~Fleming, R.~Hodges and T.~Mehen,
Strong decays of $T_{cc}^{+}$ at NLO in an effective field theory,
\href{https://doi.org/10.1103/PhysRevD.107.076001}{Phys. Rev. D \textbf{107}, 076001 (2023)}.
%doi:10.1103/PhysRevD.107.076001
%[arXiv:2301.11950 [hep-ph]].
%12 citations counted in INSPIRE as of 02 Mar 2024

%\cite{Yan:2021wdl}
\bibitem{Yan:2021wdl}
M.~J.~Yan and M.~P.~Valderrama,
Subleading contributions to the decay width of the $T_{cc}^{+}$ tetraquark,
\href{https://doi.org/10.1103/PhysRevD.105.014007}{Phys. Rev. D \textbf{105}, 014007 (2022)}.
%doi:10.1103/PhysRevD.105.014007
%[arXiv:2108.04785 [hep-ph]].
%57 citations counted in INSPIRE as of 02 Mar 2024

%\cite{Kucab:2024nkv}
\bibitem{Kucab:2024nkv}
M.~Kucab and M.~Praszalowicz,
Heavy quarkonia, heavy-light tetraquarks, and the chiral quark-soliton model,
\href{https://doi.org/10.1103/PhysRevD.109.076005}{Phys. Rev. D \textbf{109}, 076005 (2024)}.
%doi:10.1103/PhysRevD.109.076005
%[arXiv:2402.04169 [hep-ph]].
%0 citations counted in INSPIRE as of 09 Apr 2024

%\cite{Song:2023izj}
\bibitem{Song:2023izj}
Y.~Song and D.~Jia,
Mass spectra of doubly heavy tetraquarks in diquark\ensuremath{-}antidiquark picture,
\href{https://doi.org/10.1088/1572-9494/acc019}{Commun. Theor. Phys. \textbf{75}, 055201 (2023)}.
%doi:10.1088/1572-9494/acc019
%[arXiv:2301.00376 [hep-ph]].
%4 citations counted in INSPIRE as of 02 Mar 2024

%\cite{Chen:2023web}
\bibitem{Chen:2023web}
J.~K.~Chen,
Regge trajectory relations for the universal description of the heavy-light systems: diquarks, mesons, baryons and tetraquarks,
\href{https://doi.org/10.1140/epjc/s10052-024-12706-9}{Eur. Phys. J. C \textbf{84}, 356 (2024)}.
%doi:10.1140/epjc/s10052-024-12706-9
%[arXiv:2302.06794 [hep-ph]].
%5 citations counted in INSPIRE as of 02 Jun 2024

%\cite{Liu:2019yye}
\bibitem{Liu:2019yye}
Y.~Liu, M.~A.~Nowak and I.~Zahed,
Holographic tetraquarks and the newly observed $T_{cc}^{+}$ at LHCb,
\href{https://doi.org/10.1103/PhysRevD.105.054021}{Phys. Rev. D \textbf{105}, 054021 (2022)}.
%doi:10.1103/PhysRevD.105.054021
%[arXiv:1909.02497 [hep-ph]].
%10 citations counted in INSPIRE as of 02 Mar 2024

%\cite{Sonnenschein:2024rzw}
\bibitem{Sonnenschein:2024rzw}
J.~Sonnenschein and M.~M.~Green,
Taming the Zoo of Tetraquarks and Pentaquarks using the HISH Model,
\href{https://arxiv.org/abs/2401.01621}{arXiv:2401.01621}.
%2 citations counted in INSPIRE as of 02 Mar 2024

%\cite{Jin:2021cxj}
\bibitem{Jin:2021cxj}
Y.~Jin, S.~Y.~Li, Y.~R.~Liu, Q.~Qin, Z.~G.~Si and F.~S.~Yu,
Color and baryon number fluctuation of preconfinement system in production process and $T_{cc}$ structure,
\href{https://doi.org/10.1103/PhysRevD.104.114009}{Phys. Rev. D \textbf{104}, 114009 (2021)}.
%doi:10.1103/PhysRevD.104.114009
%[arXiv:2109.05678 [hep-ph]].
%31 citations counted in INSPIRE as of 01 Mar 2024

%\cite{Qin:2020zlg}
\bibitem{Qin:2020zlg}
Q.~Qin, Y.~F.~Shen and F.~S.~Yu,
Discovery potentials of double-charm tetraquarks,
\href{https://doi.org/10.1088/1674-1137/ac1b97}{Chin. Phys. C \textbf{45}, 103106 (2021)}.
%doi:10.1088/1674-1137/ac1b97
%[arXiv:2008.08026 [hep-ph]].
%61 citations counted in INSPIRE as of 01 Mar 2024

%\cite{Hu:2021gdg}
\bibitem{Hu:2021gdg}
Y.~Hu, J.~Liao, E.~Wang, Q.~Wang, H.~Xing and H.~Zhang,
Production of doubly charmed exotic hadrons in heavy ion collisions,
\href{https://doi.org/10.1103/PhysRevD.104.L111502}{Phys. Rev. D \textbf{104}, L111502 (2021)}.
%doi:10.1103/PhysRevD.104.L111502
%[arXiv:2109.07733 [hep-ph]].
%40 citations counted in INSPIRE as of 01 Mar 2024

%\cite{Godfrey:1985xj}
\bibitem{Godfrey:1985xj}
S.~Godfrey and N.~Isgur,
Mesons in a Relativized Quark Model with Chromodynamics,
\href{https://doi.org/10.1103/PhysRevD.32.189}{Phys. Rev. D \textbf{32}, 189-231 (1985)}.
%doi:10.1103/PhysRevD.32.189
%3012 citations counted in INSPIRE as of 27 Aug 2022

%\cite{Godfrey:2016nwn}
\bibitem{Godfrey:2016nwn}
S.~Godfrey, K.~Moats and E.~S.~Swanson,
$B$ and $B_s$ Meson Spectroscopy,
\href{https://doi.org/10.1103/PhysRevD.94.054025}{Phys. Rev. D \textbf{94}, 054025 (2016)}.
%doi:10.1103/PhysRevD.94.054025
%[arXiv:1607.02169 [hep-ph]].
%67 citations counted in INSPIRE as of 27 Aug 2022

%\cite{Godfrey:2015dva}
\bibitem{Godfrey:2015dva}
S.~Godfrey and K.~Moats,
Properties of Excited Charm and Charm-Strange Mesons,
\href{https://doi.org/10.1103/PhysRevD.93.034035}{Phys. Rev. D \textbf{93}, 034035 (2016)}.
%doi:10.1103/PhysRevD.93.034035
%[arXiv:1510.08305 [hep-ph]].
%135 citations counted in INSPIRE as of 27 Aug 2022

%\cite{Godfrey:2004ya}
\bibitem{Godfrey:2004ya}
S.~Godfrey,
Spectroscopy of $B_c$ mesons in the relativized quark model,
\href{https://doi.org/10.1103/PhysRevD.70.054017}{Phys. Rev. D \textbf{70}, 054017 (2004)}.
%doi:10.1103/PhysRevD.70.054017
%[arXiv:hep-ph/0406228 [hep-ph]].
%188 citations counted in INSPIRE as of 27 Aug 2022

%\cite{Barnes:2005pb}
\bibitem{Barnes:2005pb}
T.~Barnes, S.~Godfrey and E.~S.~Swanson,
Higher charmonia,
\href{https://doi.org/10.1103/PhysRevD.72.054026}{Phys. Rev. D \textbf{72}, 054026 (2005)}.
%doi:10.1103/PhysRevD.72.054026
%[arXiv:hep-ph/0505002 [hep-ph]].
%708 citations counted in INSPIRE as of 27 Aug 2022

%\cite{Godfrey:2015dia}
\bibitem{Godfrey:2015dia}
S.~Godfrey and K.~Moats,
Bottomonium Mesons and Strategies for their Observation,
\href{https://doi.org/10.1103/PhysRevD.92.054034}{Phys. Rev. D \textbf{92}, 054034 (2015)}.
%doi:10.1103/PhysRevD.92.054034
%[arXiv:1507.00024 [hep-ph]].
%109 citations counted in INSPIRE as of 27 Aug 2022

%\cite{Capstick:1986ter}
\bibitem{Capstick:1986ter}
S.~Capstick and N.~Isgur,
Baryons in a relativized quark model with chromodynamics,
\href{https://doi.org/10.1103/physrevd.34.2809}{Phys. Rev. D \textbf{34}, 2809-2835 (1986)}.
%doi:10.1103/physrevd.34.2809
%1388 citations counted in INSPIRE as of 27 Aug 2022

%\cite{Lu:2016ctt}
\bibitem{Lu:2016ctt}
Q.~F.~L\"u, Y.~Dong, X.~Liu and T.~Matsuki,
Puzzle of the $\Lambda_c$ Spectrum,
\href{https://doi.org/10.11804/NuclPhysRev.35.01.001}{Nucl. Phys. Rev. \textbf{35}, 1-4 (2018)}.
%doi:10.11804/NuclPhysRev.35.01.001
%[arXiv:1610.09605 [hep-ph]].
%17 citations counted in INSPIRE as of 27 Aug 2022

%\cite{Lu:2017meb}
\bibitem{Lu:2017meb}
Q.~F.~L\"u, K.~L.~Wang, L.~Y.~Xiao and X.~H.~Zhong,
Mass spectra and radiative transitions of doubly heavy baryons in a relativized quark model,
\href{https://doi.org/10.1103/PhysRevD.96.114006}{Phys. Rev. D \textbf{96}, 114006 (2017)}.
%doi:10.1103/PhysRevD.96.114006
%[arXiv:1708.04468 [hep-ph]].
%65 citations counted in INSPIRE as of 28 Aug 2022

%\cite{Bedolla:2019zwg}
\bibitem{Bedolla:2019zwg}
M.~A.~Bedolla, J.~Ferretti, C.~D.~Roberts and E.~Santopinto,
Spectrum of fully-heavy tetraquarks from a diquark+antidiquark perspective,
\href{https://doi.org/10.1140/epjc/s10052-020-08579-3}{Eur. Phys. J. C \textbf{80}, 1004 (2020)}.
%doi:10.1140/epjc/s10052-020-08579-3
%[arXiv:1911.00960 [hep-ph]].
%79 citations counted in INSPIRE as of 22 Aug 2022

%\cite{Lu:2019ira}
\bibitem{Lu:2019ira}
Q.~F.~L\"u, K.~L.~Wang and Y.~B.~Dong,
The $ss \bar s \bar s$ tetraquark states and the structure of $X(2239)$ observed by the BESIII Collaboration,
\href{https://doi.org/10.1088/1674-1137/44/2/024101}{Chin. Phys. C \textbf{44}, 024101 (2020)}.
%doi:10.1088/1674-1137/44/2/024101
%[arXiv:1903.05007 [hep-ph]].
%14 citations counted in INSPIRE as of 28 Aug 2022

%\cite{Lu:2016zhe}
\bibitem{Lu:2016zhe}
Q.~F.~L\"u and Y.~B.~Dong,
Masses of open charm and bottom tetraquark states in a relativized quark model,
\href{https://doi.org/10.1103/PhysRevD.94.094041}{Phys. Rev. D \textbf{94}, 094041 (2016)}.
%doi:10.1103/PhysRevD.94.094041
%[arXiv:1603.06417 [hep-ph]].
%39 citations counted in INSPIRE as of 28 Aug 2022

%\cite{Lu:2016cwr}
\bibitem{Lu:2016cwr}
Q.~F.~L\"u and Y.~B.~Dong,
$X(4140)$, $X(4274)$, $X(4500)$, and $X(4700)$ in the relativized quark model,
\href{https://doi.org/10.1103/PhysRevD.94.074007}{Phys. Rev. D \textbf{94}, 074007 (2016)}.
%doi:10.1103/PhysRevD.94.074007
%[arXiv:1607.05570 [hep-ph]].
%49 citations counted in INSPIRE as of 28 Aug 2022

%\cite{Dong:2022sef}
\bibitem{Dong:2022sef}
W.~C.~Dong and Z.~G.~Wang,
Going in quest of potential tetraquark interpretations for the newly observed $T_{\psi\psi}$ states in light of the diquark-antidiquark scenarios,
\href{https://doi.org/10.1103/PhysRevD.107.074010}{Phys. Rev. D \textbf{107}, 074010 (2023)}.
%doi:10.1103/PhysRevD.107.074010
%[arXiv:2211.11989 [hep-ph]].
%16 citations counted in INSPIRE as of 11 Mar 2024

%\cite{Ferretti:2019zyh}
\bibitem{Ferretti:2019zyh}
J.~Ferretti,
Effective Degrees of Freedom in Baryon and Meson Spectroscopy,
\href{https://doi.org/10.1007/s00601-019-1483-2}{Few Body Syst. \textbf{60}, 17 (2019)}.
%doi:10.1007/s00601-019-1483-2
%9 citations counted in INSPIRE as of 28 Aug 2022

%\cite{Landau:1991wop}
\bibitem{Landau:1991wop}
L.~D.~Landau and E.~M.~Lifshitz,
\textit{Quantum Mechanics: Non-Relativistic Theory}, 3rd ed.
(\href{https://doi.org/10.1016/C2013-0-02793-4}{Pergamon Press, Oxford, 1977}).
%ISBN 978-0-08-020940-1
%41 citations counted in INSPIRE as of 01 Sep 2022

%\cite{Ali:2017wsf}
\bibitem{Ali:2017wsf}
A.~Ali, L.~Maiani, A.~V.~Borisov, I.~Ahmed, M.~Jamil Aslam, A.~Y.~Parkhomenko, A.~D.~Polosa and A.~Rehman,
A new look at the Y tetraquarks and $\Omega _c$ baryons in the diquark model,
\href{https://doi.org/10.1140/epjc/s10052-017-5501-6}{Eur. Phys. J. C \textbf{78}, 29 (2018)}.
%doi:10.1140/epjc/s10052-017-5501-6
%[arXiv:1708.04650 [hep-ph]].
%40 citations counted in INSPIRE as of 02 Sep 2022

%\cite{Varshalovich:1988krb}
\bibitem{Varshalovich:1988krb}
D.~A.~Varshalovich, A.~N.~Moskalev and V.~K.~Khersonskii,
\textit{Quantum Theory of Angular Momentum}
(\href{https://doi.org/10.1142/0270}{World Scientific, Singapore, 1988}).
%ISBN 978-981-4415-49-1, 978-9971-5-0107-5
%doi:10.1142/0270
%107 citations counted in INSPIRE as of 02 Sep 2022

%\cite{Vijande:2009ac}
\bibitem{Vijande:2009ac}
J.~Vijande and A.~Valcarce,
Tetraquark Spectroscopy: A Symmetry Analysis,
\href{https://doi.org/10.3390/sym1020155}{Symmetry \textbf{1}, 155-179 (2009)}.
%doi:10.3390/sym1020155
%[arXiv:0912.3605 [hep-ph]].
%29 citations counted in INSPIRE as of 06 Sep 2022

%\cite{Anselmino:1992vg}
\bibitem{Anselmino:1992vg}
M.~Anselmino, E.~Predazzi, S.~Ekelin, S.~Fredriksson and D.~B.~Lichtenberg,
Diquarks,
\href{https://doi.org/10.1103/RevModPhys.65.1199}{Rev. Mod. Phys. \textbf{65}, 1199-1234 (1993)}.
%doi:10.1103/RevModPhys.65.1199
%410 citations counted in INSPIRE as of 26 Nov 2021

%\cite{Jaffe:2004ph}
\bibitem{Jaffe:2004ph}
R.~L.~Jaffe,
Exotica,
\href{https://doi.org/10.1016/j.physrep.2004.11.005}{Phys. Rept. \textbf{409}, 1-45 (2005)}.
%doi:10.1016/j.physrep.2004.11.005
%[arXiv:hep-ph/0409065 [hep-ph]].
%544 citations counted in INSPIRE as of 27 Nov 2021

%\cite{Barabanov:2020jvn}
\bibitem{Barabanov:2020jvn}
M.~Y.~Barabanov, M.~A.~Bedolla, W.~K.~Brooks, G.~D.~Cates, C.~Chen, Y.~Chen, E.~Cisbani, M.~Ding, G.~Eichmann and R.~Ent, \textit{et al.}
Diquark correlations in hadron physics: Origin, impact and evidence,
\href{https://doi.org/10.1016/j.ppnp.2020.103835}{Prog. Part. Nucl. Phys. \textbf{116}, 103835 (2021)}.
%doi:10.1016/j.ppnp.2020.103835
%[arXiv:2008.07630 [hep-ph]].
%30 citations counted in INSPIRE as of 26 Nov 2021

%\cite{Born:1989iv}
\bibitem{Born:1989iv}
K.~D.~Born, E.~Laermann, N.~Pirch, T.~F.~Walsh and P.~M.~Zerwas,
Hadron Properties in Lattice {QCD} With Dynamical Fermions,
\href{https://doi.org/10.1103/PhysRevD.40.1653}{Phys. Rev. D \textbf{40}, 1653-1663 (1989)}.
%doi:10.1103/PhysRevD.40.1653
%130 citations counted in INSPIRE as of 13 Sep 2022

%\cite{Li:2009zu}
\bibitem{Li:2009zu}
B.~Q.~Li and K.~T.~Chao,
Higher Charmonia and $X,Y,Z$ states with Screened Potential,
\href{https://doi.org/10.1103/PhysRevD.79.094004}{Phys. Rev. D \textbf{79}, 094004 (2009)}.
%doi:10.1103/PhysRevD.79.094004
%[arXiv:0903.5506 [hep-ph]].
%265 citations counted in INSPIRE as of 13 Sep 2022

%\cite{Mezoir:2008vx}
\bibitem{Mezoir:2008vx}
E.~H.~Mezoir and P.~Gonzalez,
Is the spectrum of highly excited mesons purely coulombian?,
\href{https://doi.org/10.1103/PhysRevLett.101.232001}{Phys. Rev. Lett. \textbf{101}, 232001 (2008)}.
%doi:10.1103/PhysRevLett.101.232001
%[arXiv:0810.5651 [hep-ph]].
%23 citations counted in INSPIRE as of 17 Mar 2024

%\cite{Song:2015nia}
\bibitem{Song:2015nia}
Q.~T.~Song, D.~Y.~Chen, X.~Liu and T.~Matsuki,
Charmed-strange mesons revisited: mass spectra and strong decays,
\href{https://doi.org/10.1103/PhysRevD.91.054031}{Phys. Rev. D \textbf{91}, 054031 (2015)}.
%doi:10.1103/PhysRevD.91.054031
%[arXiv:1501.03575 [hep-ph]].
%64 citations counted in INSPIRE as of 13 Sep 2022

%\cite{Pang:2017dlw}
\bibitem{Pang:2017dlw}
C.~Q.~Pang, J.~Z.~Wang, X.~Liu and T.~Matsuki,
A systematic study of mass spectra and strong decay of strange mesons,
\href{https://doi.org/10.1140/epjc/s10052-017-5434-0}{Eur. Phys. J. C \textbf{77}, 861 (2017)}.
%doi:10.1140/epjc/s10052-017-5434-0
%[arXiv:1705.03144 [hep-ph]].
%31 citations counted in INSPIRE as of 13 Sep 2022

%\cite{Song:2015fha}
\bibitem{Song:2015fha}
Q.~T.~Song, D.~Y.~Chen, X.~Liu and T.~Matsuki,
Higher radial and orbital excitations in the charmed meson family,
\href{https://doi.org/10.1103/PhysRevD.92.074011}{Phys. Rev. D \textbf{92}, 074011 (2015)}.
%doi:10.1103/PhysRevD.92.074011
%[arXiv:1503.05728 [hep-ph]].
%55 citations counted in INSPIRE as of 13 Sep 2022

%\cite{Wang:2018rjg}
\bibitem{Wang:2018rjg}
J.~Z.~Wang, Z.~F.~Sun, X.~Liu and T.~Matsuki,
Higher bottomonium zoo,
\href{https://doi.org/10.1140/epjc/s10052-018-6372-1}{Eur. Phys. J. C \textbf{78}, 915 (2018)}.
%doi:10.1140/epjc/s10052-018-6372-1
%[arXiv:1802.04938 [hep-ph]].
%27 citations counted in INSPIRE as of 13 Sep 2022

%\cite{Eichten:1979ms}
\bibitem{Eichten:1979ms}
E.~Eichten, K.~Gottfried, T.~Kinoshita, K.~D.~Lane and T.~M.~Yan,
Charmonium: Comparison with experiment,
\href{https://doi.org/10.1103/PhysRevD.21.203}{Phys. Rev. D \textbf{21}, 203 (1980)}.
%doi:10.1103/PhysRevD.21.203
%1863 citations counted in INSPIRE as of 19 Sep 2022

%\cite{Kim:2020imk}
\bibitem{Kim:2020imk}
Y.~Kim, E.~Hiyama, M.~Oka and K.~Suzuki,
Spectrum of singly heavy baryons from a chiral effective theory of diquarks,
\href{https://doi.org/10.1103/PhysRevD.102.014004}{Phys. Rev. D \textbf{102}, 014004 (2020)}.
%doi:10.1103/PhysRevD.102.014004
%[arXiv:2003.03525 [hep-ph]].
%19 citations counted in INSPIRE as of 30 Mar 2024

%\cite{Kim:2021ywp}
\bibitem{Kim:2021ywp}
Y.~Kim, Y.~R.~Liu, M.~Oka and K.~Suzuki,
Heavy baryon spectrum with chiral multiplets of scalar and vector diquarks,
\href{https://doi.org/10.1103/PhysRevD.104.054012}{Phys. Rev. D \textbf{104}, 054012 (2021)}.
%doi:10.1103/PhysRevD.104.054012
%[arXiv:2105.09087 [hep-ph]].
%15 citations counted in INSPIRE as of 30 Mar 2024

%\cite{Yoshida:2015tia}
\bibitem{Yoshida:2015tia}
T.~Yoshida, E.~Hiyama, A.~Hosaka, M.~Oka and K.~Sadato,
Spectrum of heavy baryons in the quark model,
\href{https://doi.org/10.1103/PhysRevD.92.114029}{Phys. Rev. D \textbf{92}, 114029 (2015)}.
%doi:10.1103/PhysRevD.92.114029
%[arXiv:1510.01067 [hep-ph]].
%209 citations counted in INSPIRE as of 30 Mar 2024

%\cite{Kawanai:2011xb}
\bibitem{Kawanai:2011xb}
T.~Kawanai and S.~Sasaki,
Interquark potential with finite quark mass from lattice QCD,
\href{https://doi.org/10.1103/PhysRevLett.107.091601}{Phys. Rev. Lett. \textbf{107}, 091601 (2011)}.
%doi:10.1103/PhysRevLett.107.091601
%[arXiv:1102.3246 [hep-lat]].
%72 citations counted in INSPIRE as of 20 Apr 2024

%\cite{Amsler:2018zkm}
\bibitem{Amsler:2018zkm}
C.~Amsler,
\textit{The Quark Structure of Hadrons: An Introduction to the Phenomenology and Spectroscopy}
(\href{https://doi.org/10.1007/978-3-319-98527-5}{Springer, Cham, 2018}).
%Lect. Notes Phys. \textbf{949}, pp.1-277 (2018)
%ISBN 978-3-319-98526-8, 978-3-319-98527-5
%doi:10.1007/978-3-319-98527-5
%2 citations counted in INSPIRE as of 26 Sep 2022

%\cite{Bi:2015ifa}
\bibitem{Bi:2015ifa}
Y.~Bi, H.~Cai, Y.~Chen, M.~Gong, Z.~Liu, H.~X.~Qiao and Y.~B.~Yang,
Diquark mass differences from unquenched lattice QCD,
\href{https://doi.org/10.1088/1674-1137/40/7/073106}{Chin. Phys. C \textbf{40}, 073106 (2016)}.
%doi:10.1088/1674-1137/40/7/073106
%[arXiv:1510.07354 [hep-ph]].
%23 citations counted in INSPIRE as of 06 May 2024

%\cite{Harada:2019udr}
\bibitem{Harada:2019udr}
M.~Harada, Y.~R.~Liu, M.~Oka and K.~Suzuki,
Chiral effective theory of diquarks and the $U_A(1)$ anomaly,
\href{https://doi.org/10.1103/PhysRevD.101.054038}{Phys. Rev. D \textbf{101}, 054038 (2020)}.
%doi:10.1103/PhysRevD.101.054038
%[arXiv:1912.09659 [hep-ph]].
%19 citations counted in INSPIRE as of 06 May 2024

%\cite{Afonin:2014nya}
\bibitem{Afonin:2014nya}
S.~S.~Afonin and I.~V.~Pusenkov,
Universal description of radially excited heavy and light vector mesons,
\href{https://doi.org/10.1103/PhysRevD.90.094020}{Phys. Rev. D \textbf{90}, 094020 (2014)}.
%doi:10.1103/PhysRevD.90.094020
%[arXiv:1411.2390 [hep-ph]].
%64 citations counted in INSPIRE as of 05 Jan 2025

%\cite{Jia:2018vwl}
\bibitem{Jia:2018vwl}
D.~Jia and W.~C.~Dong,
Regge-like spectra of excited singly heavy mesons,
\href{https://doi.org/10.1140/epjp/i2019-12474-8}{Eur. Phys. J. Plus \textbf{134}, 123 (2019)}.
%doi:10.1140/epjp/i2019-12474-8
%[arXiv:1811.04214 [hep-ph]].
%19 citations counted in INSPIRE as of 05 Jan 2025

%\cite{Pan:2022egn}
\bibitem{Pan:2022egn}
J.~Pan and J.~H.~Pan,
Remarks on the $S$-wave masses of singly heavy mesons,
\href{https://arxiv.org/abs/2209.14948}{arXiv:2209.14948}.
%0 citations counted in INSPIRE as of 05 Jan 2025

%\cite{Chen:2017fcs}
\bibitem{Chen:2017fcs}
K.~Chen, Y.~Dong, X.~Liu, Q.~F.~L\"u and T.~Matsuki,
Regge-like relation and a universal description of heavy\textendash{}light systems,
\href{https://doi.org/10.1140/epjc/s10052-017-5512-3}{Eur. Phys. J. C \textbf{78}, 20 (2018)}.
%doi:10.1140/epjc/s10052-017-5512-3
%[arXiv:1709.07196 [hep-ph]].
%25 citations counted in INSPIRE as of 05 Jan 2025

%\cite{Jia:2019see}
\bibitem{Jia:2019see}
D.~Jia, W.~C.~Dong and A.~Hosaka,
Regge-Like Mass Relation of Singly Heavy Hadrons,
\href{https://doi.org/10.7566/JPSCP.26.022021}{JPS Conf. Proc. \textbf{26}, 022021 (2019)}.
%doi:10.7566/JPSCP.26.022021
%2 citations counted in INSPIRE as of 05 Jan 2025

%\cite{Chen:2014nyo}
\bibitem{Chen:2014nyo}
B.~Chen, K.~W.~Wei and A.~Zhang,
Investigation of $\Lambda_Q$ and $\Xi_Q$ baryons in the heavy quark-light diquark picture,
\href{https://doi.org/10.1140/epja/i2015-15082-3}{Eur. Phys. J. A \textbf{51}, 82 (2015)}.
%doi:10.1140/epja/i2015-15082-3
%[arXiv:1406.6561 [hep-ph]].
%140 citations counted in INSPIRE as of 05 Jan 2025

%\cite{Jia:2019bkr}
\bibitem{Jia:2019bkr}
D.~Jia, W.~N.~Liu and A.~Hosaka,
Regge behaviors in orbitally excited spectroscopy of charmed and bottom baryons,
\href{https://doi.org/10.1103/PhysRevD.101.034016}{Phys. Rev. D \textbf{101}, 034016 (2020)}.
%doi:10.1103/PhysRevD.101.034016
%[arXiv:1907.04958 [hep-ph]].
%40 citations counted in INSPIRE as of 05 Jan 2025

%\cite{Jia:2020vek}
\bibitem{Jia:2020vek}
D.~Jia, J.~H.~Pan and C.~Q.~Pang,
A Mixing Coupling Scheme for Spectra of Singly Heavy Baryons with Spin-1 Diquarks in P-waves,
\href{https://doi.org/10.1140/epjc/s10052-021-09205-6}{Eur. Phys. J. C \textbf{81}, 434 (2021)}.
%doi:10.1140/epjc/s10052-021-09205-6
%[arXiv:2007.01545 [hep-ph]].
%17 citations counted in INSPIRE as of 05 Jan 2025

%\cite{Jakhad:2023mni}
\bibitem{Jakhad:2023mni}
P.~Jakhad, J.~Oudichhya, K.~Gandhi and A.~K.~Rai,
Identification of newly observed singly charmed baryons using the relativistic flux tube model,
\href{https://doi.org/10.1103/PhysRevD.108.014011}{Phys. Rev. D \textbf{108}, 014011 (2023)}.
%doi:10.1103/PhysRevD.108.014011
%[arXiv:2306.06349 [hep-ph]].
%22 citations counted in INSPIRE as of 05 Jan 2025

%\cite{Jakhad:2024fin}
\bibitem{Jakhad:2024fin}
P.~Jakhad, J.~Oudichhya and A.~K.~Rai,
Interpretation of recently discovered single bottom baryons in the relativistic flux tube model,
\href{https://doi.org/10.1103/PhysRevD.110.094005}{Phys. Rev. D \textbf{110}, 094005 (2024)}.
%doi:10.1103/PhysRevD.110.094005
%[arXiv:2407.01655 [hep-ph]].
%2 citations counted in INSPIRE as of 05 Jan 2025

%\cite{Pan:2023hwt}
\bibitem{Pan:2023hwt}
J.~H.~Pan and J.~Pan,
Investigation of the mass spectra of singly heavy baryons $\Sigma_Q$, $\Xi_Q'$, and $\Omega_Q$($Q=c,b$) in the Regge trajectory model,
\href{https://doi.org/10.1103/PhysRevD.109.076010}{Phys. Rev. D \textbf{109}, 076010 (2024)}.
%doi:10.1103/PhysRevD.109.076010
%[arXiv:2308.11769 [hep-ph]].
%8 citations counted in INSPIRE as of 05 Jan 2025

%\cite{Song:2022csw}
\bibitem{Song:2022csw}
Y.~Song, D.~Jia, W.~Zhang and A.~Hosaka,
Low-lying doubly heavy baryons: Regge relation and mass scaling,
\href{https://doi.org/10.1140/epjc/s10052-022-11136-9}{Eur. Phys. J. C \textbf{83}, 1 (2023)}.
%doi:10.1140/epjc/s10052-022-11136-9
%[arXiv:2204.00363 [hep-ph]].
%85 citations counted in INSPIRE as of 05 Jan 2025

%\cite{Chen:2023cws}
\bibitem{Chen:2023cws}
J.~K.~Chen, X.~Feng and J.~Q.~Xie,
Regge trajectories for the heavy-light diquarks,
\href{https://doi.org/10.1007/JHEP10(2023)052}{JHEP \textbf{10}, 052 (2023)}.
%doi:10.1007/JHEP10(2023)052
%[arXiv:2308.02289 [hep-ph]].
%5 citations counted in INSPIRE as of 05 Jan 2025

%\cite{Cahn:2003cw}
\bibitem{Cahn:2003cw}
R.~N.~Cahn and J.~D.~Jackson,
Spin-orbit and tensor forces in heavy-quark light-quark mesons: Implications of the new $D_s$ state at 2.32 GeV,
\href{https://doi.org/10.1103/PhysRevD.68.037502}{Phys. Rev. D \textbf{68}, 037502 (2003)}.
%doi:10.1103/PhysRevD.68.037502
%[arXiv:hep-ph/0305012 [hep-ph]].
%169 citations counted in INSPIRE as of 22 Jul 2024

%\cite{Matsuki:2021eck}
\bibitem{Matsuki:2021eck}
T.~Matsuki, D.~Y.~Chen, X.~Liu and Q.~F.~L\"u,
Magic mixing angles for doubly heavy baryons,
\href{https://doi.org/10.11804/NuclPhysRev.38.2021074}{Nucl. Phys. Rev. \textbf{38}, 373-379 (2021)}.
%doi:10.11804/NuclPhysRev.38.2021074
%[arXiv:2108.05642 [hep-ph]].
%0 citations counted in INSPIRE as of 22 Jul 2024

\end{thebibliography}
\end{document}